%% file: MiningAndStateCapacity.tex
\documentclass[12pt]{article}
\usepackage{adjustbox}
\usepackage{amsfonts}
\usepackage{amsmath}
\usepackage{amssymb}
\usepackage[page]{appendix}
\usepackage{bm}
\usepackage{bbm}
\usepackage{booktabs}
\usepackage{caption}
\usepackage[capposition=top]{floatrow}
\usepackage[margin=1 in]{geometry}
\usepackage{graphicx}
\usepackage{hyperref}
\usepackage[utf8]{inputenc}
\usepackage{mathtools}
\usepackage{multirow}
\usepackage{natbib}
\usepackage{pdflscape}
\usepackage{ragged2e}
\usepackage{setspace}
\usepackage{subfigure}
\usepackage{tikz}
\usepackage{tkz-graph}
\usepackage{xcolor}
\hypersetup{
    colorlinks=true,
    allcolors=blue!60!black,
    }

\def\sym#1{\ifmmode^{#1}\else\(^{#1}\)\fi}
 
\usetikzlibrary{arrows.meta,automata,positioning}
\usetikzlibrary{shapes,arrows,snakes,backgrounds,chains}

\begin{document}
\doublespacing
\title{\vspace{-20mm}\textbf{Commodity Booms,} \\ \textbf{Local State Capacity,} \textbf{and Development}\thanks{ \footnotesize  We thank Lori Beaman, Michael Best, Jonas Hjort, Dean Karlan, Chris Moser, Gabriel Natividad, Suresh Naidu, Andrea Prat, Nancy Qian,  Edoardo Teso, Chris Udry, Silvia Vannutelli, Ebonya Washington, and participants at the Columbia Development and Political Economy Colloquia and the Northwestern Development Lunch for useful comments and suggestions.\vspace{.1cm}} 
}
\author{Dafne Murillo\thanks{Columbia University, \texttt{dm3160@columbia.edu}} \quad\hspace{2cm}Sebastian Sardon\thanks{Northwestern University, \texttt{sebastiansardon2026@u.northwestern.edu}}}
\date{\today}
\maketitle

\begin{abstract}

State capacity may shape whether natural resources generate prosperity, as it determines if windfalls are effectively turned into useful projects or wasted. We test this hypothesis studying the 2004-2011 mining boom in Peru, where mines' profits are redistributed as windfall transfers to local governments. Our empirical strategy combines geological data with the central government's mining windfalls allocation formula to identify the windfalls' effects on household incomes and other measures of economic development. Proxying local state capacity with the ability to tax and relying on a triple difference strategy we uncover significant variation in treatment response, with  positive effects of windfalls limited to high state capacity localities. We find suggestive evidence that only localities with high state capacity succeed at transforming windfalls into infrastructure stocks, which in turns contributes to structural transformation and market integration. Lastly, social unrest increases in low state capacity localities that receive windfalls but fail to perceive their benefits. Our findings underscore important complementarities between investments in extractive industries and in state capacity.
\vspace{5mm}

\end{abstract}

\clearpage
\section{Introduction}
\setcounter{footnote}{0}
Resource-rich economies such as Nigeria and Venezuela often perform poorly, suggesting that natural resources may be a curse  \citep{sachs2001curse}, however Norway and Botswana appear to owe much of their prosperity to their effective usage of natural resource windfalls \citep{torvik2009some, van2011natural, venables2016using}. Why are some countries and regions  able to harness natural resources for long-term growth while others fail to do so? A possible explanation is that the governments mismanage windfalls.  At the local level, offices tasked with spending windfall transfers may lack the capacity to handle sudden, large inflows far exceeding their usual budgets.\footnote{In developing countries, where the state often lacks the capital to extract resources themselves, these windfalls are commonly transferred from private companies to the government through windfall taxes imposed on the resource extracting firms. These can then be transferred to local governments for redistribution.} We hypothesize that local state capacity, defined as a government’s ability to implement policies \citep{besley2010state}, is key in determining whether resource wealth translates into local growth.

In this paper, we examine whether state capacity, which we proxy using local governments' tax revenue\footnote{We use ``power to tax'' as our state capacity proxy following \cite{acemoglu2005politics} and \cite{besley2010state}. By ``power to tax'', we mean the ability of governments to collect taxes, which can be measured by their tax revenue.},  mediates the effects of resource windfalls in Peru.  Our setting, a resource-rich economy where minerals account for around 70\% of national exports, offers three major advantages. First, although Peru is one of the world's leading mining exporters, this industry is confined to only a subset of the country's localities, which allows us to build, within the same nation, a control group where natural resources play a smaller role. Second, Peruvian  local governments directly receive mining windfalls and are largely autonomous in deciding how to spend them, which makes it possible to assess whether local state capacity falls short of making use of these transfers to promote development. Third, during our study period, the country experienced a dramatic commodity boom (2004-2011), which provides a compelling event-study of natural resources and development.

Specifically, we investigate the effects of windfalls transfers through which the Central Government redistributes profit taxes from mines to district governments.\footnote{Peru contains over 1,800 districts, which are the smallest administrative units with a local government. Local governments are officially known as \textit{municipalidades}.} The vast majority of Peruvian districts received at least \textit{some} positive amount of windfall transfers during the boom, though transfers are concentrated on a small group of localities around mines' locations. We define ``treated'' districts as those receiving transfers above the 75th percentile during the boom---a group to which around 90\% of transfers were allocated---and we compare how district outcomes (e.g. incomes) evolve in treated and non-treated districts before and after the onset of the mining boom. To approximately measure state capacity, we rely on the baseline (pre-boom) level of taxes per capita collected by local governments for treatment and control localities.\footnote{We also complement this with a separate measure of state capacity. Following a recent literature that emphasizes the importance of human resources in the public sector \citep{acemoglu2015state, fenizia2022managers, finan2017personnel}, we use the presence of tertiary-educated bureaucrats working in each district government.} To understand heterogeneous effects, we employ a triple difference strategy, interacting exposure to large windfalls during the boom with local state capacity, and find that in high state capacity localities, these windfalls lead to gains in household income.\footnote{In Appendix \ref{app:staggered}, we implement a staggered event study to estimate the effects of a mine opening in a locality for the first time and a district receiving large windfalls for the first time in a district. To test for heterogeneous effects by state capacity, we perform this same staggered differences-in-differences exercise splitting the sample into districts with high and low state capacity. Our heterogeneous results are robust to this alternative specification.}

A potential concern with our empirical strategy is that the location of mining windfalls is endogenous. It depends on the location of mines, which profit-maximizing companies chose depending on each locality's economic prospects and potential capacity of the local governments' to use windfalls optimally. Moreover, it depends on a progressive formula that deliberately targets poor localities. To address this, we use geological exploration data which provides geo-referenced copper and gold densities\footnote{Copper and gold are the main products mined during the boom, accounting for over 75\% of the sector's output during.} to predict windfall transfers. We argue that the density of a mineral underground is exogenous   because it was unknown at the time of district settlement. With this geological data, we predict the mining output for each district and then apply the central government’s windfall assignment formula to estimate how each district’s mining output will be redistributed as windfall transfers to itself and neighboring districts. We then re-implement our triple difference strategy with treatment assignment based on these geology-predicted windfalls.\footnote{We first predict mining output per district based on the district's geology, then combine this predicted output with the windfall policy formula to predict the amount of windfalls received by each district. Then, we define treatment status using the same 75th-percentile cutoff. Section \ref{subsec:geology} outlines how we predict each locality's yearly windfalls using district geology. Conceptually, regressions based on this ``large predicted windfalls'' treatment variable can be considered reduced form estimates of an IV strategy where predicted windfalls are used as instruments for actual windfalls.  The $F-$statistic of this instrument, obtained regressing our actual windfalls treatment dummy against the predicted windfalls dummy, is in all cases above 100.}  

Again, we find that in high state capacity localities, receiving large predicted windfalls during the boom has positive and persistent effects on household income.\footnote{Our results are robust to alternative  proxies of state capacity.} Reassuringly, both measures of windfall transfers (actual and predicted) yield consistent results.   Our findings suggest local state capacity and natural resources are complementary. 

For mechanisms, we focus on local construction of public goods, specifically roads and sewerage systems, as these are infrastructure projects managed by local governments.\footnote{We use electricity as a placebo test, as this public good is provided by the central government rather than local governments.} Treated districts with high state capacity experience considerably higher growth of roads after the boom onset.  Similarly, household access to sewerage  increases differentially for high state capacity treated districts.  Indeed, in some specifications we find that roads and sewerage \textit{only} improve in treated localities with high state capacity, which is remarkable because both high and low state capacity treated districts received similar amounts of transfers. 

Moreover, structural transformation appears to be a mechanism leading to income growth: the share of households employed in the agricultural sector falls after the mining boom in high state capacity treated localities, while the share employed in the services sector increases by a similar amount in these districts. Additionally, we find some suggestive evidence of long run price convergence in high state capacity treated districts, which we take as suggestive evidence that these localities are becoming more integrated with national markets.

Lastly, we complement this analysis with measures of social unrest using district-level protest data from the Ombudsman office. A growing literature has related resource booms and state capacity with conflicts and unrest \citep{besley2010state}, however, the direction of their relationship remains unclear. Does unrest arise because of resource booms, as argued by \cite{berman2017mine}? Is it the case that low state capacity begets social unrest when resources are found in a locality because of local governments' inability or unwillingness to redistribute? This project aims to contribute to this literature by using social unrest as an left-hand variable in the analysis. We find that social unrest increases only  in low state capacity localities that receive large windfall transfers during the mining boom, thereby creating a subnational local resource curse. 

 A limitation of our analysis is that state capacity is not randomly assigned to districts. In particular, state capacity may be correlated with unobservables that cause high economic growth and so treated localities with high state capacity may be outperforming the rest during the mining boom for reasons unrelated to the boom.  To assuage this concern, we show that high and low state capacity localities in the low windfalls (control) group seem to follow parallel trends during the mining boom. Therefore, during the mining boom, state capacity only seems to drive economic growth in treatment localities and not elsewhere. 
 
 Further, the interpretation of our results would be invalid if localities with higher state capacity receive more mining windfalls: they could be experiencing better outcomes simply because they received more resources. But we find that localities with higher state capacity have, if anything, less mining output and get smaller windfalls.

This paper mostly relates to two strands of literature. First, a growing body of research examines the role of state capacity in economic development, largely building on the theoretical work of \cite{besleypersson2009, besley2010state}. Prominent empirical contributions include \cite{acemoglu2015state}, \cite{dinceccokatz2016}, \cite{sanchez2020origins},\cite{weigel2020participation}, and \cite{bergeron2024state}. It is now well established that state capacity is an important component of development, and much of the literature is focused on which policies are  effective in building it (see for example the reviews by \citeauthor{finan2017personnel}, \citeyear{finan2017personnel}, and \citeauthor{besleyetal2022}, \citeyear{besleyetal2022}). Our contribution to this literature is showing that state capacity complements natural resources by enabling local governments to manage windfalls received during booms. This finding has an important implication: the returns to building state capacity may be especially high in resource-rich localities. Our result also echoes recent findings by \cite{moscona2023management}, who shows that international aid projects are only helpful when implemented by high-ability bureaucrats.

Second, this paper belongs to a vast literature on the relation between natural resources and economic growth. This relation remains elusive. While some studies show positive effects of resources on growth \citep{aragon2013natural, allcott2018dutch}, others find no effects \citep{caselli2013oil, aragon2023long} or even negative effects \citep{sachs2001curse, jacobsen2016economic}. These effects seem to be heterogeneous across settings.\footnote{Previous work has found largely positive impacts on growth-relevant of the latest Peruvian mining boom. Positive effects have been documented for living standards \citep{aragon2013natural, loayza2016local}, human capital \citep{aguero2021value}, credit markets \citep{garmaise2021financial}, and firms' performance \citep{garmaise2023fiscal}. However, \cite{aragon2023long} argue that the long-run effects of the boom may be null, and \cite{maldonado2023natural} show that the windfalls generated by the boom are mismanaged by many local governments.}  Although  \cite{mehlum2006institutions} argue that natural resources only cause growth in places with institutions of high quality, most of the evidence on institutions mediating the relation between natural resources and growth comes from cross-country correlations, which suffer from  identification problems.\footnote{An important exception is \cite{konte2021}, who use a triple-differences strategy similar to ours' to study the effects of mining in Africa. They rely on various assessments of  institutions' quality (e.g., citizens' corruption perceptions and distrust in the government). Their findings are that mine openings lead to the largest improvements in the state of the local economy, as  assessed by survey respondents, in localities where institutions are perceived as better-functioning.} We contribute to this literature by studying this heterogeneity of resource booms'  effects within a relatively well-identified within-country study, and by explicitly focusing on local state capacity -- an important dimension of institutions' quality that has received less attention in previous research.

The rest of the paper is organized as follows: Section \ref{sec:setting} provides an overview of our setting. Section \ref{sec:data} describes our data sources and explains how we measure key variables. Section \ref{sec:strategy} outlines our empirical strategy. Section \ref{sec:results} presents our results. Section \ref{sec:robustness} addresses potential threats to identification and presents robustness checks. Finally, Section \ref{sec:conclusion} concludes.

\section{Setting}\label{sec:setting}
\begin{figure*}[h!]
    \caption{Commodity prices and Peruvian mining exports} 
     \label{fig:exports}
     \begin{center}\subfigure[Commodity prices before and after boom]{\includegraphics[height=8cm]{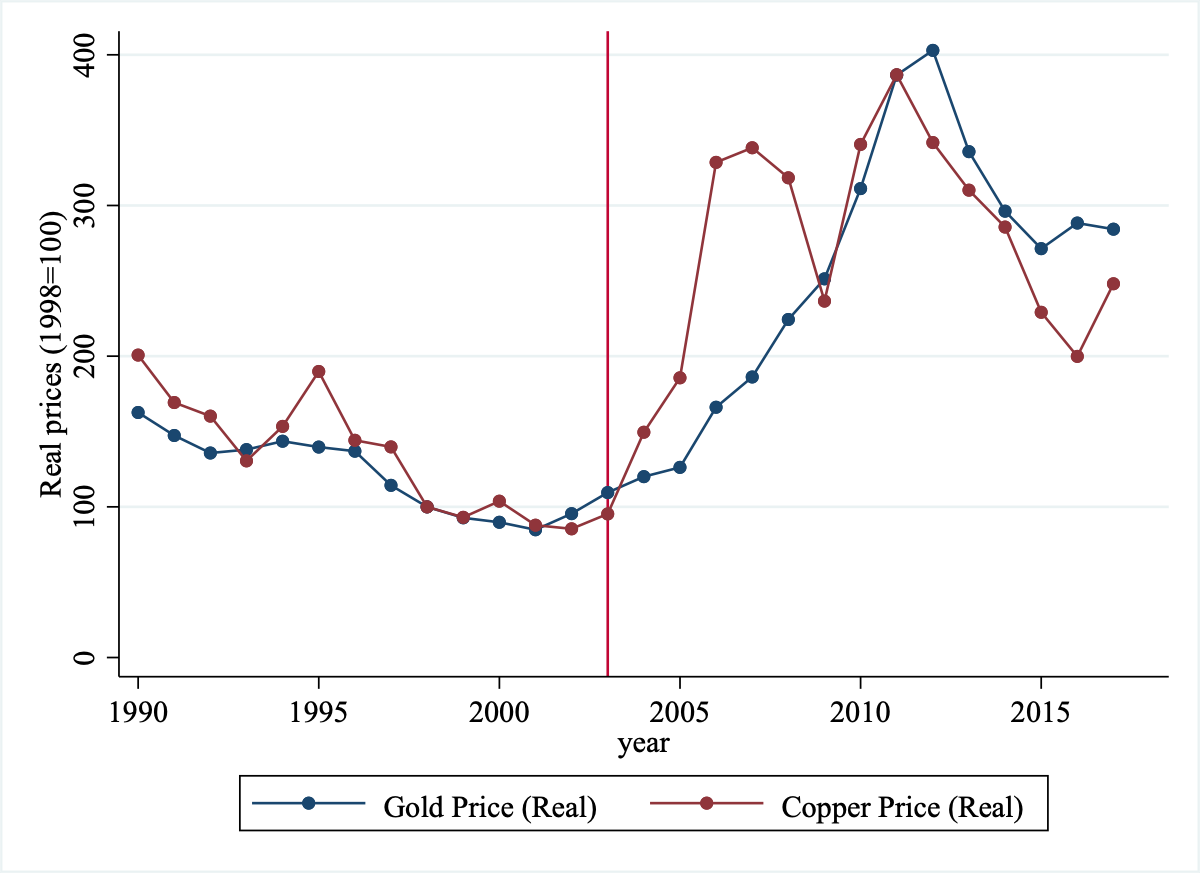}} 
     \vspace{.5cm}\centering\subfigure[Total mining exports during boom]{\includegraphics[height=8.2cm]{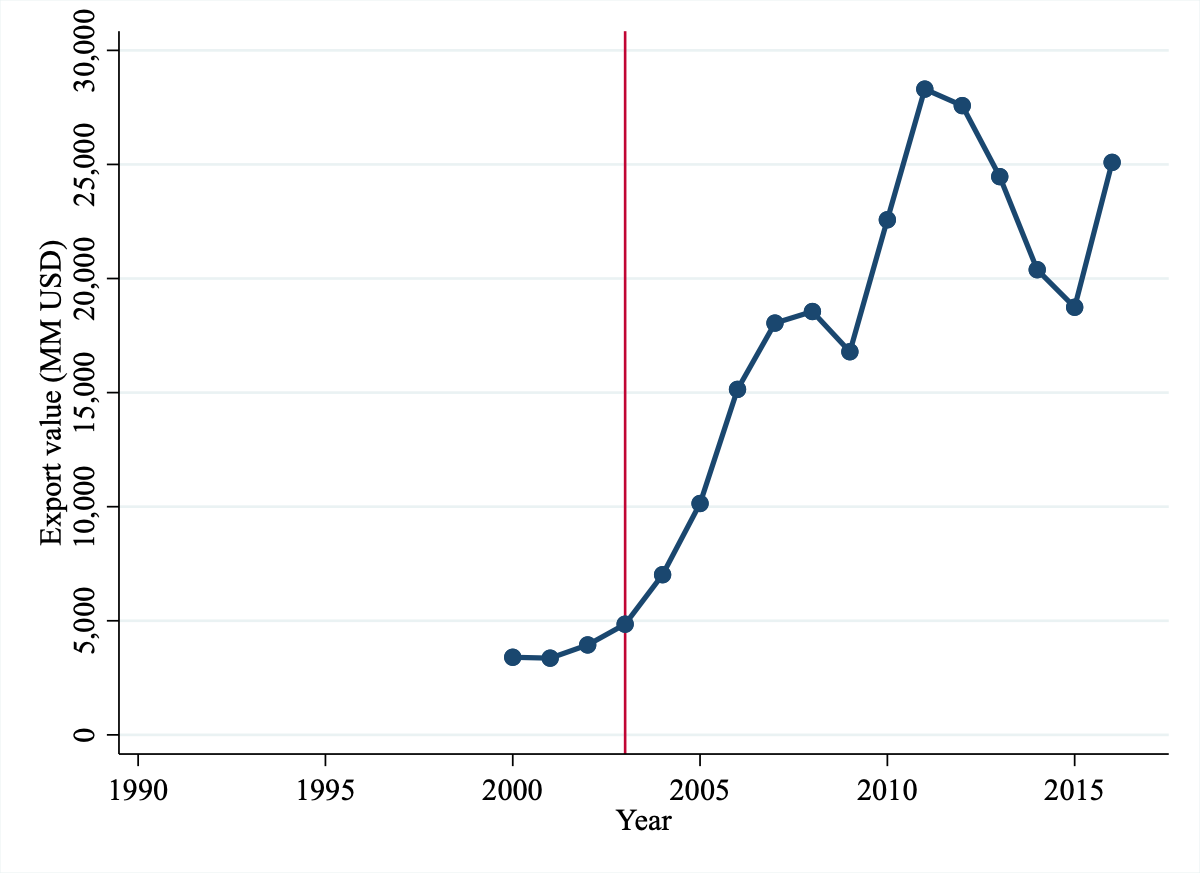}}
    \quad
    \end{center}
    \end{figure*}

At the turn of the century, Peru experienced a mining boom: the prices of Peru's main exports---copper and gold---rose by a factor of four, leading to an  increase in the value of the country's mining exports (Figure \ref{fig:exports}). The mining sector is largely comprised of large formal firms.\footnote{
These firms are often multi-million Multinational Corporations (MNCs). The sector is highly concentrated: nine mines represent 50\% of mining GDP (Figure \ref{fig:9mines}). }

Peru has a central government and is divided into 25 regions, which are analogous to states in the U.S. These contain provinces ($n=196$), which in turn contain districts ($n=1,832$).  Districts  have local governments -- each of these is run by a locally elected government which has decentralized autonomy over its budget. Local governments can invest their budget in social and economic development public work projects within their constituency. They are also in charge of collecting local taxes, as well as local contribution rates and fees.

An advantage of our setting is that local governments have limited influence over mining activities.  Local governments cannot for instance  give mines permission to open, nor can they limit, regulate, or sanction mining operations. These tasks are instead managed by the central government through the Ministry of Energy and Mines. The central government is also in charge of collecting mining windfalls and setting windfall tax rates.\footnote{See Appendix \ref{app:localgovt} for an extended outline on what local government can and cannot do.} This assuages concerns of ``good'' local governments selecting ``good'' mining companies.

Peruvian local governments play a key role in redistributing the collected mining windfalls to the local population due to the \textit{Canon Minero}, a large public investment program with the goal of promoting the sustainability of boom-driven gains. Established in a 2001 national law, the \textit{Canon} program directed 50\% of mining companies' profit taxes to local governments located inside resource-exporting regions.\footnote{See Figure \ref{fig:canonpolicy} and Section \ref{subsec:geology} for details on  the windfalls  allocation policy.} 

This program required the recipients to invest the transfers on public investment projects.\footnote{The central government, through the Ministry of Finance, is in charge of taxing the mining firms and distributing the windfall to local governments. The Ministry of Finance also closely monitors the expenditure of these transfers in an attempt to prevent politicians from pocketing the money. This monitoring might reduce the scope for shirking and corruption in the use of these windfalls by local politicians but at the same time could create red-tape. This, in turn, might affect local governments' ability to spend these transfers in public goods.} Crucially, these transfers represented over 10 times some of these local governments' taxation income, which underscores the magnitude of the resource windfalls.  The set of districts that benefit from the \textit{Canon} windfall transfers are not the same as those that actually have a mine within their borders. The \textit{Canon} policy transfers a percentage of the windfalls to all districts within the same province and region where the mine is located.\footnote{10\% goes to the districts where the natural resource is located. 25\% goes to the districts in the same province as the district with a mine. Another 40\% went to the districts sharing a region with the mining district and the remaining 25\% is transferred to the regional government. The precise amounts of windfalls are split between neighboring districts according to population and unmet basic needs weights. } The windfalls policy sought to promote infrastructure development in resource-rich localities and their neighboring communities, thereby reaping benefits after the boom ended. %

We restrict attention to the 1997-2017 period because after 2017 a separate resource boom cycle begins. It thus becomes harder to distinguish between the long-run effects of the resource boom we are studying and the short-run effects of this subsequent boom.

\section{Data}\label{sec:data}
 
Our empirical analysis relies on a dataset built from a wide array of sources. First, we obtain yearly  district-level mining production and investment from the Ministry of Energy and Mines as well as windfall transfers data from the Ministry of Finance. Second,  we obtain population counts at the district  level from the National Institute of Statistics' (INEI) official census. Third, we build state capacity measures at the district level from the Municipality Survey (RENAMU) that INEI conducts, which all local governments must comply with. Using the same survey, we obtain public goods (roads) build by each local government.  Fourth, we measure living standards, access to public goods (sewerage and electricity), and employment shares with the National Household Survey's (ENAHO) 1997-2017 waves, also published by INEI.\footnote{We obtain an alternative measure of living standards from night lights as reported by the DMSP (1992-2012) and VIIRS (2013-2017) datasets} Sixth, we rely on the Ministry of the Economy's administrative database of public investment projects undertaken by all local governments in Peru. Lastly, we obtain yearly district-level agricultural prices data from the Ministry of Agriculture.

\subsection{Mining output, windfall transfers and geology}\label{subsec:miningdata}

 \textbf{Output.} We use georeferenced mining production data available for 2001--2018 from the Ministry of Energy and Mining (MINEM). We take advantage of the within-country variation in regions' exposure to the commodity boom relying on the fact that  commodity production varies across Peru's localities  (Figure \ref{fig:map1}). Mining output, which determines windfalls' location, is particularly concentrated in a relatively small number of districts and most localities' output is exactly zero. 
 
 \textbf{Windfall transfers.} We use the Ministry of Finance's district-level data on the amount of windfall transfers from 1997-2017. The windfall policy stipulates that 50\% of all mining profits taxes get redistributed to local governments as mining windfalls. These profit taxes are collected and redistributed by the central government, using a progressive formula that prioritizes poor districts. The policy is designed so that the windfall transfers go to the district that contains the mine and its neighboring localities (those in the same province and region as the district with the mine).\footnote{Section \ref{subsec:geology} explains the windfalls allocation formula in detail.} 
 
Around 90\% of districts receive a non-zero amount of mining windfalls at some point in time.\footnote{For instance, some may receive a small amount of canon by virtue of being inside the same region as a mining producing district.} We consider a district as ``treated'' with windfalls if it receives a sizable amount of windfalls. Using a district-year balanced panel, we compute the upper quartile of per capita transfers during the resource boom. Specifically, we define \textit{treated localities} as those whose windfall transfers per capita during the 2004-2011 period are above the population-weighted 75th percentile, which received 93\% of total windfalls during the boom period (2004-2011). This cutoff is around 451 PEN per capita (approximately 133 USD).\footnote{We are using the exchange rate of 2003 --the year before the onset of the boom, which is 3.4 PEN to 1 USD. Henceforth, all monetary variables will be expressed in USD using this exchange rate.} 

\textbf{Geology.} The Peruvian Geological, Mining and Metallurgical Institute (INGEMMET), managed by MINEM, provides a large set of over 20,000 copper and gold density measures scattered across the country (parts per million -ppm). These are collected from sampling soil and sediment. Figure \ref{fig:mapgeology} shows how this variable varies within one particular region in Southern Peru. Each point on the map shows variation in mineral density ppm, across measurement sites. We use this geological data  to predict mining output per district and combine this with the windfall allocation formula to construct geology-predicted transfers per district, which will play the role of an instrumental variable. We validate that predicted and actual windfalls are strongly correlated using the windfalls data described above. We then use this instrument to build an additional treatment assignment dummy, classifying districts with predicted windfall transfers above the 75th percentile as treated.

\subsection{State capacity}\label{subsec:SC}
We obtain state capacity proxies from the Municipality Survey (RENAMU), available for the years 1994, 1997, and over the 2002-2022 period. Our main proxy is the total tax collected by local governments during the baseline (pre-boom) period.\footnote{To complement this measure, we build a baseline local human resources variable  using the number of tertiary-educated employees in the local government staff, which we use for robustness checks. We divide local bureaucracy size by local population size. Figure \ref{fig:corr} plots our relative tax state capacity proxy against our bureaucracy proxy. Reassuringly, there is a strong positive correlation between two state capacity proxies we will use. }  To standardize our proxy, we divide tax by local population. We obtain \textit{baseline} state capacity proxies from the 2002-2003 waves of the RENAMU, which measures  each government's  revenue sources.\footnote{In this survey, we can observe the total incomes collected by the local government. We thus pinpoint three types of taxes collected by the local government (\textit{local taxes and local contribution rates.}), there by disentangling our proxy from other income generating activities that included rents, fines, and sales of capital and financial assets --which are less directly related to the state's ability to tax. We focus on  2002-2003 instead of using the 1994 and 1996 waves because these 90s surveys had  many zero values for our outcomes of interest -- many of which were missing values miscoded as zero. Moreover, these zeros were not evenly distributed across treatment and control groups: $\sim$44\% districts were missing or zeros for the control group and $\sim$25\% for the treatment group. We therefore use instead the 2002-2003 Municipality Survey, which has almost complete district coverage ($\geq$96\% of districts have non-missing above zero values).} 

We classify localities into high state capacity and low state capacity groups, based on whether districts score above or below the (population weighted) median on each measure.\footnote{Figure \ref{fig:determinants} shows  some determinants for our  proxy, as measured by bivariate regressions of Z-scores. We validate our state capacity proxy using post-boom public investment data in Appendix \ref{subsec:SC_validate}. This validation assumes that our state capacity measure is proxying the government's ability to implement public work projects and therefore should be correlated with higher public expenditure. } 

\subsection{Living standards}

We rely on the ENAHO household survey, available over 1997--2022, and focus on the 1997--2016 period. Our main outcome variable is real household income, which intends to capture changes in living standards.  This survey also includes a wide array of observable characteristics available, which we use as controls. Moreover, we use the survey sections on employment to classify households by industry depending on the occupation of the household head. Such industry shares allow us to test whether structural transformation is happening more rapidly in mining areas. 

\subsection{Public goods data}\label{subsec:invdata}

We focus on the provision of two main public goods: roads and sewage. First, using the RENAMU, we obtain the number of square meters of roads built every year by the local government. Using this data, we compute the stock of roads per district as our main outcome variable for public goods. Second, from the ENAHO, we obtain the share of households that report having access to sewerage, a public good facilitated by the district government.  The windfalls program described in Section \ref{sec:setting} forces recipients of windfalls to invest these in infrastructure, so we expect this outcome to improve among treated districts if such investments are successful. As a placebo outcome, we also use the share of households with access to electricity from the ENAHO, since this is a public good that is provided by the central government.

\subsection{Social unrest data}\label{subsec:conflictdata}

We obtain the number of protests per district each year from the Peruvian Ombudsman Office records for the 2004-2022 period.\footnote{The dataset was provided under a confidentiality agreement and is currently not publicly available.} The Ombudsman Office registers every instance of social unrest occurring in Peru since 2004. According to their website, the Ombudsman defines these events as processes in which sectors of society or the State ``are in conflict'', potentially escalating into violence. The complexity of these ``conflicts'' is influenced by factors such as the number of parties involved, cultural, economic, social, and political diversity, the types of violence that may emerge, or institutional weaknesses in managing the situation, among other elements \citep{defensoria2020}. They also track and record the events' length and resolution. These social unrest episodes are categorized according to the reason they occurred. Over half of the events in this dataset are related to natural resources.

\subsection{Summary statistics}

Table \ref{tab:balance} shows balance tests between the treatment and control group and between high and low state capacity districts for the pre-boom (1997-2003) period.\footnote{Table \ref{tab:ds} shows summary statistics (e.g. mean, standard deviation and number of observations) for the outcome variables (living standards) and the state capacity proxies we work with throughout our study period (1997-2017), as well as some controls (head of household's age, education, and gender) and potential mechanisms (public goods, sector employment).} Prior to the boom, treated localities were smaller in terms of their population and had lower living standards, as proxied by real household income.  They also score lower in the tax per capita state capacity proxy  described above. That said, local taxation per capita seems  low across the board: the average district had an average tax revenue of  around 14 USD per capita, which is roughly equivalent to 1\% of per capita incomes. This might reflect the large informal economy of Peru, which accounted for over 80\% of the labor force in the 1990s.

Treated districts, as expected, received more windfall transfers than the control group even before the boom began; however, the transfer amounts before the boom were quite small: 1.2 USD per capita on average, compared to around 22 USD per capita in the post-boom period. During this  2004-2017 timeframe, treated districts receive on average 80 USD per capita while control districts receive a mean amount close to  average pre-boom levels of windfalls (2.7 USD per capita).

High state capacity localities receive on average less windfalls, both before and after the mining boom onset. By construction, high state capacity local governments collect more local taxes (26 USD per capita) than low state capacity districts (1 USD per capita) in the pre-boom period. We verify that the ranking of high and low state capacity localities is constant. Figure \ref{fig:sc_time} plots the time series of our state capacity measurement over time. We observe that those districts that started off as high/low state capacity remain high/low over time. This is suggestive of a low state capacity trap: those places that start off with low state capacity in the pre-boom period, remain so during the boom.\footnote{Note that district tax revenue in the pre- and post-boom period are highly correlated ($R^2\geq 0.6$), so the localities that score above on these baseline measures are more likely to have a high-quality state capacity at the start of the windfalls' period. }

\section{Empirical Strategy}\label{sec:strategy}

\subsection{Triple difference (OLS)}\label{subsec:ols}

We divide our  sample into three time periods: the \textit{pre-boom period} runs from 1997 to 2002, the \textit{base period} (2003), and the \textit{post-boom period} includes 2004-2016, which will capture both short and longer run effects of the mining boom's windfalls. 

State capacity is, as discussed in Section \ref{subsec:SC}, measured by tax per capita.\footnote{In Section \ref{sec:robustness} we show our results are robust to alternative proxies of state capacity, including a measure of bureaucracy quality: the per capita local government employees with tertiary education.} We  use our state capacity proxy  to divide districts into two groups: those with above-median baseline (pre-boom) state capacity and their below-median counterparts.  

 We estimate the triple-differences equation (\ref{eq:HTE_TWFE}), where the outcome variable $Y_{idt}$ denote log incomes for household $i$ in district $d$ during period $t$ for our main specification. The treatment dummy $W_{d}$ equals 1 if a district receives ``large'' windfalls during the boom (above the population-weighted 75 percentile). We construct $W_{d}$ using actual windfall transfers. Figure \ref{fig:treatment_map} shows how the treatment status looks on the map.\footnote{This figure shows the spatial distribution of mining output and windfalls during the mining boom (2004-2011). As can be seen in Panel \ref{fig:treatment_map}(a), only a few districts in Peru have mining production. However, because of the windfall policy (Figure \ref{fig:canonpolicy}) allocates windfalls not only to the district that contains the mine (and therefore have mining output) but also to its province and regional neighbors, almost all districts in the country perceive some amount of windfalls. This can be seen in Panel \ref{fig:treatment_map}(b), where only the Amazon regions in east of the country have 0 windfalls. All other districts have non-zero windfall transfers during the boom. However, there is variation in how much windfalls districts receive, with the highest receivers obtaining over 300 USD per capita transfers, while the lowest receiving under 50 USD per capita. Panel \ref{fig:treatment_map}(c) shows the treatment status of each district according to our treatment definition: districts that perceive per capita windfalls above the 75th percentile during the mining boom. As can be observed, in many cases, entire regions are treated. This is unsurprising given the windfall allocation policy, as a single district with a very profitable mine within its borders will result in large windfalls being allocated to its province and regional neighbors according to the windfall policy. Because treatment status varies relatively little within regions, our specifications with region-year fixed effects ought to be interpreted with caution.  } 
 
 The treatment dummy is interacted with a post-2003 dummy, as well as a state capacity dummy $I^{SC}_{d}$ equal to one if district $d$'s local government collected above-median taxes per capita before the boom. This leads to the following estimating equation:
\begin{align}
     Y_{idt}=\alpha&+\beta_{Post}^W \big(W_{ d}\times \bm{1}_{[t>2003]}\big)+\beta_{Post}^I\big(I^{SC}_{d}\times \bm{1}_{[t>2003]}\big)+\beta_{{Post}}^{WI}\big(W_{d}\times I^{SC}_{d}\times \bm{1}_{[t>2003]}\big) \notag\\
     &+\beta_{Pre}^W \big(W_{ d}\times \bm{1}_{[t\leq 2002]}\big)+\beta_{Pre}^I\big(I^{SC}_{d}\times \bm{1}_{[t\leq 2002]}\big)+\beta_{{Pre}}^{WI}\big(W_{d}\times I^{SC}_{d}\times \bm{1}_{[t\leq 2002]}\big) \label{eq:HTE_TWFE}\\
     &+X'_{idt}\bm{\gamma} + \phi_{d}  + \phi_{t} + \varepsilon_{idt}\notag
\end{align}
Equation (\ref{eq:HTE_TWFE}) has district and time fixed effects, $\phi_{d}$ and $\phi_{t}$, and a vector of covariates $X_{idt}$ including household heads' gender and age (squared), as well as districts' baseline poverty (measured using districts' unsatisfied basic needs index) and altitude interacted with time fixed effects. To test for pre-trends, we include the treatment and state capacity dummies  interacted with a pre-boom dummy equal to 1 for years 1997-2002. In some specifications, we include administrative and geographic region-time fixed effects instead of time fixed effects.\footnote{Peru has 25 administrative regions known as \textit{departamentos} and 8 natural geographic regions known as \textit{dominios} created by INEI based on common topographical and meteorological characteristics.}

We denote the mining boom's large windfalls on low state capacity districts by $\beta_{Post}^{W}$ and the mining boom's windfalls additional effect on high state capacity districts by $\beta_{Post}^{WI}$.   Equation (\ref{eq:HTE_TWFE}) will also be augmented into a specification where these coefficients all vary year-to-year and where we also test for parallel trends in the pre-period.\footnote{Specifically, we will rely on the following event study specification: \begin{equation}\label{eq:TWFE_byyear}
    Y_{idt} = \alpha +  \sum_{h\neq 2003}\beta_{h}^{W}\big( W_{d} \times\bm{1}_{[t=h]}\big) +  \sum_{h\neq 2003}\beta_{h}^{WI}\big( W_{d}\times I^{SC}_d \times\bm{1}_{[t=h]}\big)+   \mathbf{X}^{'}_{idt}\gamma + \phi_{d} + \phi_{t} + \varepsilon_{idt}
\end{equation} for all years $ h\in[1997, 2016]$. $\mathbf{X}^{'}_{idt}$ includes  baseline state capacity interacted with time fixed effects, as well as the covariates discussed above. The remaining terms are defined as in equation (\ref{eq:HTE_TWFE}). } 

This strategy successfully recovers the effect of the mining boom's windfall transfers under the assumption that, absent the boom, treatment and control districts would have evolved in a similar fashion.\footnote{This assumption might seem strong, but, reassuringly, before the boom (i.e. prior to 2004) treatment and control districts were evolving along roughly parallel trends, which makes it plausible that they would have continued to do so absent the boom. We use equation (\ref{eq:TWFE_byyear} to verify that this is the case for the pre-boom years.} An additional assumption required by this strategy is the stable unit treatment value assumption (SUTVA), which, in this setting, requires assuming that the boom \textit{only} affects living standards in the high windfalls localities. Ex-ante, if the mining boom's windfalls raise living standards in treatment districts, we could expect SUTVA violations to manifest as positive effects in control (low windfalls) districts because these localities may trade with each other and so when one becomes richer so do its neighbors. Under effects of this type, our  estimates would only provide a lower bound on the total effects of the windfalls policy.

\subsection{Triple difference (Geology IV)}\label{subsec:geology}

During the mining boom, the presence of mining operations might be endogenous. Mining firms could be selecting districts with the best development outcomes to start their operations. Since windfalls assignment is largely based on which district has a mine, this means that our treatment districts as defined in Section \ref{subsec:ols} might be positively selected. Ideally, we would randomly assign windfalls to districts to study their impact, but this is not feasible. However, the presence of mining depends on geological suitability, which is exogenous to development outcomes. Different districts have varying levels of copper and gold density in their mountains and soil. This metal density is given and not influenced by economic activity. 

Recent explorations conducted by the Peruvian Geological, Mining, and Metallurgical Institute (INGEMMET) measured gold and copper density by taking soil and rock samples across the country. Thus, we can use metal density to predict mining output, then combine this with the windfall allocation formula to assign predicted windfalls to districts. This predicted windfalls treatment assignment serves as an instrumental variable (IV) for actual windfalls treatment assignment therefore we can interpret results using this strategy as reduced form estimates of an IV strategy. This design is similar in spirit to the one used by \cite{cassidy2019long} to study the effects of oil wealth. 

 We  verify that a district's suitability for mining is closely linked to the metal wealth of the \textit{richest} areas of its territory, as that is where mining companies chose to operate.\footnote{We focus attention on the highest ppm score of each district, since a district does not need to score high across all its territory for a mine to operate --the existence of a suitable area suffices.} Thus we measure each district's suitability for mining as the 90-th percentile of metal density among all the INGEMMET measures available for that district. 

Figure \ref{fig:1ststage} plots the kernel densities of mining and non-mining localities over mineral density, suggesting that districts with mineral-rich geologies are more likely to have active mines during our study period. 

Given the we have established the relationship between geological features and mining output, we use geology to predict mining output for each locality through Ordinary Least Squares (OLS). We first run (\ref{eq:IV1_1})
\begin{equation}\label{eq:IV1_1}
    \log m_{dt} = \theta + \beta^{G}_t \log\text{Geology}^{p90}_{d} + \varepsilon_{dt} 
\end{equation}
where $m_{dt}$ represents the real mining output in district $d$ in year $t$, $\text{Geology}^{p90}_{d}$ denotes district $d$'s 90th percentile copper and gold density measured in parts per million (ppm). Table \ref{tab:firststage} confirms that, as expected, mineral wealth predicts actual mining activity, with an F-statistic of around 100.

 Then we use the estimated coefficients $\widehat{\beta}^G_t$ from (\ref{eq:IV1_1}) to predict mining output in district $d$ in year $t$: $ \widehat{m}_{dt}= \exp\big(\theta + \widehat{\beta}^G_t\log\text{Geology}^{p90}_d\big).$

 We will use this predicted output, $\widehat{m}_{dt}$, combined with the windfall allocation formula (Figure \ref{fig:canonpolicy}) to estimate the windfalls that will be distributed among a locality and its neighbors. We first need to map output onto windfall rents.\footnote{Note that the windfalls \textit{rents} from each mine, collected by the central government by taxing mining profits, do not correspond one-to-one to the windfall \textit{transfers} received by each district, because transfer allocation depends on the policy formula. } Since know the actual output ($m_{dt}$) and windfall transfers  ($w_{dt}$) per district $d$ in year $t$,  we can infer the parameter $\alpha_t$ that maps total output $m_t=\sum_d m_{td}$ onto total windfall rents $R_t$ for every year $t$. As shown in Figure \ref{fig:canonpolicy}, the policy distributes 10\% of the windfalls rents obtained from any mine to the district where the mine is located, 25\% of these go to districts in the province where the mine is located and 40\%  to the districts that share a region with the mine. The 25\% that goes to the regional government does not show up in our district windfall transfers data, so we have $\sum_d w_{dt}=.75R_t$. Putting this together we find: $    \alpha_t = \frac{R_{t}}{m_{t}}=\frac{\frac{4}{3}\sum_d{w_{dt}}}{\sum_d{m_{dt}}}
$.

Assuming that this $\alpha_t$ parameter that maps national rents to national output also approximately maps district rents to district output we can define predicted rents generated from each district $d$ in year $t$ as: $\widehat{R}_{dt} = \alpha_{t}\widehat{m}_{dt}$.

As the next step, we  use the windfall formula to distribute predicted rents ($\widehat{R}_{dt}$), which yields the local predicted windfall ($\widehat{w_{dt}}$). Each district $d$ that generates $\widehat{R_{dt}}$  allocates 10\% of $\widehat{R_{dt}}$ to district $d$. Moreover, 25\% of $\widehat{R_{dt}}$ goes districts in its same province $p(d)$, with the total predicted windfall allocated to $p(d)$ given by $\widehat{w}_{p(d)t}=.25\sum_{j\in p(d)}{\widehat{R}_{jt}}\text{ for all districts } j\in p(d)$.

Additionally, 40\% of $\widehat{R_{dt}}$ goes to districts in the same region $r(d)$, with the total predicted windfall allocated to $r(d)$ given by $\widehat{w}_{r(d)t}=.4\sum_{j\in r(d)}{\widehat{R}_{jt}}\text{ for all districts } j\in r(d)$.

We use the windfall formula to determine how districts get a share of their province and region's allocated windfall based on a weight determined by their unsatisfied basic needs ($\text{UBN}_d$) index and population ($\text{Pop}_{dt}$).\footnote{The weight is computed as a ratio of the population of the locality $d$ where the mine is located times its UBN divided by the sum of this interaction for all districts $j$ in the province (or region) $p(d)$ (or $r(d)$) where the mine is located. UBN measures are taken from the 1993-2007 Census since these are the only instances in which this index is recoreded for all districts. We rely on the 1993 UBN index to construct the instrument since the 2007 one will already be affected by the mining boom.} The local predicted windfalls for district $d$ in year $t$ are given by:
\begin{equation*}
    \widehat{w}_{dt} = .1\widehat{R}_{dt} + \widehat{w}_{p(d)t}\left(\frac{\text{Pop}_{dt} \times \text{UBN}_{d}}{\sum_{j\in p(d)}[\text{Pop}_{jt} \times \text{UBN}_{j}]}\right) + \widehat{w}_{r(d)t}\left(\frac{\text{Pop}_{dt} \times \text{UBN}_{d}}{\sum_{j\in r(d)}[\text{Pop}_{jt} \times \text{UBN}_{j}]}\right)
\end{equation*}

Figure \ref{fig:scatter_pred} shows the correlation between actual (true) windfalls $w_{dt}$ and predicted windfalls $\widehat w_{dt}$. Let $\widehat W_{d}$ be our new treatment dummy constructed from predicted windfalls, which equals 1 if district $d$ has a predicted windfall $\widehat{w_{dt}}$ greater than the 75th percentile, and 0 otherwise.\footnote{Using the 75th percentile cutoff ensures we have a similar number of districts in the treatment and control groups as in our OLS specification.} Table \ref{tab:firstIV} shows the first stage of our  reduced form geology IV, resulting from estimating equations (\ref{eq:1S}) and (\ref{eq:1S_2}):
\begin{equation}\label{eq:1S}
    \log(\widehat w_{dt}) = \theta + \beta\log(w_{dt}) +  \sum_h \gamma_h\big(\text{UBN}_d \times\bm{1}_{[t=h]} \big) + \phi_{r} + \varepsilon_{dt}
\end{equation}
\begin{equation}\label{eq:1S_2}
    \widehat W_{d} = \theta + \beta W_{d} +  \sum_h \gamma_h\big(\text{UBN}_d \times\bm{1}_{[t=h]} \big) + \phi_{r} + \varepsilon_{dt}
\end{equation}
where we regress actual windfalls on predicted windfalls (in logarithms) and  the treatment dummy specified in Section \ref{subsec:ols} (i.e. ``large'' actual windfalls, above the 75th percentile) on the new treatment dummy (``large'' predicted windfalls). We control for each district's 1993 unsatisfied basic needs index (UBN$_d$) interacted with time $h\in [1997, 2016]$ and add region-time fixed effects $\phi_{rt}$. Table \ref{tab:firstIV} Columns (1) and (2) show the coefficients $\beta$ estimated from (\ref{eq:1S}) while Columns (3) and (4) show results for (\ref{eq:1S_2}). Even numbered columns include region-time fixed effects. We can interpret these results as our instrument's first stage.\footnote{The F-statistic in all cases except Column (4) exceeds 100, suggesting that the instrument is sufficiently strong.}

Finally, we estimate the effect of local windfalls on local development by implementing the  specification (\ref{eq:HTE_TWFE}),  but using as treatment dummy $\widehat W_{d}$ instead of $W_{d}$. Thus, we assign  treatment status to districts based on their predicted instead of actual windfalls per capita.\footnote{This equation looks like this: \begin{align}
     Y_{idt}=\alpha&+\beta_{Post}^{W} \big(\widehat{W}_{ d}\times \bm{1}_{[t>2003]}\big)+\beta_{Post}^I\big(I^{SC}_{d}\times \bm{1}_{[t>2003]}\big)+\beta_{{Post}}^{WI}\big(\widehat W_{d}\times I^{SC}_{d}\times \bm{1}_{[t>2003]}\big) \notag\\
     &+\beta_{Pre}^{W} \big(\widehat W_{ d}\times \bm{1}_{[t\leq 2002]}\big)+\beta_{Pre}^I\big(I^{SC}_{d}\times \bm{1}_{[t\leq 2002]}\big)+\beta_{{Pre}}^{ WI}\big(\widehat W_{d}\times I^{SC}_{d}\times \bm{1}_{[t\leq 2002]}\big) \label{eq:HTE_TWFE_IV}\\
     &+X'_{idt}\bm{\gamma} + \phi_{d}  + \phi_{t} + \varepsilon_{idt}\notag
\end{align}} We also estimate a version of the by-year event study specified in equation \ref{eq:TWFE_byyear} but using this reduced form IV  treatment variable.\footnote{This event study specification looks like equation (\ref{eq:TWFE_byyear}) but with $\widehat{W}_d$ as treatment: \begin{equation}\label{eq:TWFE_byyear_IV}
    Y_{idt} = \alpha +  \sum_{h\neq 2003}\beta_{h}^{W}\big( \widehat W_{d} \times\bm{1}_{[t=h]}\big) +  \sum_{h\neq 2003}\beta_{h}^{WI}\big( \widehat W_{d}\times I^{SC}_d \times\bm{1}_{[t=h]}\big)+   \mathbf{X}^{'}_{idt}\gamma + \phi_{d} + \phi_{t} + \varepsilon_{idt}
\end{equation} for all $h$ in $[1997, 2016]$.}

\vspace{.3cm}

\section{Results}\label{sec:results}

\subsection{Living standards}
 Table \ref{tab:main_hte} shows our results for living standards, measured in log household income per capita. For Columns (1) and (2), we use  the ``large'' windfalls treatment computed from actual windfall transfers during the mining boom (2004-2011) while for Columns (3) and (4) we utilize as treatment assignment the predicted windfalls from geology  as outlined in Section \ref{subsec:geology}. With both treatment definitions, we estimate a positive and significant triple interaction coefficient $\beta_{Post}^{WI}$ and a negative double interaction coefficient $\beta^W_{Post}$.\footnote{The  double interaction betas are  smaller than those for the triple difference in all but one  specification. For Columns (1) and (2) we estimate equation (\ref{eq:HTE_TWFE}) while for (3) and (4) we estimate regression (\ref{eq:HTE_TWFE_IV}). Even-numbered columns include region-time fixed effects.} Our results suggest gains in household income from exposure to mining boom windfalls in districts with high state capacity, and losses elsewhere.

 Figure \ref{fig:results_logy_HH} shows year-by-year event study coefficients of a modified specification (\ref{eq:TWFE_byyear}) where instead of having three periods we set use 2003 as the base year and interact  year fixed effects with treatment and high state capacity dummies to estimate year-by-year $\beta^{WI}_t$ and $\beta^W_t$ coefficients. The results shown in Panels (a) and (b) in this event study figure are comparable to  those in Table \ref{tab:main_hte} Columns (1) and (3).

\subsection{Mechanisms}\label{subsec:mechanisms}

We propose public goods as the main mechanism driving our results. Specifically, we hypothesize that high state capacity localities will be more able to channel the gains (e.g. mining windfalls) from the boom into better public expenditure, in particular, public goods that will promote development. Thereby, the local government will be able to redistribute the benefits of the mining boom back to their constituency. These hypotheses are illustrated in Figure \ref{fig:tikz}. 

\begin{figure}[h!]
    \centering
    \includegraphics[height=8cm,trim={0cm .5cm 0cm 1.5cm},clip]{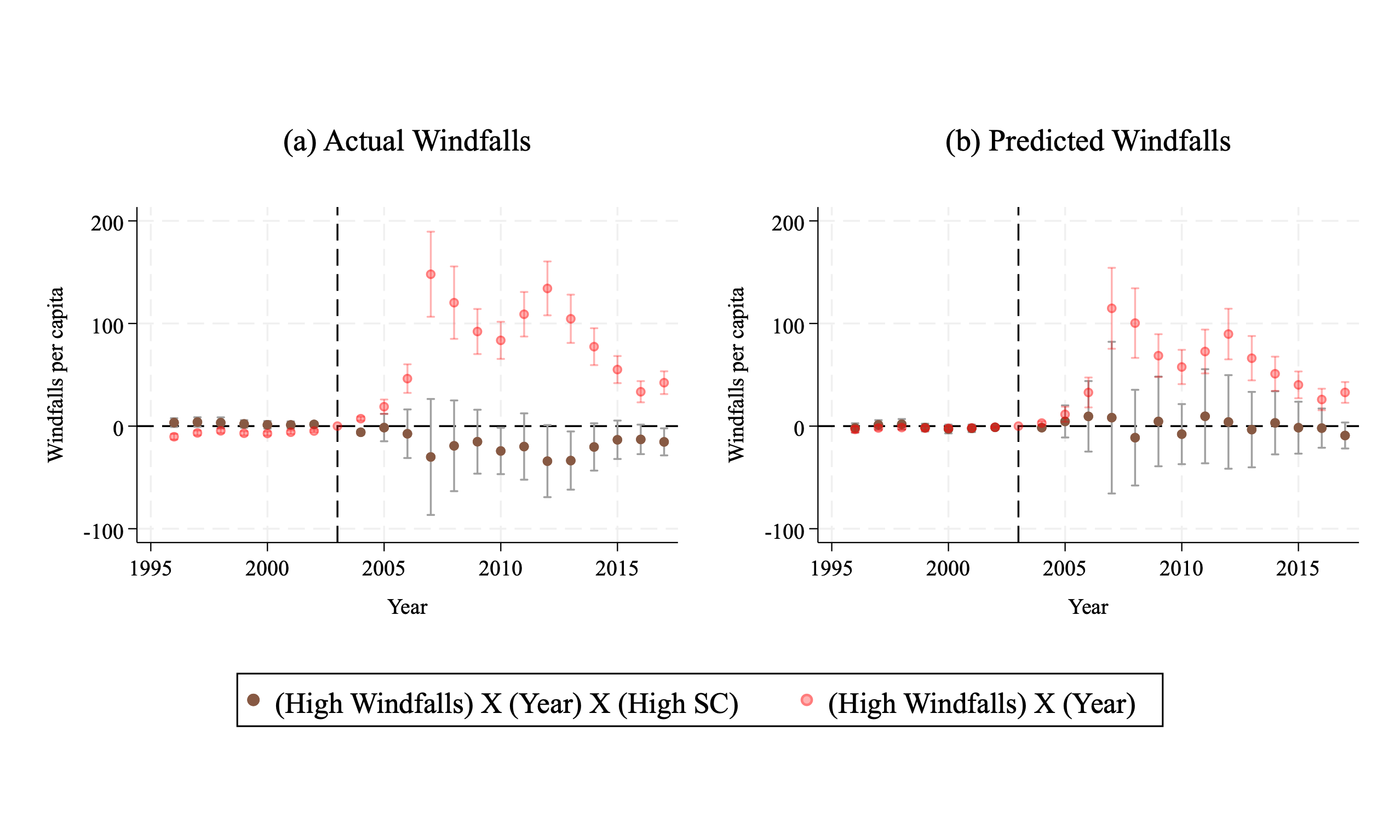}
    \caption{Windfalls per capita (USD)}
    \label{fig:mecha_canon}
        \vspace{-.6cm}
     \begin{changemargin}{0.8cm}{0.6cm} 
\begin{spacing}{0.125}
{\footnotesize \setstretch{.125} \textit{High Windfalls}: locality with large (above the 75th percentile) windfall transfers during boom. \textit{High State Capacity (SC)}: locality with above median tax per capita collected in pre-boom period.}  
\end{spacing}
\end{changemargin}
\end{figure}

Tables \ref{tab:mecha1} and \ref{tab:mecha2} show our results from estimating (\ref{eq:HTE_TWFE}) to test for these mechanisms behind our results on household income.\footnote{Many of the outcome variables in this section come from district-year panel data, instead of the ENAHO household survey. For outcomes at the district-level instead of at the household-level, we run a modified version of equation (\ref{eq:HTE_TWFE}), where we remove the $i$ subscript, e.g: \begin{align}
     Y_{dt}=\alpha&+\beta_{Post}^{W} \big({W}_{ d}\times \bm{1}_{[t>2003]}\big)+\beta_{Post}^I\big(I^{SC}_{d}\times \bm{1}_{[t>2003]}\big)+\beta_{{Post}}^{WI}\big( W_{d}\times I^{SC}_{d}\times \bm{1}_{[t>2003]}\big) \notag\\
     &+\beta_{Pre}^{W} \big( W_{ d}\times \bm{1}_{[t\leq 2002]}\big)+\beta_{Pre}^I\big(I^{SC}_{d}\times \bm{1}_{[t\leq 2002]}\big)+\beta_{{Pre}}^{ WI}\big( W_{d}\times I^{SC}_{d}\times \bm{1}_{[t\leq 2002]}\big) \label{eq:HTE_TWFE_district}\\
     &+X'_{dt}\bm{\gamma} + \phi_{d}  + \phi_{t} + \varepsilon_{dt}\notag
\end{align}} Odd numbered columns use the actual windfalls treatment; even numbered columns use predicted windfalls. The mining windfalls policy aimed to provide large transfers to be invested on public work projects, which in turn were expected to bring about local development. We first verify that windfalls transfers are not larger for high state capacity districts by using windfalls per capita as our outcome variable, to ensure that heterogeneous effects by state capacity are not driven by high state capacity treated districts simply receiving bigger transfers than their low state capacity counterparts.  We find that both high and low state capacity localities receive approximately the same amount of windfall transfers, as the triple interaction coefficients are statistically insignificant in Table \ref{tab:mecha1}, Columns 1 and 2.\footnote{Figure \ref{fig:mecha_canon} panel (a) and (b) show our dynamic specification's coefficients comparable to Table \ref{tab:mecha1} Columns 1 and 2 respectively.} This assuages our concern that heterogeneous sizes of mining windfalls are mechanically driving our results.\footnote{Figure \ref{fig:shocksize} plots the raw mining output and windfall transfers per capita data and shows that, if anything, high state capacity districts have less mining  production and receive fewer windfalls, relative to the size of their population. } 

\begin{figure}[h!]
    \centering
    \includegraphics[height=8cm,trim={0cm .5cm 0cm 1.5cm},clip]{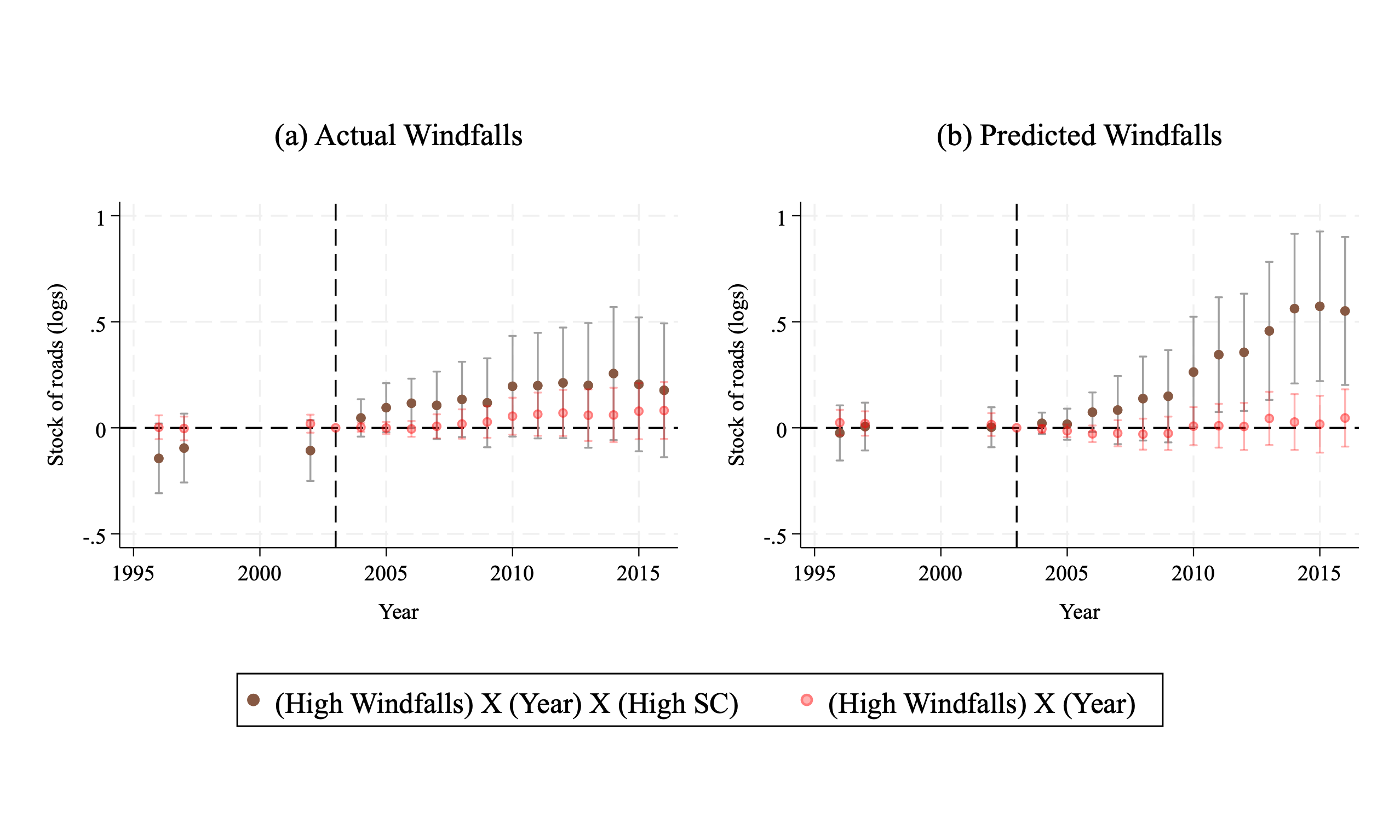}
    \caption{Stock of Roads (log cumulative investment per capita) }
    \label{fig:mecha_roads}
    \vspace{-.6cm}
   \begin{changemargin}{0.8cm}{0.6cm} 
\begin{spacing}{0.125}
{\footnotesize \setstretch{.125} \textit{High Windfalls}: locality with large (above the 75th percentile) windfall transfers during boom. \textit{High State Capacity (SC)}: locality with above median tax per capita collected in pre-boom period.}  
\end{spacing}
\end{changemargin}
\end{figure}

We then look at specific public goods that could plausibly be related to increases in income. Transportation projects have the largest share on local governments' investment budgets\footnote{See Figure \ref{fig:obratypes}.}, so we first investigate the windfalls' effects on each district's stock of roads constructed by the local government.\footnote{We obtain amount of square meters built every year by the local government from the Municipality Surveys of 1996-1997 and 2002-2016.} Table \ref{tab:mecha1}, Columns 3 and 4 show our results: using cumulative stock of roads per capita as outcome, we find that this public good increases significantly in treated high state capacity districts.\footnote{See figure \ref{fig:mecha_roads} panels (a) and (b) for corresponding dynamic specifications results.} We then explore sewage provision, which the local government is in charge of providing, using as the outcome variable a dummy on whether the household has access to sewage at home, also built from the household survey with which we measure incomes.\footnote{Sewerage access can be related to inputs of worker productivity such as health as well as reduce household healthcare costs \citep{galianietal2005, bancalari2024can}} We find positive and significant windfall effects on sewage provision, which are much larger in high state capacity districts (Table \ref{tab:mecha1}: Columns 5 and 6). Finally, as a placebo test, we use as outcome an analogous indicator for households' access to electricity, which is a public good provided by the central rather than the local government. We find positive but not significant effects on this outcome variable. 

\begin{figure}[h!]
    \centering
    \includegraphics[height=6cm,trim={0cm 0cm 0cm 9cm},clip]{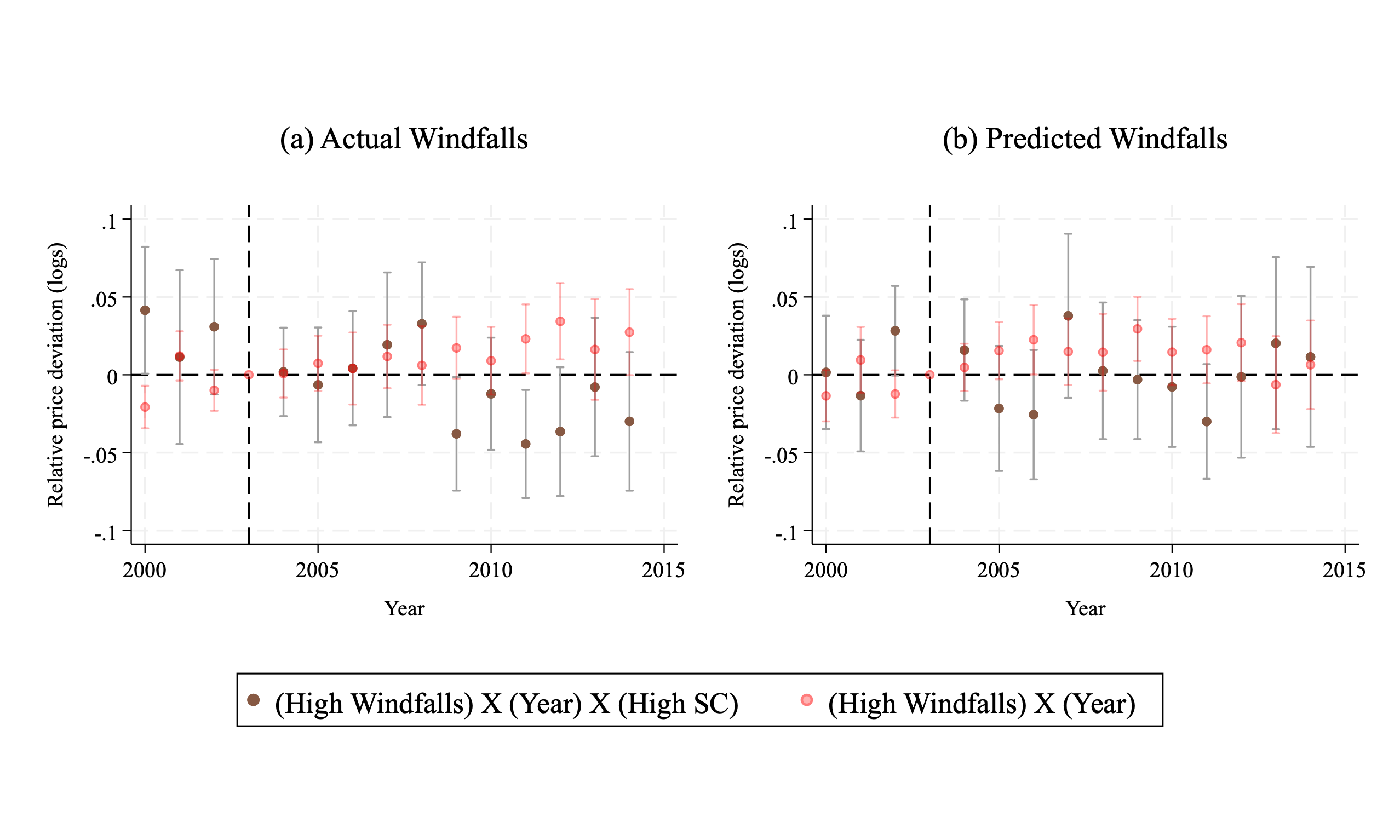}
    \caption{Price convergence of main  crops (log deviation from national average)} 
    \label{fig:mecha_pricez}
    \vspace{-.6cm}
    \begin{changemargin}{0.8cm}{0.6cm} 
\begin{spacing}{0.125}
{\footnotesize \setstretch{.125} \textit{High Windfalls}: locality with large (above the 75th percentile) windfall transfers during boom. \textit{High State Capacity (SC)}: locality with above median tax per capita collected in pre-boom period.}  
\end{spacing}
\end{changemargin}
\end{figure}

\begin{figure}[h!]
    \centering
    \includegraphics[height=6.5cm,trim={0cm 0cm 0cm 4cm},clip]{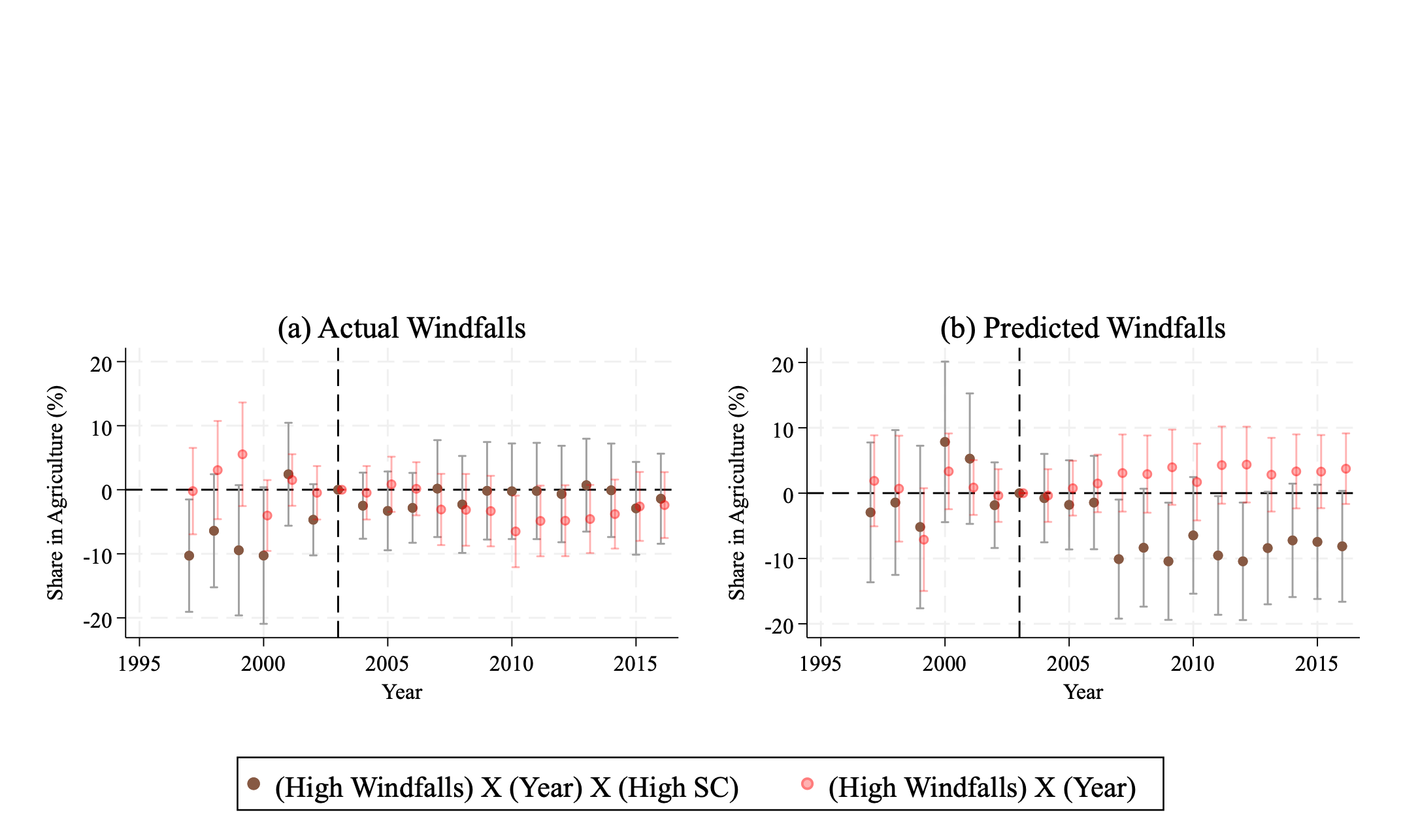}
    \caption{Structural Transformation: Decline of Agriculture Share}
    \label{fig:mecha_agro}
        \vspace{-.6cm}
     \begin{changemargin}{0.8cm}{0.6cm} 
\begin{spacing}{0.125}
{\footnotesize \setstretch{.125} \textit{High Windfalls}: locality with large (above the 75th percentile) windfall transfers during boom. \textit{High State Capacity (SC)}: locality with above median tax per capita collected in pre-boom period.}  
\end{spacing}
\end{changemargin}
\end{figure}

\begin{figure}[h!]
    \centering
    \includegraphics[height=6.5cm,trim={0cm 0cm 0cm 4cm},clip]{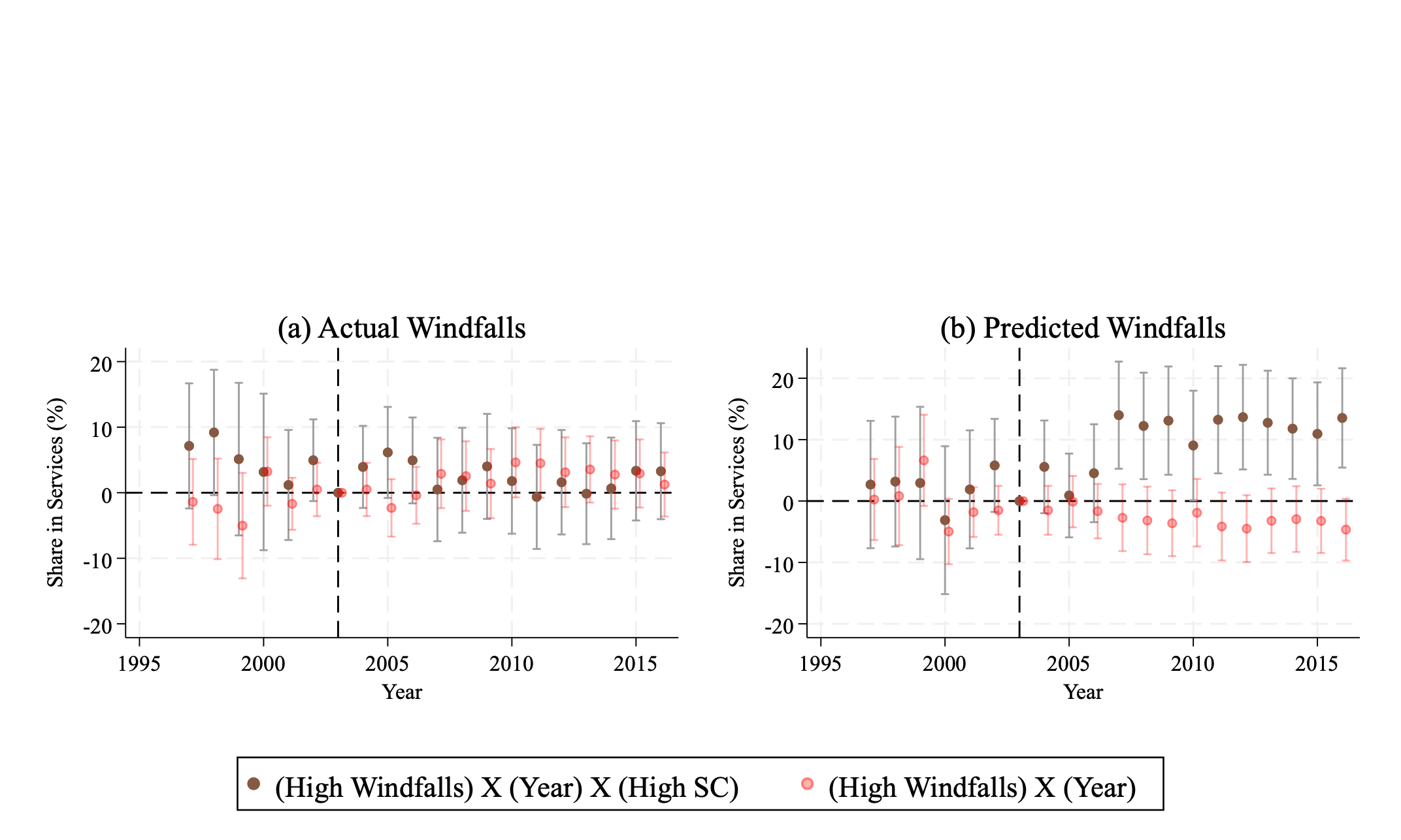}
    \caption{Structural Transformation: Rise of Services Share}
    \label{fig:mecha_serv}
    \vspace{-.6cm}
     \begin{changemargin}{0.8cm}{0.6cm} 
\begin{spacing}{0.125}
{\footnotesize \setstretch{.125} \textit{High Windfalls}: locality with large (above the 75th percentile) windfall transfers during boom. \textit{High State Capacity (SC)}: locality with above median tax per capita collected in pre-boom period.}  
\end{spacing}
\end{changemargin}
\end{figure}

Infrastructure, and in particular roads, may increase household income by increasing trade opportunities \citep{donaldson2018railroads} and accelerating structural transformation \citep{asher2020rural}. We cannot directly test for the trade opportunities channel because we do not observe intranational trade flows at the district level. We instead conduct an indirect test based on \cite{sotelo2020domestic}: we find suggestive evidence that agricultural prices are converging to national averages in treated districts, and specifically in those with high state capacity.\footnote{Our outcome variable is a relative price deviation index based on a district-crop-year panel. We (i) obtain the average yearly price for each crop, (ii) compute the by-crop price deviation of each district from the national mean (in relative terms), and (iii) collapse the data from the crop-district-year level to the district-year level using crop's planted hectares as weights.} We find that crop prices converge to national averages in high state capacity districts during the boom in the long-run (this can be observed in Figure \ref{fig:mecha_pricez}), indicating that these districts are becoming more integrated with national markets.\footnote{Table \ref{tab:mecha2}, Columns 1 and 2, show these results for the entire post-boom period, which averages across short and long-run, therefore we get negative but not significant coefficients.} Furthermore, reducing transportation costs via road construction might increase access to non-agricultural, typically higher paying, jobs.\footnote{High transportation costs can make it difficult for workers to move between different areas and industries \citep{jacoby2000, bryanetal2014, bryanmorten2019}.} We find evidence of structural transformation occurring in treated districts with high state capacity, with labor shifting from agriculture to services (Table \ref{tab:mecha2}: Columns 3-6).\footnote{Compare results in columns 3-6 to results for figures \ref{fig:mecha_agro} and \ref{fig:mecha_serv}. } 

\subsection{Social unrest and the resource curse}

\begin{figure}[h!]
    \centering
    \includegraphics[height=8.5cm,trim={0cm 0cm 0cm 1cm},clip]{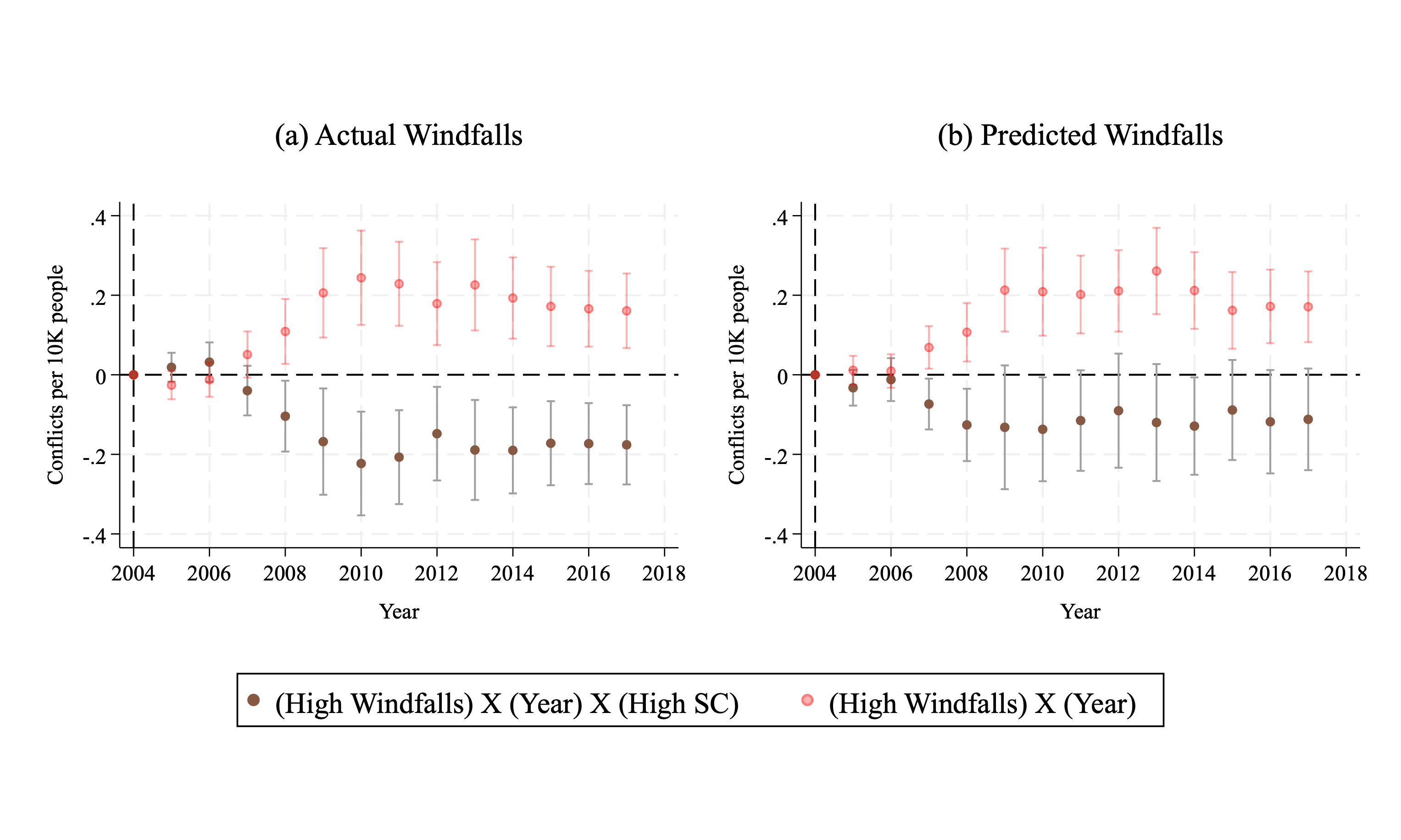}
    \caption{Social unrest events per 10,000 people}
    \label{fig:conflicts}
    \vspace{-.6cm}
     \begin{changemargin}{0.8cm}{0.6cm} 
\begin{spacing}{0.125}
{\footnotesize \setstretch{.125} \textit{High Windfalls}: locality with large (above the 75th percentile) windfall transfers during boom. \textit{High State Capacity (SC)}: locality with above median tax per capita collected in pre-boom period.}  
\end{spacing}
\end{changemargin}
\end{figure}

We use our Ombudsman data on protests across the country to investigate social unrest as a potential explanation as to why the low state capacity localities fare worse after the mining boom. Figure \ref{fig:conflicts} shows our results after estimating equations (\ref{eq:TWFE_byyear}) and (\ref{eq:TWFE_byyear_IV}). Protests per capita increase  by around 50\% of the baseline mean (.2 per 10,000 people in the medium-run) in the low state capacity high-windfalls group.  Meanwhile, the high state capacity treated group does not suffer from an increase in social unrest. 

The effects for low state capacity localities are in line with the literature suggesting that extractive industries can generate a resource curse due to the social unrest that follow \citep{adhvaryu2021resources}. However, it seems that these ``conflicts'' only take place in the localities in which local governments are unable to reap the benefits of the boom.  Social unrest thus ensues as a response to resource booms absent tangible benefits to the community. 

The increase in per capita protests in low state capacity places can illuminate the mechanism behind the potential decrease in household income in low state capacity localities. Table \ref{tab:main_hte} row 3 shows negative (Windfalls $\times$ Post-Boom) coefficients, which  suggest that these localities perceive negative (and, in some specifications, significant) effects on household income after the windfall transfers begin to enter the district.\footnote{These negative effects are of similar magnitude as the positive effects for high state capacity localities shown in the first row (Windfalls $\times$Post-Boom$\times$State Capacity).}

We conducted a series of interviews to stakeholders in the mining industry to further investigate the effect of social unrest on income.\footnote{The names and firms of the interviewees are kept confidential. We interviewed three employees. They belong to consulting firms that specializes in mining firms, and a the top 15 mining companies in Peru.} Anecdotally, investors take into account whether a locality has had documented social unrest when deciding to continue investments in a mining enterprise. Therefore, (violent) protests ensuing could deter investments in  localities, which could potentially explain why local people perceive lower household income in the post-mining period. Moreover, according to the Ombudsman office, social unrest should be mediated by a combination of local and regional government, whereby the participation of local governments in organizing a dialogue between  stakeholders is paramount; therefore, the local government's state capacity is a relevant observable when mediating and preventing future social unrest \citep{defensoria2020}. 

\section{Robustness and Threats to Identification }\label{sec:robustness}

A threat to identification for the observed positive relationship between  state capacity and windfalls' effects is that high state capacity localities could outperform others for reasons unrelated to windfall management. We verify that this is not the case by comparing high and low state capacity localities in our control group with low windfall transfers. 

Another potential concern with our analysis is that our state capacity proxy might not be capturing all dimensions of state capacity, or it may be spuriously higher in some localities even if the local government is weak. We therefore test robustness to an alternative proxy related to the quality of local governments' bureaucracy. 

Furthermore, an issue with our current main outcome variable (household income) is that it comes from a nationally representative household survey (ENAHO) covering a subset of the countries' districts which is not stable over time. Moreover, the survey underwent  methodological changes over the 2001-2004 period which may confound our estimates. We test robustness to an alternative outcome variable: nightlights, which is available for all districts in Peru throughout our study  period. The underlying methodology for the nightlight variable also changes, but at a different time than ENAHO. For the 1992-2012 period, the variable is based on imagery  from the Defense Meteorological Satellite Program (DMSP), while for the 2013-2018 period it is based on the Visible Infrared Imaging Radiometer Suite (VIIRS).

An additional consideration is that our treatment cutoff (districts with windfalls above the 75th percentile during the mining boom) is largely arbitrary.\footnote{We mention in Section \ref{subsec:miningdata} that the average district receives some non-zero amount of windfalls. Hence, defining a treatment status requires a cutoff. }  Therefore we check for robustness to two alternative cutoffs: the median and the 90th percentile. 

Finally, a limitation of our current design is that districts do not start receiving large transfers all at once during the boom. Windfalls arrive  after mining production begins and several mines begin operations at different points in our study period.  To address the staggered nature of mining windfalls and mine openings, in Appendix \ref{app:staggered} we check robustness to an alternative design: a staggered event study. 

\subsection{Placebo test: High vs. Low state capacity in control group}
It is possible that districts with high state capacity districts outperform their low state capacity counterparts regardless of the mining boom, and this may be (spuriously) driving the heterogeneous effects of the mining boom documented above. To test for this possibility, we conduct a placebo test comparing high and low state capacity localities in the  control group. 

We restrict our sample to localities with low windfall transfers per capita---our control group as defined in Section \ref{subsec:ols} i.e. those with below 75th percentile per capita windfall transfers-- and we codify the group with high state capacity as ``treated''\footnote{Like before, we define ``high'' state capacity localities as those with above median state capacity.}. Then, we estimate the following two-way fixed-effects (TWFE) equation:
\vspace{-.2cm}
\begin{equation}\label{eq:placebo}
 Y_{ipt} = \alpha +  \sum_{h\neq 2003}\beta_{h}^{I} I_{ip} \times\bm{1}_{[t=h]} + \mathbf{X}^{'}_{ip}\gamma + \phi_{p} + \phi_{t} + \varepsilon_{ipt}
\end{equation}
\vspace{-.2cm}
 $Y_{ipt}$ represents the outcome of interest (household income) for household $i$ in locality $p$ at time $t$. The binary variable $I_{ip}$ takes the value of 1 if $i$ is classified as a high state capacity locality (the placebo treatment group) and 0 otherwise. $\bm{1}_{[h=t]}$ is an indicator function that takes value 1 if $t$ is equal to year $h$, excluding 2003, the year before the mining boom begins. The term $\beta_h^I$ captures the effect of being a high state capacity locality during the mining boom (within our restricted low-windfalls sample) on $Y$ under parallel trends assumption. The vector $\mathbf{X_{ip}^{'}}$ includes demographic controls and pre-boom characteristics interacted with year fixed effects. The model also includes locality ($\phi_{p}$) and year ($\phi_{t}$) fixed effects, and an error term $\varepsilon_{ipt}$. We also estimate a version of this specification, where the sample is the control group defined in Section \ref{subsec:geology}: localities with low (below 75th percentile) predicted windfalls. 

If state capacity effects unrelated to the boom are driving our estimated heterogeneous treatment effects, then we would expect high state-capacity localities to outperform low state-capacity ones in the low-windfalls group. Figure \ref{fig:placebo} shows the results of this placebo test, plotting the  $\beta^I_{h}$ coefficients of equation \ref{eq:placebo}, using the sample of low actual and low predicted windfalls respectively. Reassuringly, we can rule out that high state capacity localities grow faster in the control group during this period. We conclude that our main results are not driven  by \textit{all} high state capacity localities growing  faster than their low state capacity counterparts.

\subsection{Alternative state capacity proxy}
An active literature highlights the importance of bureaucracies for public sector performance (\cite{finan2017personnel},\cite{besleyetal2022},\cite{best2023individuals}). We check whether our results withstand an alternative proxy of state capacity measuring the strength of the local bureaucracy: bureaucrats per capita. We obtain a measure of local government employees from the RENAMU Municipal Survey and compute a bureaucrats per capita measure counting as ``bureaucrats'' all workers in professional and technical occupations (i.e., in white-collar jobs). Figure \ref{fig:corr} shows the correlation between the tax per capita and bureaucracy proxies, which is strong and positive. The right panels of Figures\ref{fig:sc_time} and \ref{fig:determinants} show, respectively, that the ranking of municipalities according to the  bureaucracy proxy stays roughly constant over time and that the bureaucracy proxy has somewhat similar to the tax proxy. 

We estimate equation (\ref{eq:HTE_TWFE}) changing the definition of $I^{SC}_d$.\footnote{We run this regression using both actual and predicted  windfalls to determine treatment status. As in Table \ref{tab:main_hte}, Columns 1-2 use the actual windfalls treatment definition and Columns 3-4 use the predicted windfalls reduced form IV (equation \ref{eq:HTE_TWFE_IV}).} This state capacity indicator  will now equal 1 if the district has above median tertiary educated bureaucrats working in its government in 2002. The results we obtain from this alternative state capacity definition, shown in Table \ref{tab:robustness_pt}, are  qualitatively similar to those  in Table \ref{tab:main_hte}.  

\subsection{Alternative outcome variable: Nightlights}
 Our main results have household income per capita as outcome variable. However, this measure is based on a repeated cross-sectional representative  survey of Peruvian households, therefore does not include every district in Peru and is subject to the limitations of this kind of database. We therefore implement a robustness check a year-district panel with a different outcome variable that aims to capture living standards.
 
Our alternative measure of living standards is nightlights, which have established in the literature as a  development proxy \citep{hendersonetal2012}. Global grid (raster) data is available from the Defense Meteorological Satellite Program (DMSP) for the 1992-2013 period and from the Visible Infrared Imaging Radiometer Suite (VIIRS) for the 2013-2017 period.\footnote{These grids have a very high spatial resolution, and the typical district contains thousands of cells, each of which receives a positive value indicating the intensity of night lights detected for that cell over the course of a year. We collapse this cell-year data to the district-year level by averaging across all cells contained in each district. Then, we standardize the mean luminosity observed for each district-year by the size of each districts' population. Our preferred living standards measure built in this way is the \textit{mean luminosity per capita}, measured for each district-year. The mine itself could be emitting its own luminosity and this would be a confounder for what we intend to measure, so we crop out the pixels that correspond to the mine coordinates.} 

A potential concern with using nightlights as an outcome is that the mines themselves emit luminosity, so we might be solely picking up the light that the mines produce for their own operations. To avoid this, we leverage the fact that luminosity is averaged at the pixel level and that the mines in Peru are geo-referenced using their latitude and longitude. We can thus crop out the pixels that contain mines in order to not contaminate the nightlights measure.

Table \ref{tab:robustness_lights} shows our results estimating the district-level version of equation (\ref{eq:HTE_TWFE}) using log mean luminosity per capita as our $Y_d$ variable.\footnote{In other words, we implement equation \ref{eq:HTE_TWFE_district} for both actual and predicted windfalls.} Again these results are comparable with those in Table \ref{tab:main_hte}. Similarly, Figure \ref{fig:nightlights} shows our results from estimating the district-year panel version of equations (\ref{eq:TWFE_byyear}) and (\ref{eq:TWFE_byyear_IV}) with nightlights (mean luminosity per capita, with cropped out mines) as the left-hand side variable. These results are if anything stronger in the long run than those in Figure \ref{fig:results_logy_HH}.

Since the methodology of the ENAHO household survey changed during the mining boom onset (2003-2004), the fact that our nightlight results are in the same direction assuages the concern that our household income results were driven by the change in survey methodology.\footnote{The nightlights data also underwent a methodology change, but at a later point in time (year 2013).} 

\subsection{Alternative treatment cutoff}

 We are classifying districts as treated if they receive windfalls above the nationwide 75th percentile. We test robustness to two alternative cutoffs: the median and the 90th percentile.\footnote{For comparison, the median cutoff in actual windfalls is 3.5 USD per capita, the 75th percentile threshold is 129.8 and the 90th percentile is 441. We have 1,433 treated districts using the median as the cutoff, 872 using the 75th percentile, and 432 using the 90th percentile.} 
 
 Table \ref{tab:robustness_p50} show our results when estimating (\ref{eq:HTE_TWFE}) while relying on the 50th percentile as treatment cutoff. The results we get are like those in Table \ref{tab:main_hte}. Table \ref{tab:robustness_p90} shows the same coefficients but using the 90th percentile as treatment threshold. These results are noisier and lack statistical significance. 

We did not pick the median threshold in our original specification because the transfers per capita amount right above this cutoff (3.5 USD) is quite low and so observing effects with this treatment intensity is less plausible. Using the 90th percentile threshold, on the other hand, the treatment intensity is stronger but only $\sim$10\% of the population is in treated districts, therefore it is also less likely we detect effects. The 75th percentile was a compromise between high enough windfall transfers treatment intensity (over $\sim$130 USD per capita) and enough population (25\%) in the treatment group to estimate effects. Moreover, with the population weighted 75th percentile cutoff, we have almost 50\% of districts in the each of the treatment and control groups (47.5\% of districts are treated).

\vspace{-.25mm}
 
\section{Concluding Remarks}\label{sec:conclusion}
This paper sheds light on the synergies between natural resources and state capacity. We show that the gains from an important mining boom in the Peruvian setting are concentrated in localities with higher baseline state capacity.\footnote{These results are robust to modifications in our state capacity and living standards measurement as well as changes in our cutoff to classify districts with ``large windfalls'' treatment status. Moreover, staggered event study results reinforce these findings, showing a gradual improvement in living standards after mining operations begin, driven by high state capacity districts. } These results are consistent across our two different specifications and are not driven by high state capacity localities having higher windfall transfers. 

We explore investment on public goods provision as a potential mechanism. Both high and low state capacity treated districts invest in similar types in public goods, but high state capacity localities increase their transportation and sewerage infrastructure during the post-boom period. In addition, in high state capacity districts  perceiving large windfall transfers,  workers seem to be shifting away from the agricultural sector and into the service sector, which suggests structural transformation takes place during the boom in these localities. 

Finally, we investigate social unrest as a potential reason why households in low state capacity that receive large windfall transfers fail to perceive the gains from the boom. We find that social unrest, classified as ``conflicts'' by the Ombudsman office, increases in localities with high windfall transfers but low state capacity. This is consistent with these localities' governments failing to redistribute the windfalls and thus prompting their civil society to express their frustration through manifestations. These in turn can potentially discourage future investments in these localities, creating a local resource curse driven by low state capacity. 

The main limitations of this paper arise from the fact that the spatial distribution of mining activities and of state capacity are both endogenous. We have presented a wide array of supplementary findings to assuage concerns about our results being driven by omitted variables. Our results are robust to using soil and sediment metal density as an instrument for mining windfalls. But ultimately, our results will need to be supplemented with additional work looking at the effects of specific exogenous interventions that build state capacity in areas with high windfall transfers.\footnote{In ongoing work, we  address concerns about the endogeneity of state capacity by implementing a closed-elections regression discontinuity design comparing differently qualified bureaucrats. We leverage the fact that a large number of local elections in Peru are won by a small margin and that the runner-up and winner often have very different educational outcomes. This allows us to compare localities where a candidate with tertiary education marginally wins an election against a candidate without a diploma to localities where the opposite happens.} 

Overall, our results suggest that the existence of mining windfalls alone does not ensure development: even in the presence of large windfalls, what matters seems to be the ability of the district government to exploit the gains from the mining boom and promote development.  These results underscore the need for state capacity investments in resource-rich localities. More work remains to be done to understand the mechanisms that drive the gains from state capacity that we have documented. It also seems critical to understand which interventions build state capacity in resource-rich localities along the specific dimensions that help in windfall management.
%\section{Next steps} \label{sec:nextsteps}
%\subsection{Geology IV}

%\subsection{Antamina mine}

%\appendix
%\section{Appendix}
%\begin{appendices}{Appendix}

{\bibliography{bibliography}} 
\clearpage

\section*{Figures}

\begin{landscape}
\vspace{-.5cm}
     \begin{figure}[p]
        \centering
        \includegraphics[height=13cm]{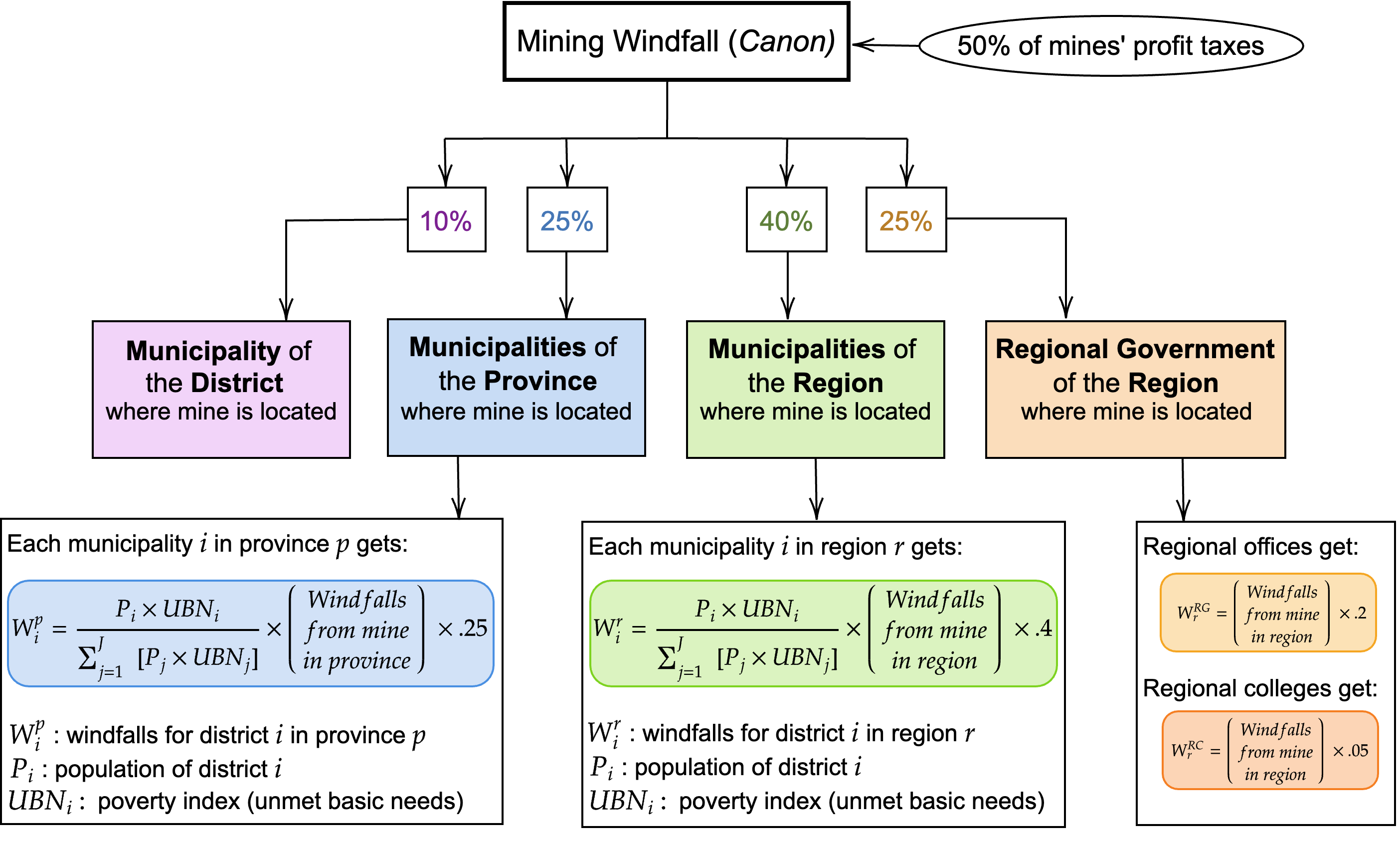}
        \caption{Windfalls Allocation Policy\vspace{.8cm}}
        \label{fig:canonpolicy}
    \end{figure}
    \clearpage
\end{landscape}

\begin{landscape}	

\vspace{3cm}

\begin{figure}[]
	\centering
	\caption{Spatial Distribution of Mining Output and Windfalls}
    \label{fig:treatment_map}
	\minipage{0.32\textwidth}
	  \includegraphics[width=\linewidth]{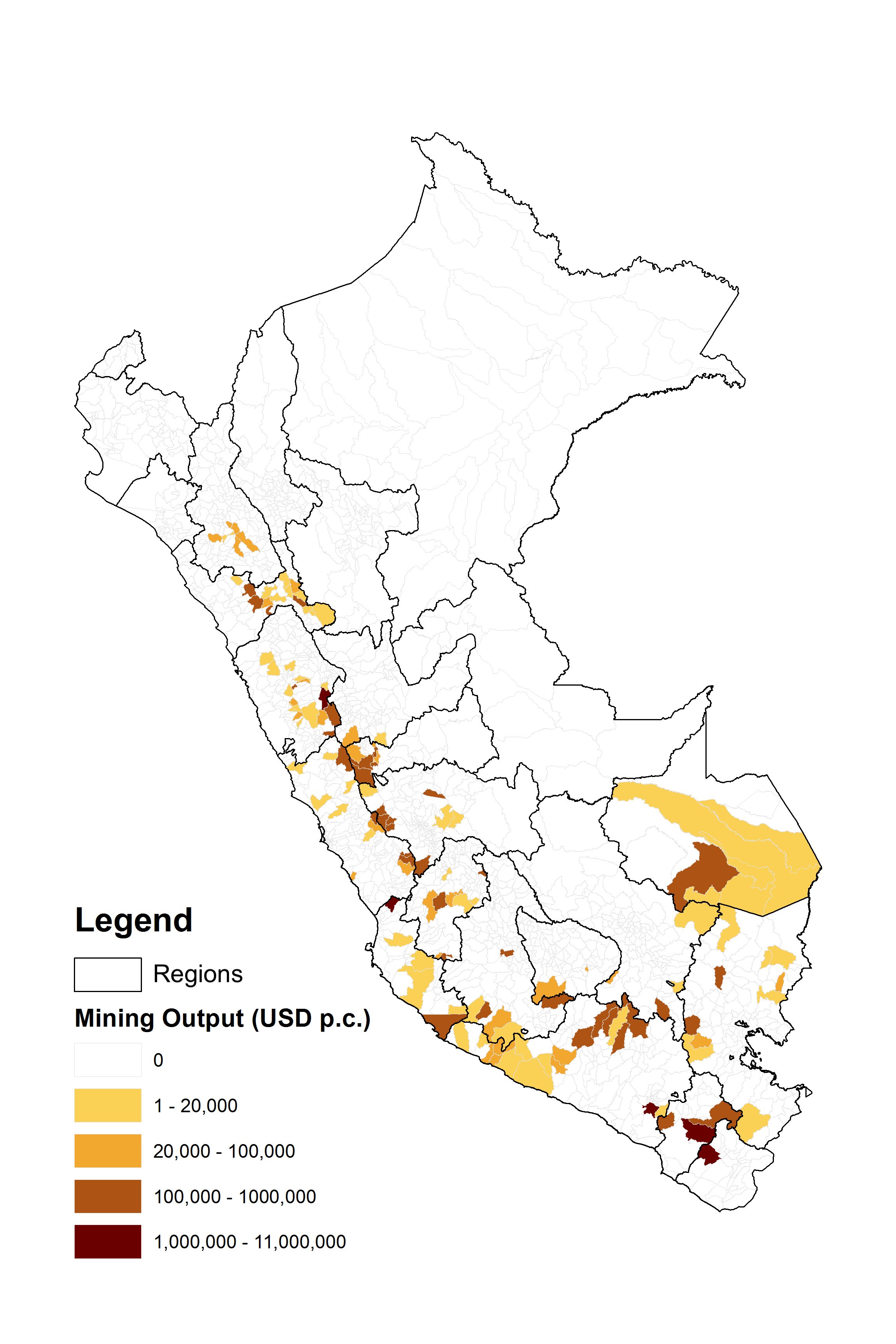}
	    \begin{minipage}[c]{7.2cm}\small
		\vspace{5mm}
  \centering
		\textbf{Panel (a)}: Total mining output in USD per capita. (2004-2011). Source: MINEM.
		\end{minipage}%    
	\endminipage\hfill
	\minipage{0.32\textwidth}
	  \includegraphics[width=\linewidth]{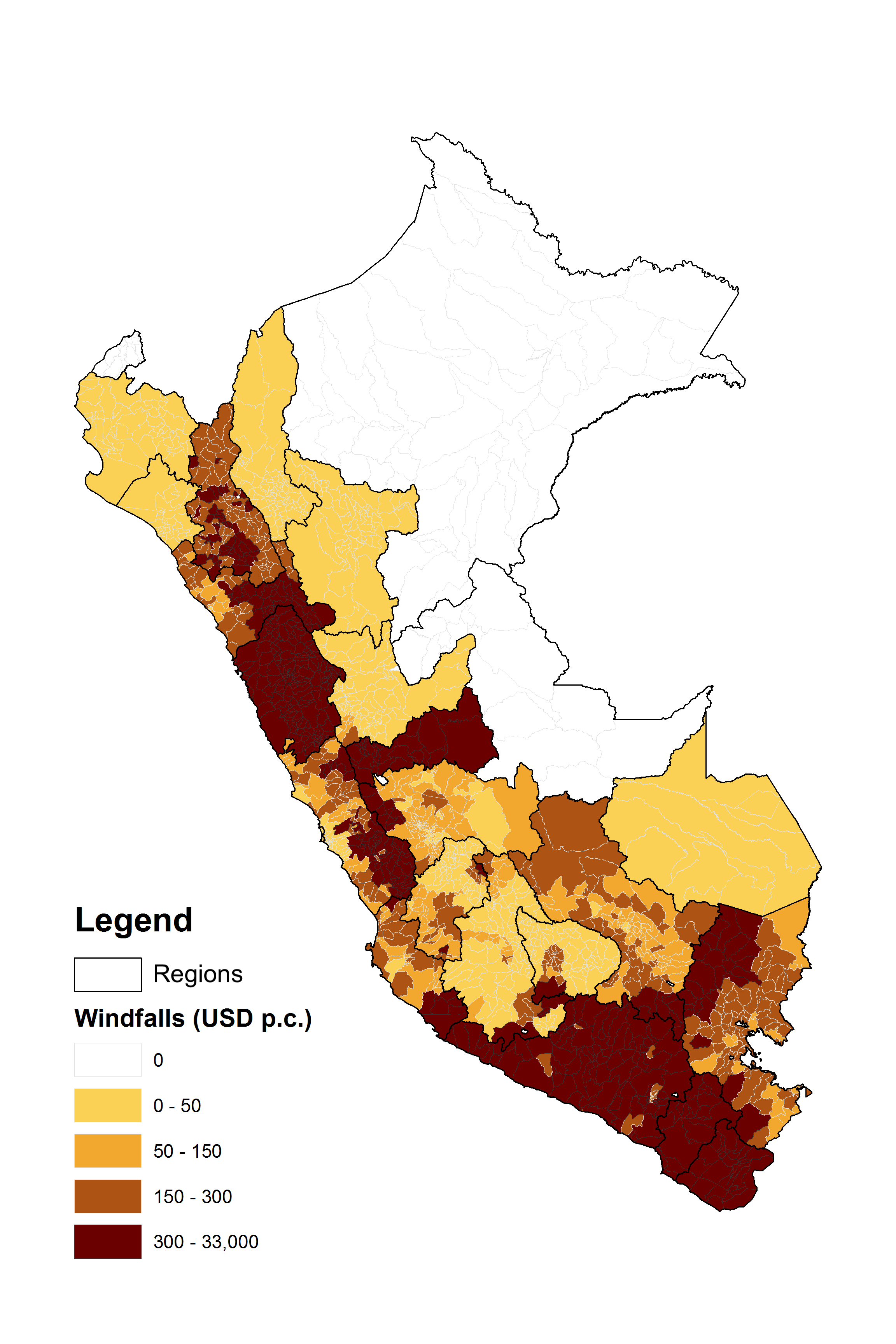}
	    \begin{minipage}[c]{7.2cm}\small
		\vspace{5mm}
		\centering
		\textbf{Panel (b)}: Total mining windfalls in USD per capita (2004-2011). Source: MINEM.
		\end{minipage}%    
	\endminipage\hfill
	\minipage{0.32\textwidth}%
	  \includegraphics[width=\linewidth]{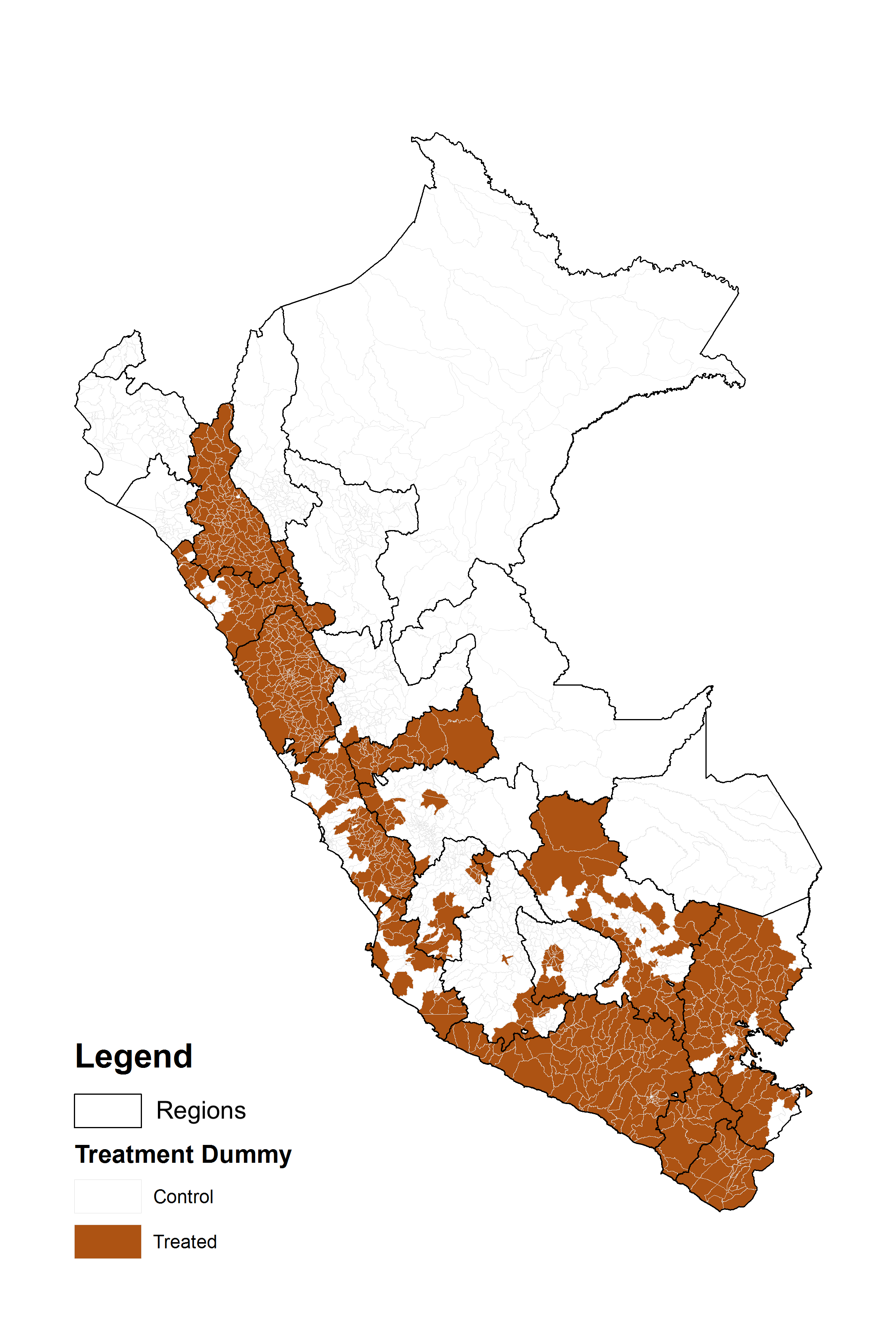}
	    \begin{minipage}[c]{7.2cm}\small
		\vspace{5mm}
  \centering
		\textbf{Panel (c)}: Treatment dummies (based on actual windfalls per capita $>$p75). 
		\end{minipage}%    
	\endminipage
 \label{fig:map1}
\end{figure}
\clearpage
\end{landscape}

%Main results HH income
\begin{figure}[h!]
    \centering
    \includegraphics[height=8cm,trim={0cm .5cm 0cm 1.5cm},clip]{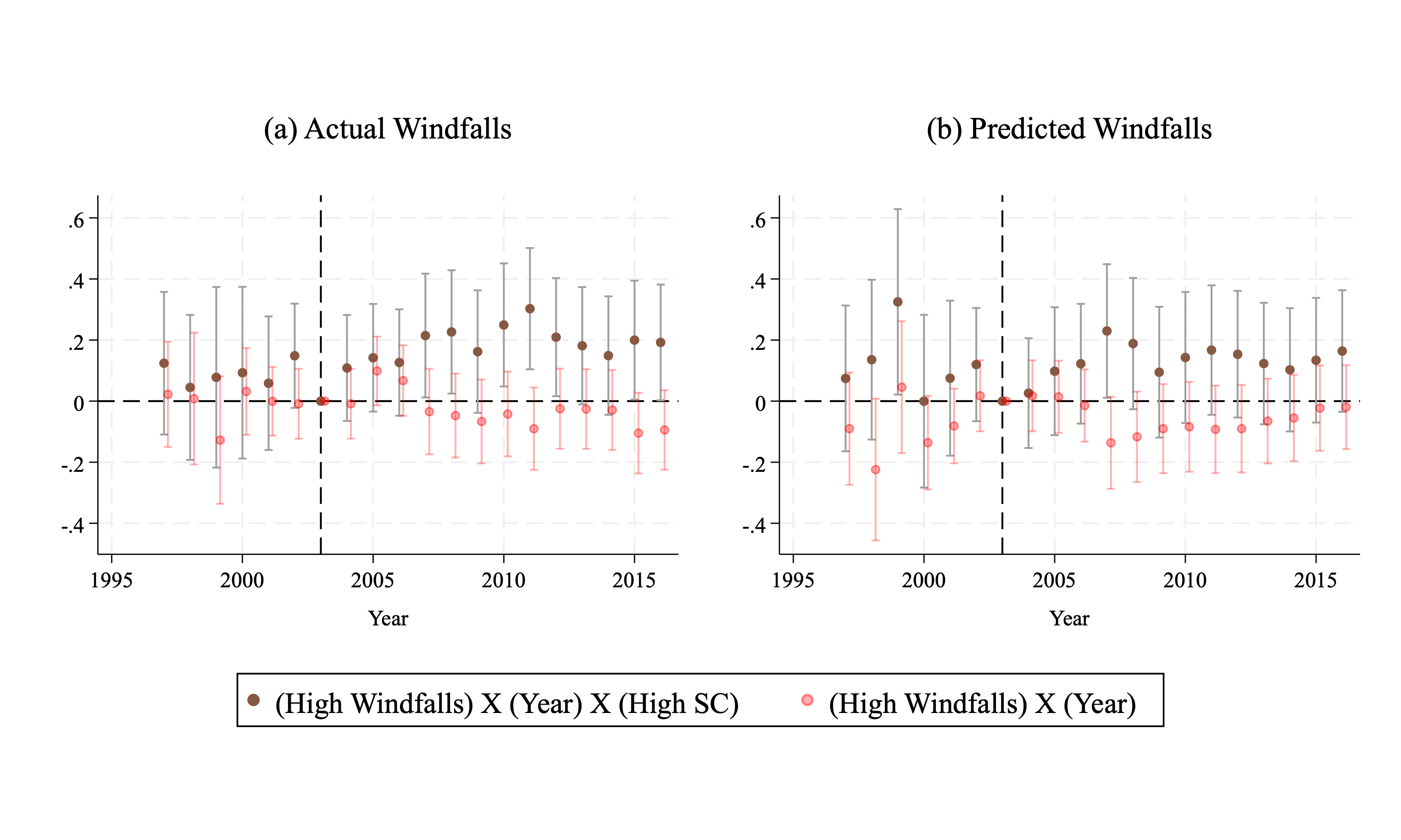}
    \caption{Household Income (log, per capita)}
    \label{fig:results_logy_HH}
    \vspace{-.6cm}
    \begin{changemargin}{0.8cm}{0.6cm} 
\begin{spacing}{0.125}
{\footnotesize \setstretch{.125} \textit{High Windfalls}: locality with large (above the 75th percentile) windfall transfers during boom. \textit{High State Capacity (SC)}: locality with above median tax per capita collected in pre-boom period.}  
\end{spacing}
\end{changemargin}
\end{figure}

\vspace{1cm}

%CANON PER CAPITA

\clearpage
%STOCK OF ROADS  

\vspace{2cm}

%PRICE CONVERGENCE 

\clearpage
%AGRICULTURE SHARE

\vspace{3cm}

%SERVICES SHARE
\clearpage

\begin{landscape}

\section*{Tables}
\begin{table}[h!]
\centering \input{tables/Table1_Balance}
    \caption{Balance Table (prior to the mining boom: 1997-2003) }
    \label{tab:balance}
\end{table}
    \end{landscape}    

\input{tables/Table2_FirstStageOutput}
\vspace{3cm}
\input{tables/Table3_FirstStageWindfalls}
\centering
\input{tables/Table4_HTE_HHinc}

\clearpage
\begin{landscape}

\input{tables/Table5_Mecha1}
\clearpage
\end{landscape}

\begin{landscape}
            \input{tables/Table6_Mecha2}

\clearpage
\end{landscape}

\appendix
\section*{Appendix}
\renewcommand*{\thesection}{\Alph{section}}
\setcounter{figure}{0}
\setcounter{table}{0}

\renewcommand{\thefigure}{A\arabic{figure}}
\renewcommand{\thetable}{A\arabic{table}}

 \section{Supplementary Figures and Tables}\label{app:suppfigs}

\vspace{3cm}

\begin{figure}[h!]
    \centering
    \includegraphics[height=10.5cm]{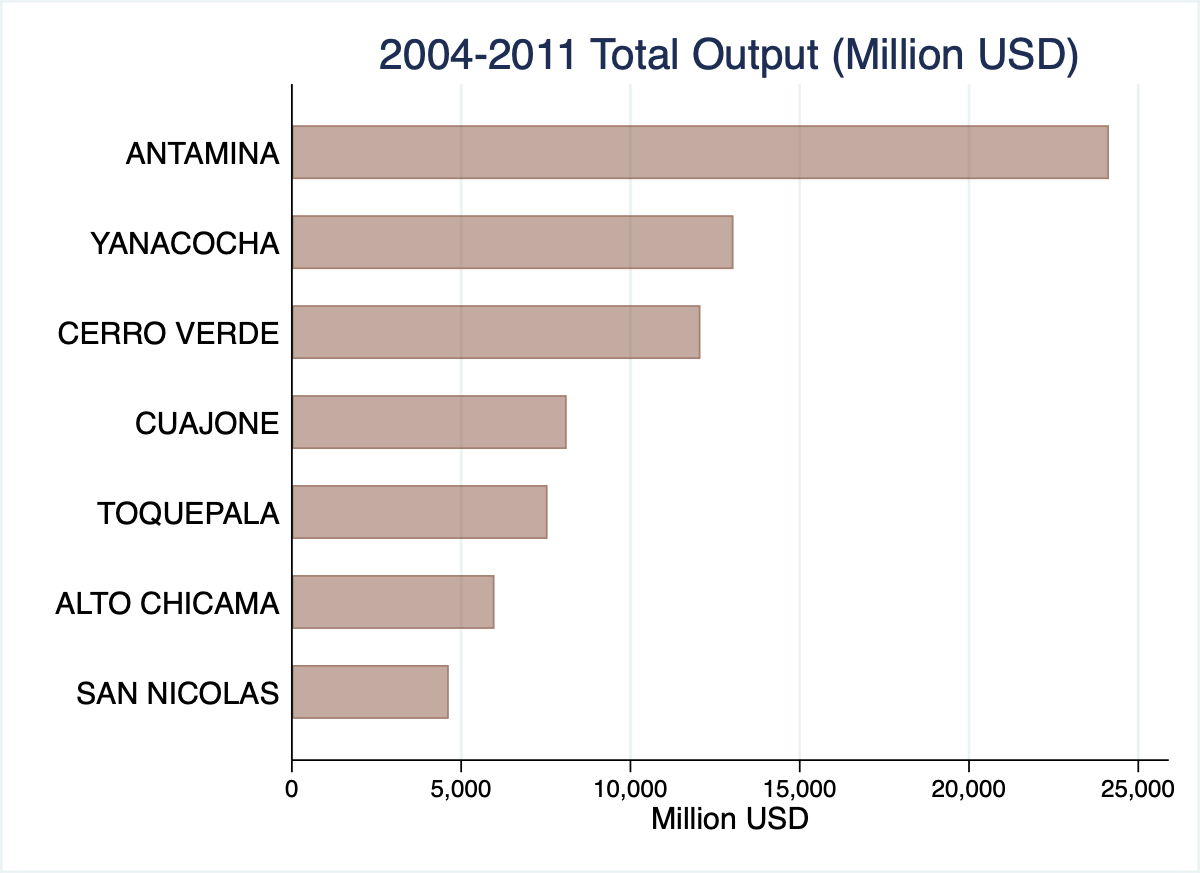}
    \caption{Largest mines in Peru}
    \label{fig:9mines}
      \vspace{-.4cm}
\begin{changemargin}{1.05cm}{1.1cm} 
\begin{spacing}{0.125}
{\footnotesize \setstretch{.125} There are the nine largest mines in Peru in terms of mining output during the 2004-2011 period. Together, these represent 54\% of total mining output in Peru during the boom.  }  
\end{spacing}
\end{changemargin}
\end{figure}

\clearpage
\begin{landscape}
\input{tables/TableA1_DescriptiveStatistics_ALL}
\end{landscape}

    \begin{figure}[h!]
        \centering
        \includegraphics[height=9cm]{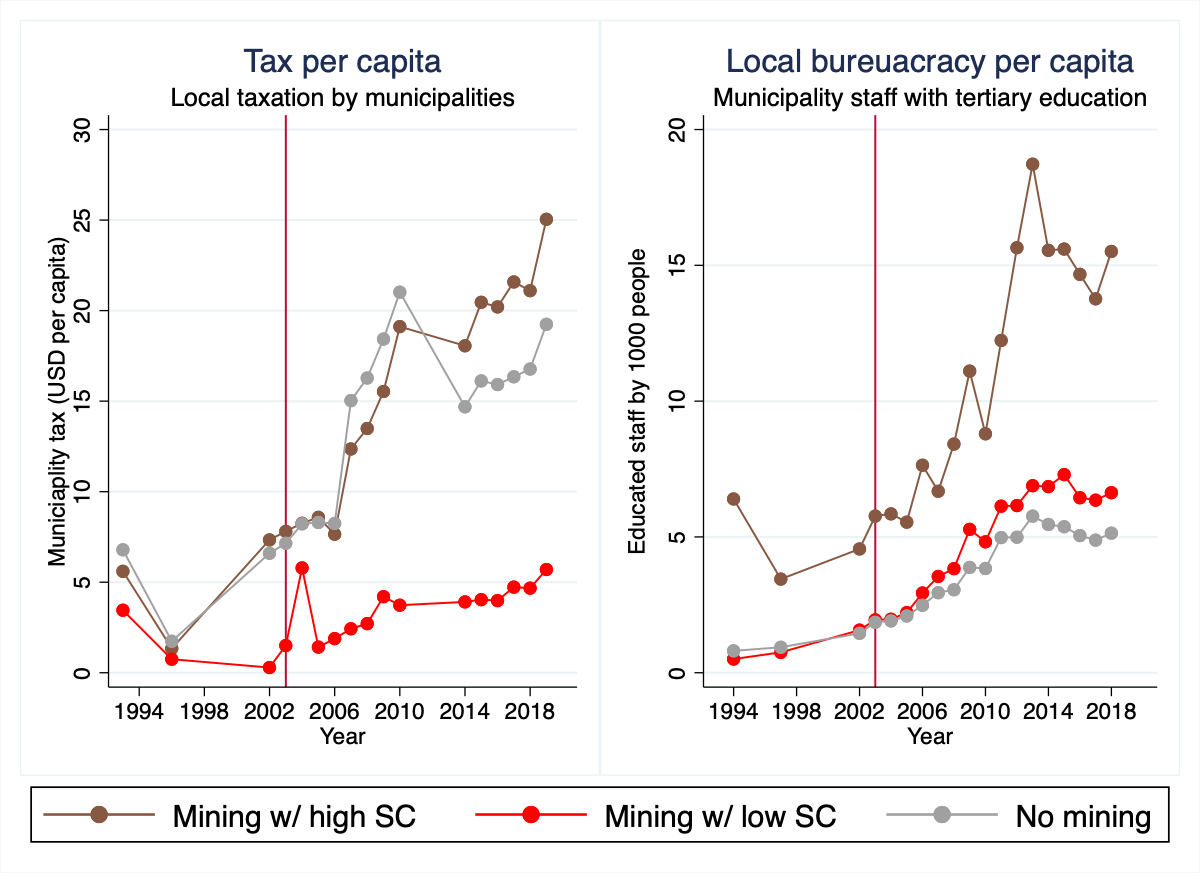}
        \caption{Time series of state capacity}
        \label{fig:sc_time}
    \end{figure}

    \begin{figure}[h!]
    \centering
    \includegraphics[height=8.5cm,trim={0cm 0cm 0cm 3cm},clip]{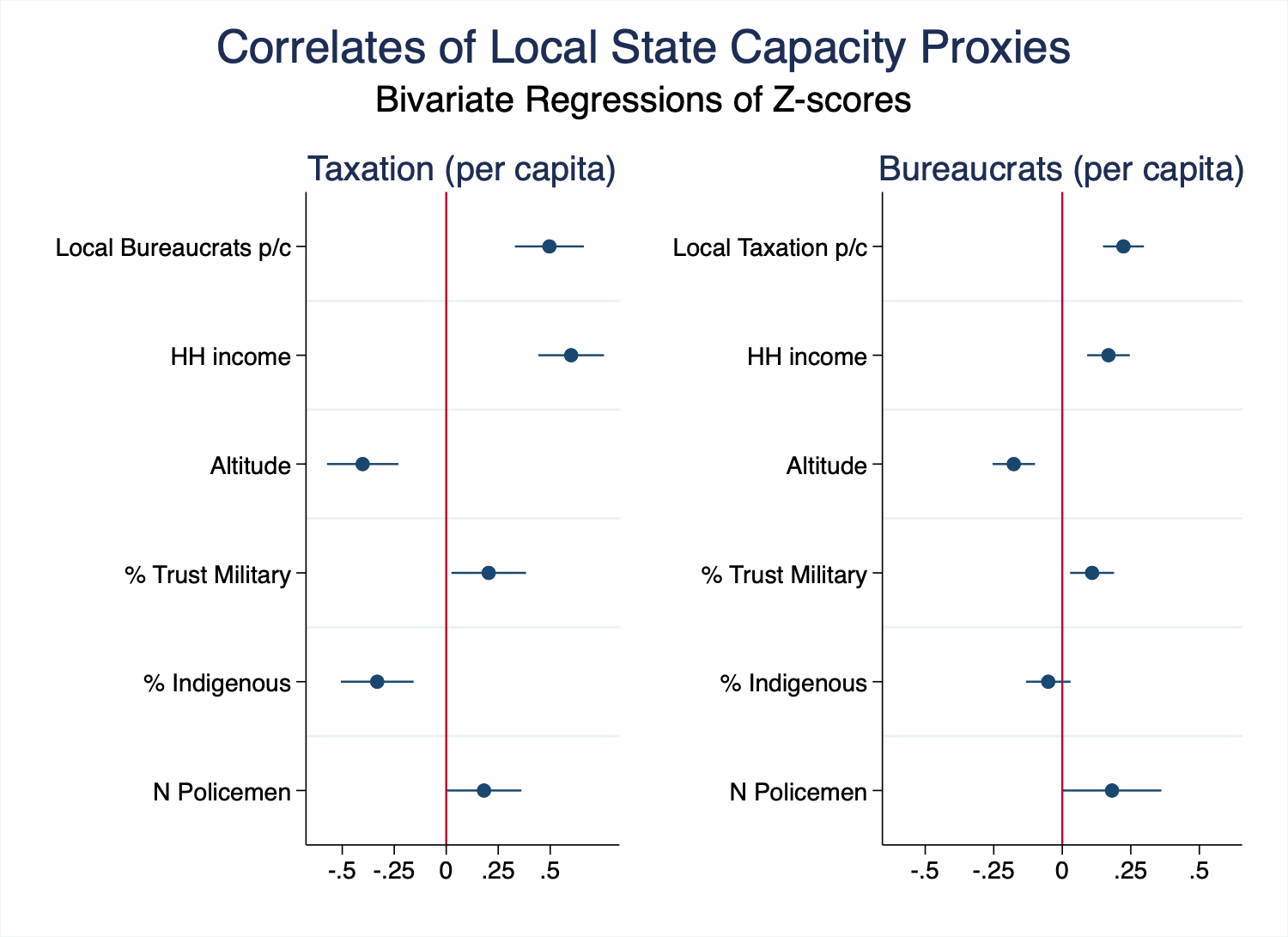}
    \caption{Correlates of local state capacity (2002)}
    \label{fig:determinants}
\end{figure}

% DESCRIPTIVES
\begin{figure}[h!]
    \centering
    \includegraphics[height=9cm]{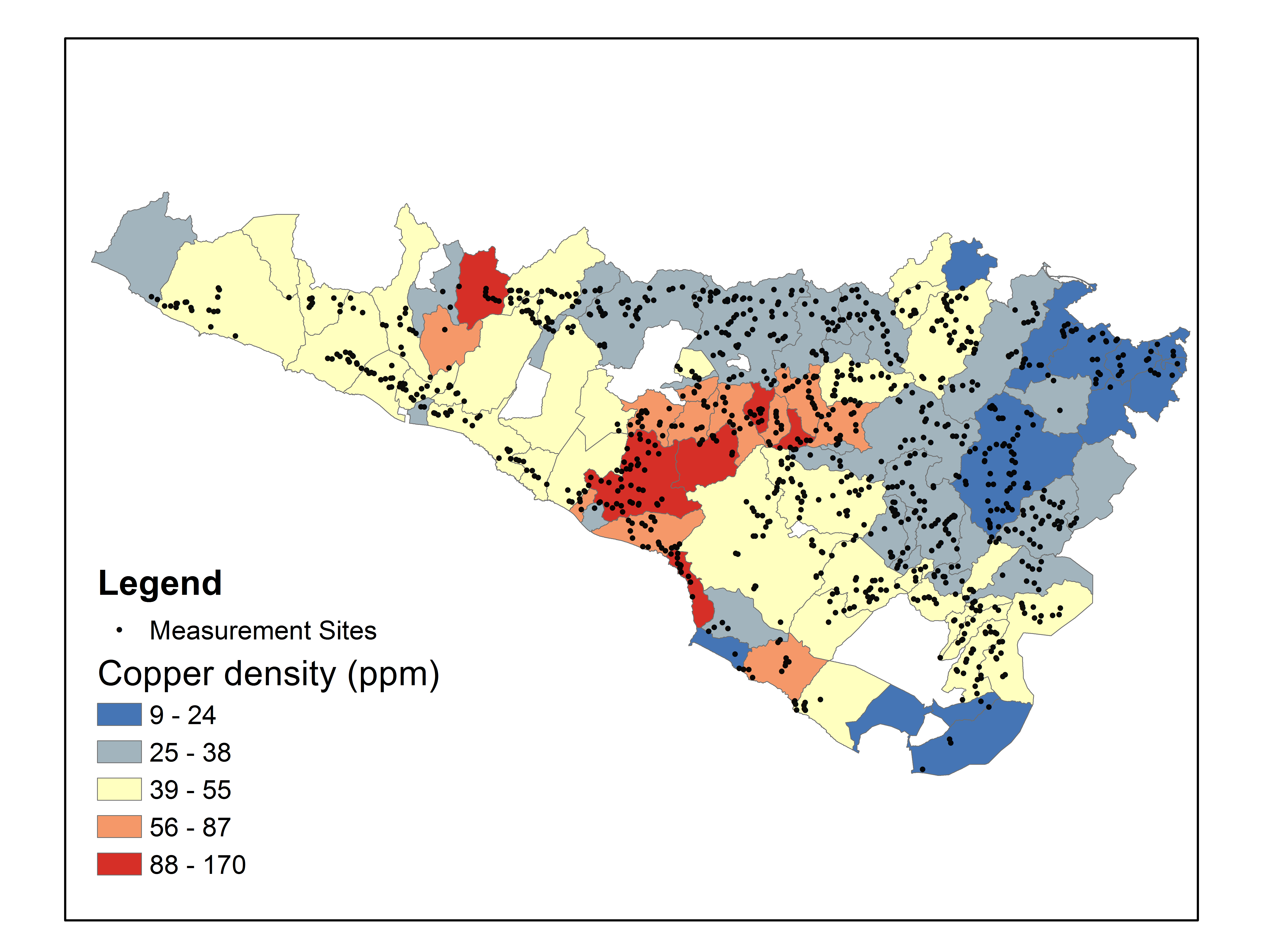}
    \caption{Map of southern region of Peru showing variation in mineral wealth}
    \label{fig:mapgeology}
\end{figure}

\begin{figure}
    \centering
    \includegraphics[height=8.5cm]{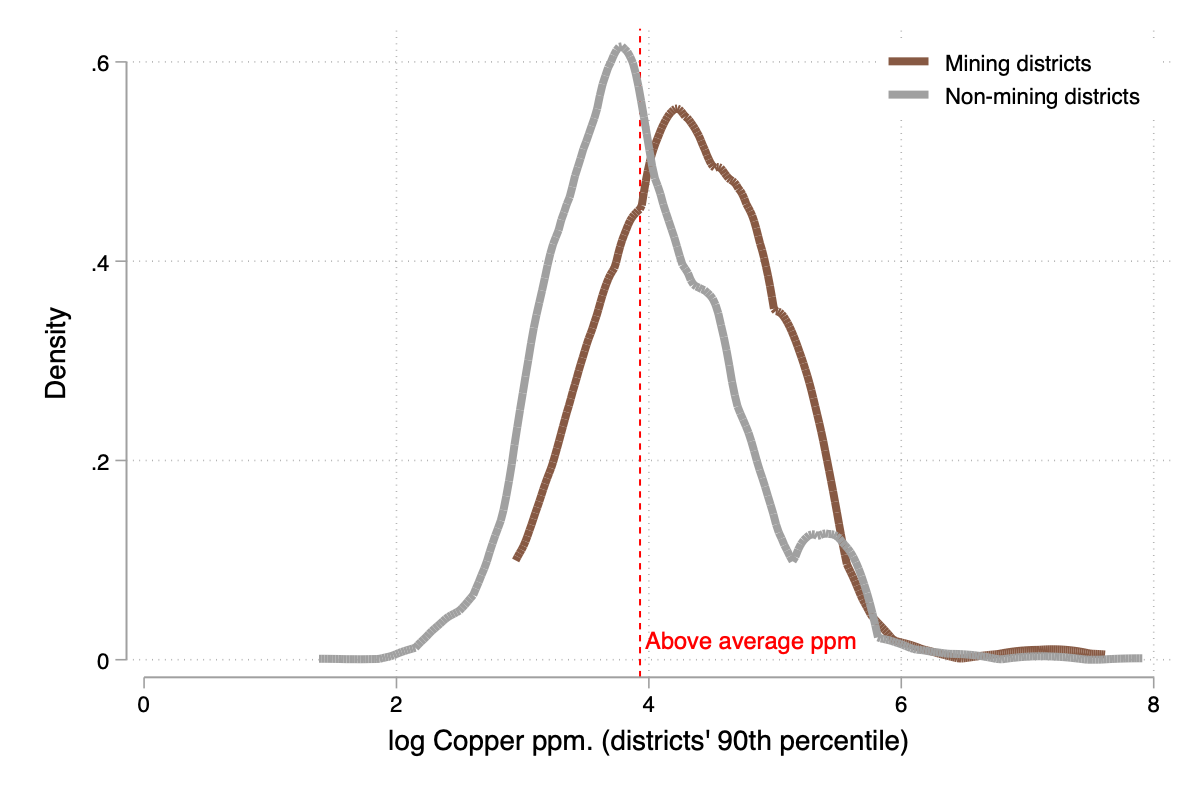}
    \caption{Mineral wealth predicts mining activity}
        \label{fig:1ststage}
\end{figure}
\begin{figure}
    \centering
    \caption{Correlation between true windfalls and predicted windfalls}
    \label{fig:scatter_pred}
    \includegraphics[height=7cm]{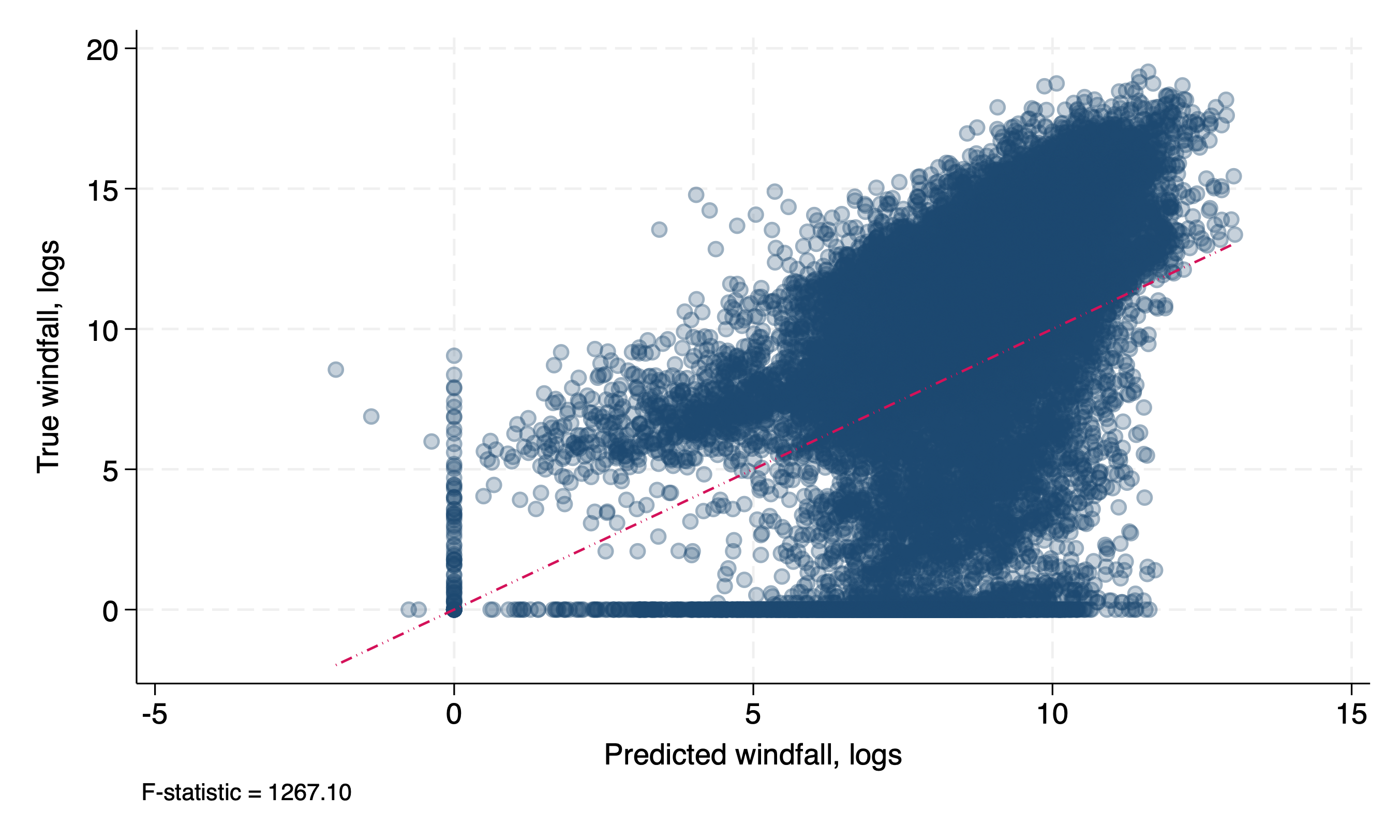}
\end{figure}

    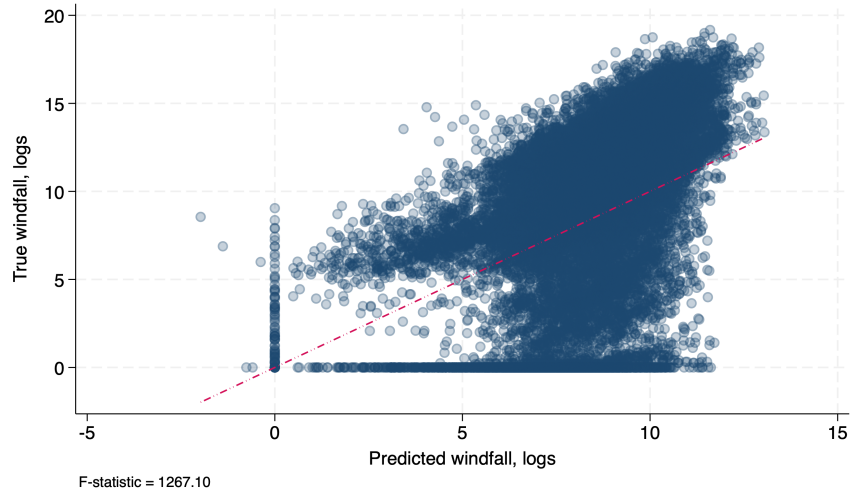
\begin{figure}[p]
    \centering
\input{figures/FigureA7_hypothesized_mech.tex}    \caption{Hypothesized Mechanisms for HTE}
    \label{fig:tikz}
\end{figure}
\begin{figure*}[h!]
    \captionsetup{justification=centering}
    \caption{Mining GDP and windfall  transfers for high and low state capacity districts} \includegraphics[height=9cm,trim={0cm 2cm 0cm 0cm},clip]{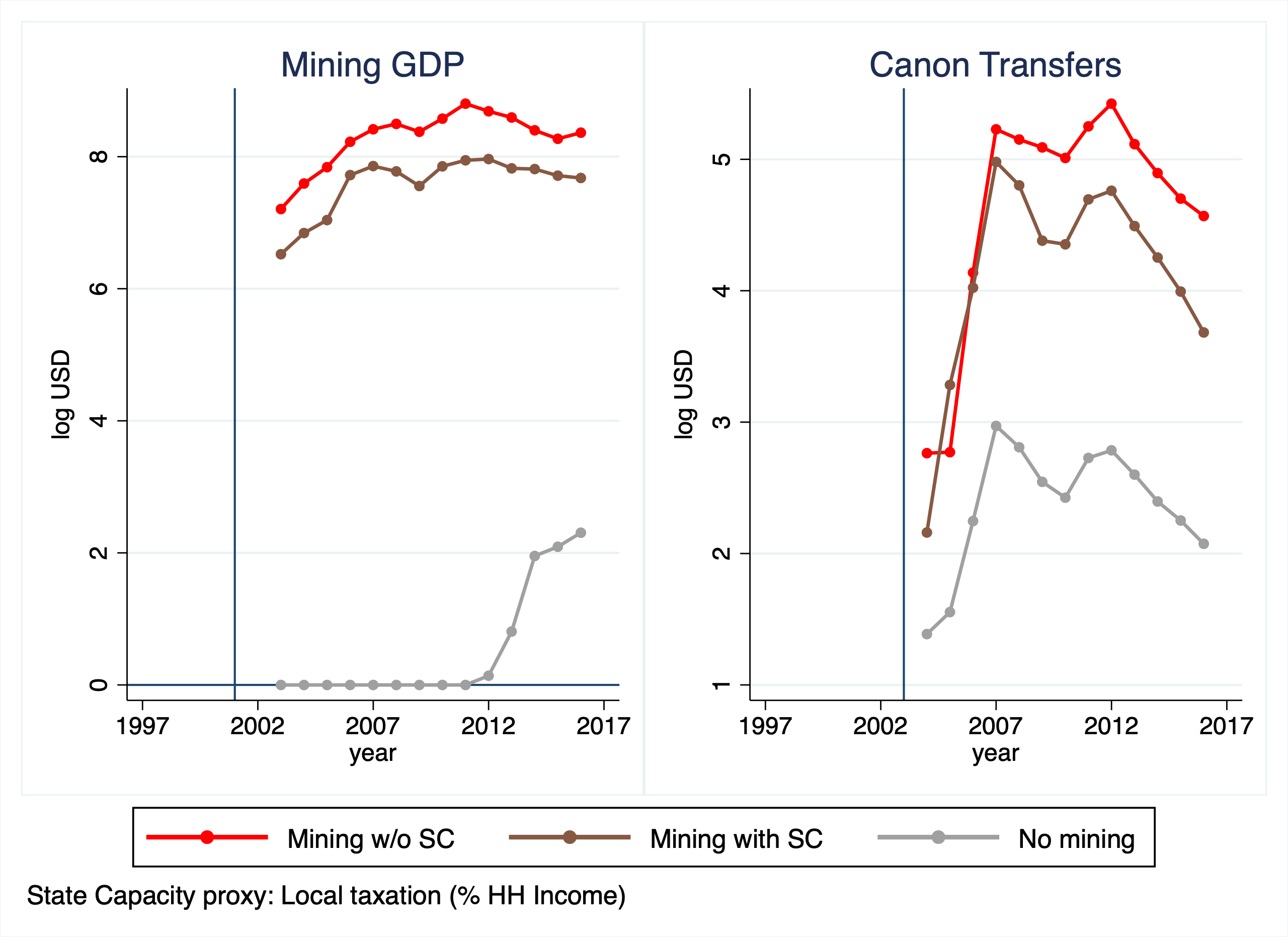}
    \label{fig:shocksize}
    \end{figure*} 
\begin{figure}[h!]
    \centering
    \includegraphics[height=9cm]{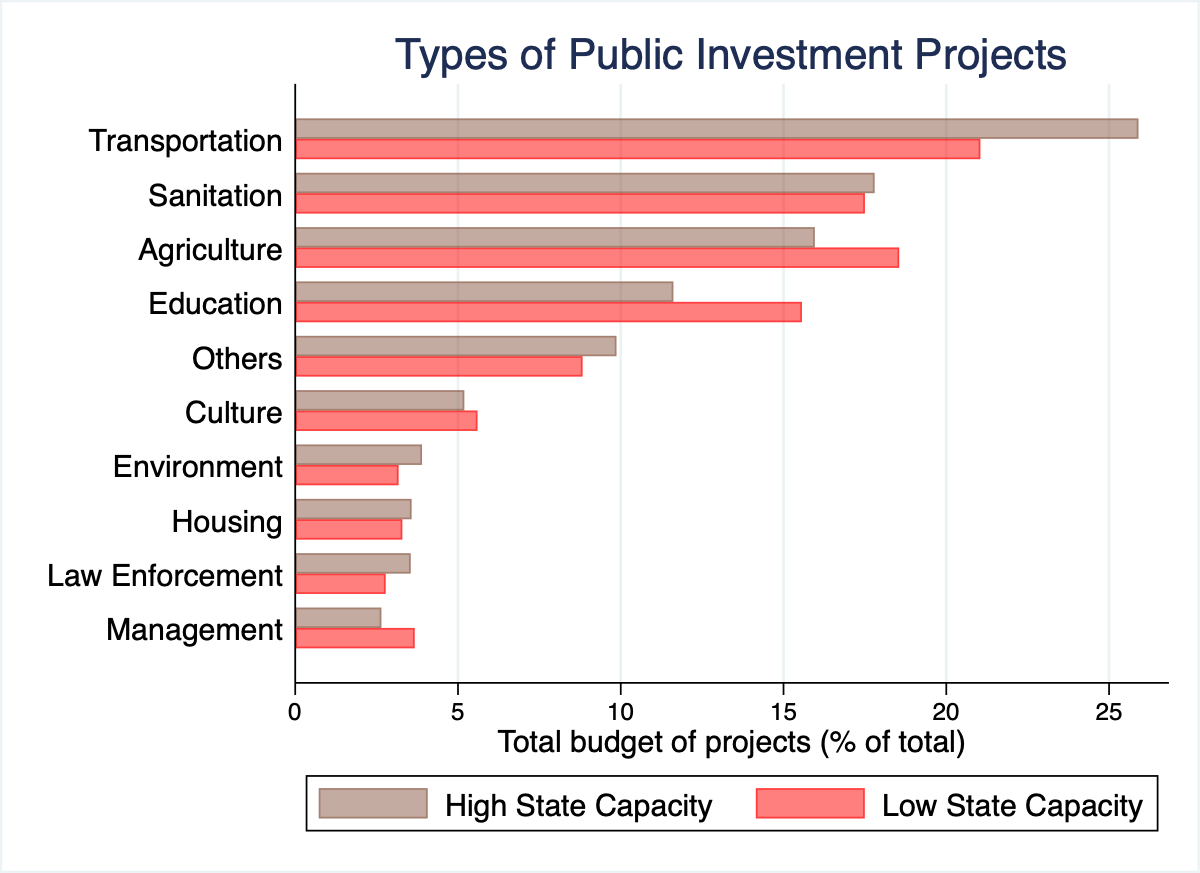}
    \caption{Types of Public Investment Projects by State Capacity}
    \label{fig:obratypes}
\end{figure}

\begin{figure}[h!]
    \centering
    \includegraphics[height=8cm]{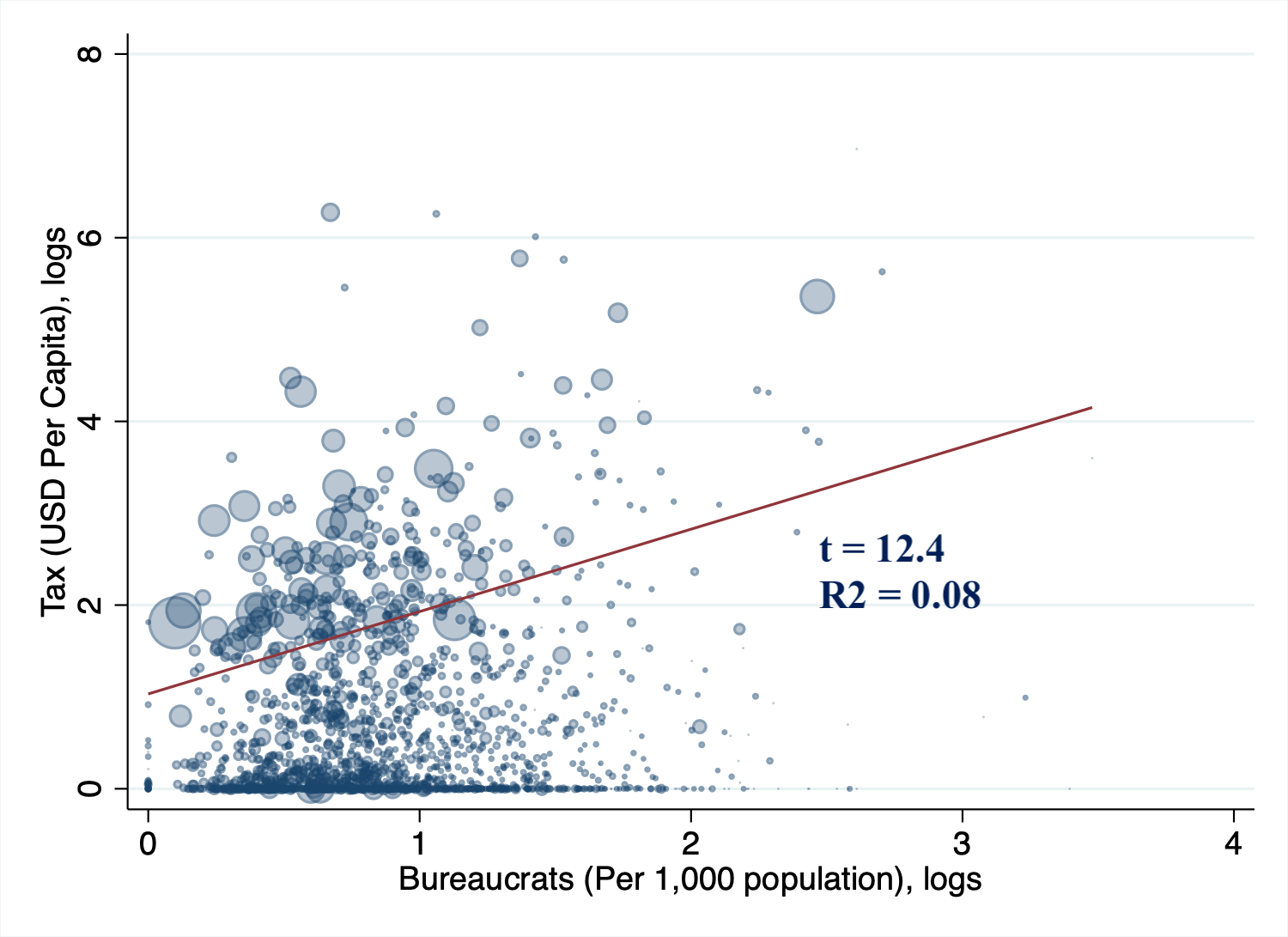}
    \caption{Correlation between two proxies of state capacity (2002)}
    \label{fig:corr}
\end{figure}

%PLACEBO TEST
\begin{figure}[h!]
    \centering
    \includegraphics[height=8.2cm,trim={0cm .5cm 0cm 1cm},clip]{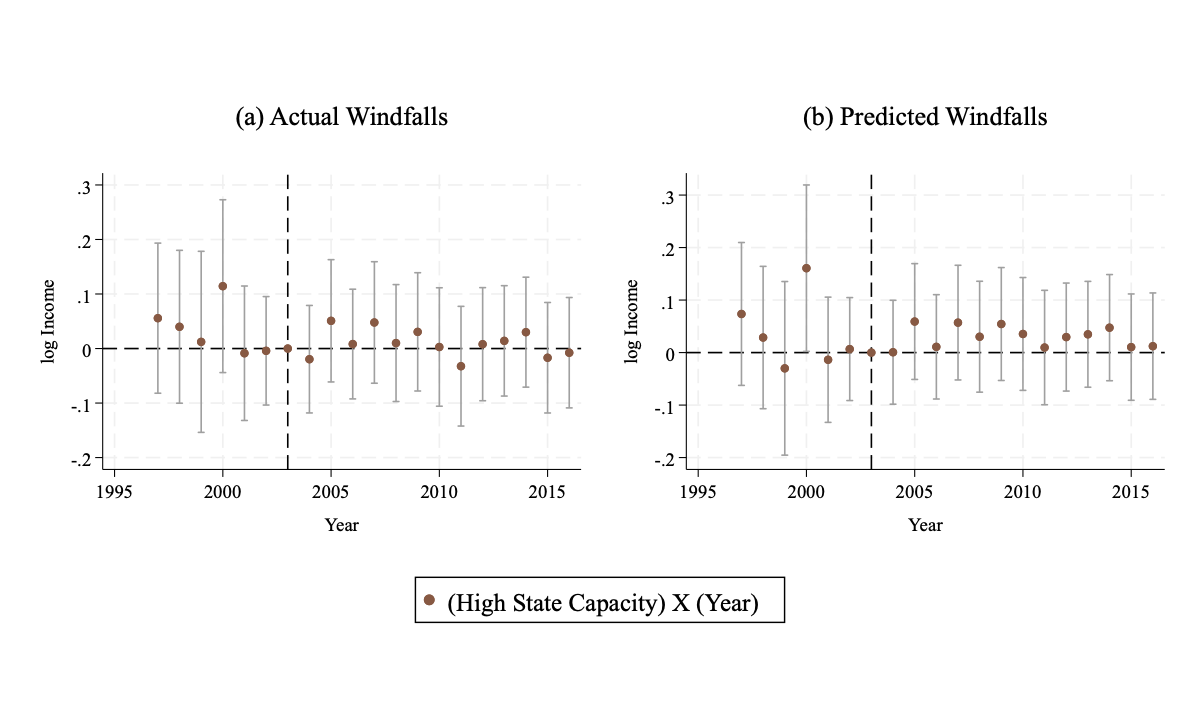}
    \caption{Placebo Test. Outcome: Household Income (log, per capita)}
    \label{fig:placebo}
    \vspace{-.9cm}
    \begin{changemargin}{0.8cm}{0.6cm} 
\begin{spacing}{0.125}
{\footnotesize \setstretch{.125} \textit{High State Capacity}: locality with above median tax per capita collected in pre-boom period.}  
\end{spacing}
\end{changemargin}
\end{figure}

%ROBUSTNESS: BUREAUCRACY PROXY

\begin{table}[p]
\centering
\footnotesize
\captionsetup{justification=centering}
\caption{Robustness Check: Triple difference  using alternative State Capacity  proxy (tertiary educated bureaucrats per capita). Dep. Var.: Household Income (logs, USD)}
\label{tab:robustness_pt}
\input{tables/TableA2_ptpop_HHinc}

 \vspace{-.3cm}
\begin{changemargin}{.93cm}{.93cm} 
\begin{spacing}{0.125}
{\footnotesize \setstretch{.125}Standard errors are clustered at the level of the primary sampling unit (enumeration areas known as conglomerados). Region-time fixed effects include geographic and administrative regions. Columns (1) and (3) show results from estimating equation (\ref{eq:HTE_TWFE}) where $I^{SC}_d$ is defined as districts with above median tertiary educated bureaucrats per capita. Columns (2) and (4) show the modified specification including region-time fixed effects. The 3rd row shows the coefficient $\beta^W$ (Large Windfalls $\times$ Post-Boom Period) which captures the effect of the mining boom's large windfalls on low state capacity (defined with our alternative proxy) districts. The 1st row shows our coefficient of interest, $\beta^{WI}$, the triple interaction coefficient (Large Windfalls $\times$ Post-Boom Period $\times$ High State Capacity), which  measures the windfalls' additional effects on high state capacity districts. The 2nd and 4th rows test for pre-trends in the pre-boom period. }  
\end{spacing}
\end{changemargin}
\end{table}
\clearpage

%ROBUSTNESS: NIGHTLIGHTS
\begin{table}[p]
\centering
\footnotesize
\captionsetup{justification=centering}
\caption{Robustness Check: Triple difference results using alternative outcome variable. Dep. Var.: Nightlights (mean luminosity p.c., cropping mines and expressed in $Z$-scores)}
\label{tab:robustness_lights}
\input{tables/TableA3_Lights}

 \vspace{-.3cm}
\begin{changemargin}{.93cm}{.93cm} 
\begin{spacing}{0.125}
{\footnotesize \setstretch{.125}  Standard errors are clustered at the district level (level of treatment assignment). Region-time fixed effects include geographic and administrative regions. Columns (1) and (3) show results from estimating equation (\ref{eq:HTE_TWFE}). Columns (2) and (4) show the modified specification including region-time fixed effects. The 3rd row shows the coefficient $\beta^W$ (Large Windfalls $\times$ Post-Boom Period) which captures the effect of the mining boom's large windfalls on low state capacity (defined with our alternative proxy) districts. The 1st row shows our coefficient of interest, $\beta^{WI}$, the triple interaction coefficient (Large Windfalls $\times$ Post-Boom Period $\times$ High State Capacity), which  measures the windfalls' additional effects on high state capacity districts. The 2nd and 4th rows test for pre-trends in the pre-boom period. }  
\end{spacing}
\end{changemargin}
\end{table}

\begin{figure}[h!]
    \centering
    \includegraphics[height=8cm,trim={0cm .5cm 0cm 1.5cm},clip]{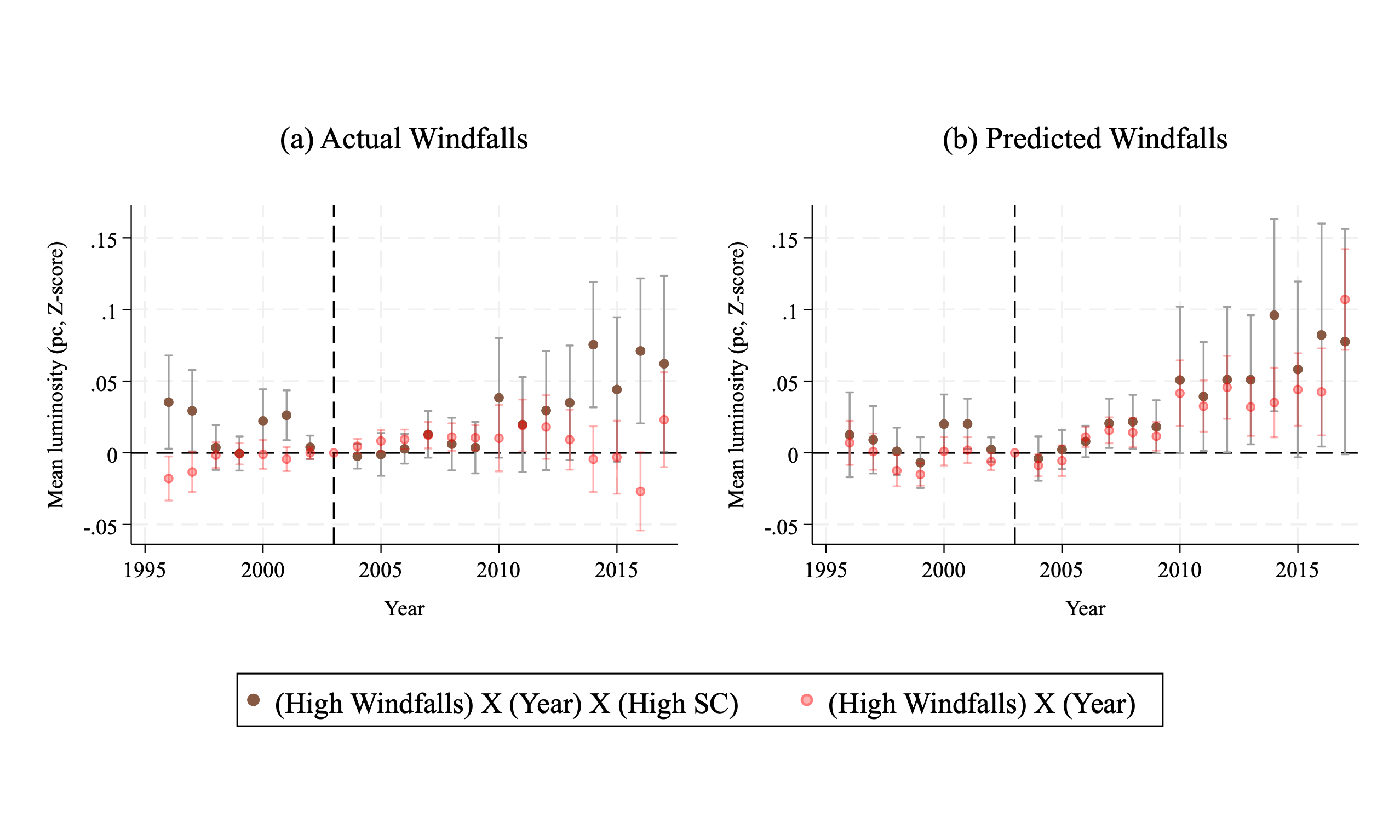}
    \caption{Nightlights (mean luminosity p.c., cropping out mines, in $Z$ scores)}
    \label{fig:nightlights}
        \vspace{-.6cm}
     \begin{changemargin}{0.8cm}{0.6cm} 
\begin{spacing}{0.125}
{\footnotesize \setstretch{.125} \textit{High Windfalls}: locality with large (above the 75th percentile) windfall transfers during boom. \textit{High State Capacity (SC)}: locality with above median tax per capita collected in pre-boom period.}  
\end{spacing}
\end{changemargin}
\end{figure}

\clearpage

%ROBUSTNESS: MEDIAN CUTOFF
\begin{table}[p]
\centering
\footnotesize
\captionsetup{justification=centering}
\caption{Robustness Check: Triple difference results using alternative treatment: districts with above median windfalls per capita (population-weighted 50th percentile) during mining boom. Dep. Var.: Household Income (logs, USD)}
\label{tab:robustness_p50}
\input{tables/TableA4_AltcutoffWindfall50}

 \vspace{-.3cm}
\begin{changemargin}{.93cm}{.93cm} 
\begin{spacing}{0.125}
{\footnotesize \setstretch{.125}Standard errors are clustered at the level of the primary sampling unit (enumeration areas known as conglomerados). Region-time fixed effects include geographic and administrative regions. Columns (1) and (3) show results from estimating equation \ref{eq:HTE_TWFE} (or its predicted windfalls version, equation \ref{eq:HTE_TWFE_IV}) where $W_d$ (or $\widehat W_d$) equals 1 if districts have above median actual or predicted windfalls per capita during 2004-2011. Columns (2) and (4) show the modified specification including region-time fixed effects. The 3rd row shows the coefficient $\beta^W$ (Large Windfalls $\times$ Post-Boom Period) which captures the effect of the mining boom's large windfalls on low state capacity (defined with our alternative proxy) districts. The 1st row shows our coefficient of interest, $\beta^{WI}$, the triple interaction coefficient (Large Windfalls $\times$ Post-Boom Period $\times$ High State Capacity), which  measures the windfalls' additional effects on high state capacity districts. The 2nd and 4th rows test for pre-trends in the pre-boom period. }  
\end{spacing}
\end{changemargin}
\end{table}
\clearpage

%ROBUSTNESS: 90th PERCENTILE CUTOFF
\begin{table}[p]
\centering
\footnotesize
\captionsetup{justification=centering}
\caption{Robustness Check: Triple difference results using alternative treatment: districts with  windfalls per capita above the (population weighted) 90th percentile during mining boom. Dep. Var.: Household Income (logs, USD)}
\label{tab:robustness_p90}
\input{tables/TableA5_AltcutoffWindfall90}

 \vspace{-.3cm}
\begin{changemargin}{.93cm}{.93cm} 
\begin{spacing}{0.125}
{\footnotesize \setstretch{.125}Standard errors are clustered at the level of the primary sampling unit (enumeration areas known as conglomerados). Region-time fixed effects include geographic and administrative regions. Columns (1) and (3) show results from estimating equation \ref{eq:HTE_TWFE} (or its predicted windfalls version, equation \ref{eq:HTE_TWFE_IV}) where $W_d$ (or $\widehat W_d$) equals 1 if districts have actual or predicted windfalls per capita above the 90th percentile during 2004-2011. Columns (2) and (4) show the modified specification including region-time fixed effects. The 3rd row shows the coefficient $\beta^W$ (Large Windfalls $\times$ Post-Boom Period) which captures the effect of the mining boom's large windfalls on low state capacity (defined with our alternative proxy) districts. The 1st row shows our coefficient of interest, $\beta^{WI}$, the triple interaction coefficient (Large Windfalls $\times$ Post-Boom Period $\times$ High State Capacity), which  measures the windfalls' additional effects on high state capacity districts. The 2nd and 4th rows test for pre-trends in the pre-boom period. }  
\end{spacing}
\end{changemargin}
\end{table}
\clearpage

\section{State Capacity Validation}\label{subsec:SC_validate}
\setcounter{figure}{0}
\setcounter{table}{0}
\renewcommand{\thefigure}{B\arabic{figure}}
\renewcommand{\thetable}{B\arabic{table}}
\RaggedRight
 To test the validity of our state capacity proxy we obtain data on all public investment projects undertaken by each local government in Peru from the Ministry of the Economy's website during the boom period (starting in 2008). From this source, we can calculate the total investment budget of each local government in each year, which provides a measure of public good provision in each district. However, we cannot use this as an outcome variable due to the lack of pre-boom data. 
 
We combine our data on public works projects expenditure   for the  2008-2017 period with local windfall transfers  to create a ratio of actual and scheduled expenditure over windfall transfer. We also compute the share of windfall expenditure (e.g. projects finances exclusively with transfers) over total transfers for a particular locality.\footnote{We focus on this period since it is the period for which we have complete data on public goods at the local government level.} Using these shares as outcome variables,  we estimate:
\begin{equation}\label{eq:SC_validation}
    Y_{it}=\alpha + \beta\text{State Capacity}_{i} +\mathbf{X}^{'}_{i}\gamma +  \phi_{t}
\end{equation}
where $Y_{it}$ is the share of windfall transfers spent by local government $i$. We add controls for pre-boom local government share of unmet basic needs and add year fixed effects.  

Table \ref{tab:SC_validation} shows the result of our validation exercise.  We confirm that state capacity predicts higher local government expenditure on public goods. Localities with high state capacity have between $\sim$6\% and $\sim$2\% higher investments shares  than their lower state capacity peers.

\input{tables/TableB1_SCValidation}

\clearpage
\section{Staggered Event Study}\label{app:staggered}
\setcounter{figure}{0}
\setcounter{table}{0}
\renewcommand{\thefigure}{C\arabic{figure}}
\renewcommand{\thetable}{C\arabic{table}}

A limitation of our triple difference strategy outlined in Section \ref{sec:strategy} is that district do not start receiving large windfalls all at once during the boom. Several mines start operating during the mining boom and therefore the districts where they are located, and their neighbor districts, start perceiving windfalls during our study period. To address the staggered nature of mining and windfalls, we rely on a staggered event study design. We define event onset as the first time a district perceives windfalls above the third quartile (75th percentile).\footnote{We choose the third quartile because most districts receive some amount of canon even if small, so the median is low and not as informative since what matters for public investment is receiving sufficiently large windfalls. Moreover, this threshold makes our results directly comparable to our main treatment definition in equation (\ref{eq:HTE_TWFE}).} 

\begin{figure}[h!]
    \centering
    \includegraphics[height=9cm]{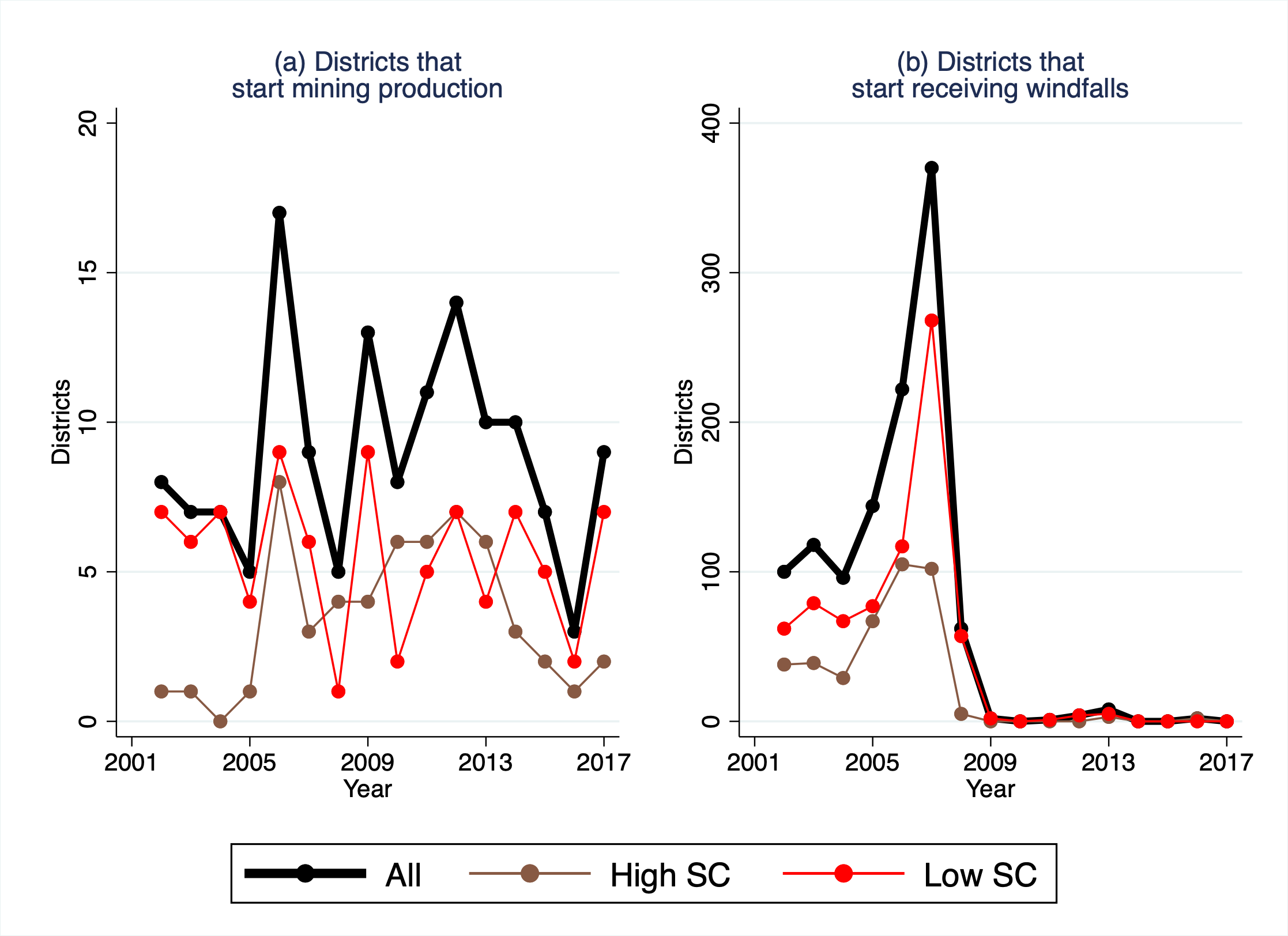}
    \caption{Event timing for staggered event-study design}
    \label{fig:events}
\end{figure}

Figure \ref{fig:events} shows the number of event onsets across time during the resource boom period. Panel (a) shows that the peak of fist-time production happens in 2006, but there are other peaks in 2009 and 2012. On the other hand, Panel (b) shows that there is a single peak in first-time large windfall transfers, in 2007. This might be because, although mines continue to start first-time operations throughout the boom period, the larger (and most productive) mines---that result in highest windfalls---open in the first half of the mining boom. 

The basic specification for the event of a mining  receiving windfalls above the 75th percentile for the first time during the mining boom period is:
\begin{equation}\label{eq:staggered}
    Y_{i d t}=\alpha+\sum_{k=\underline{C}, k\neq -1}^{\overline{C}} \theta_k D_{id t}^k+\phi_d+\phi_{t}+\varepsilon_{id t}
\end{equation}
where $Y_{i d t}$ is log household income of a household $i$ living in district $d$ in year $t, \phi_d$ is a district fixed effect, and $\phi_{t}$ is year fixed effect. We define the event time dummies as $D_{id t}^k:=\bm{1}\left[t=\tau_d+ k\right]$  $\forall k \in(\underline{C},\overline{C}), D_{id t}^{\overline{C}}=\bm{1}\left[t \geqslant \tau_d+\overline{C}\right]$, and $D_{id t}^C=\bm{1}\left[t \leqslant \tau_i+\underline{C}\right]$, where $\bm{1}$ [.] is the indicator function and $\tau_d$ is the first year when district $d$ has mining windfalls above 75th percentile (Q3). We use $\underline{C}=-3$ and $\overline{C}=+4$ since most event onsets for large transfers happen in the first half of the boom. $\varepsilon_{id t}$ is an error term; we cluster standard errors at the district level.\footnote{We exclude from our analysis those districts that had large mining windfalls prior to 2003, since a limitation of our data is it starts in 2001 and thus, we cannot rule out that these districts' first-time event was before 2001.}

For our study of heterogeneous effects by state capacity we estimate (\ref{eq:staggered}) splitting the sample into above and below median state capacity. A potential concern  is that high and low state capacity districts experience event onsets at different points in time. However, Figure \ref{fig:events} shows  that the pattern of event onsets for these two sets of districts is similar.

\begin{figure}
    \centering    
        \captionsetup{justification=centering}
    \includegraphics[height=9cm,trim={0cm 1cm 0cm 0cm},clip]{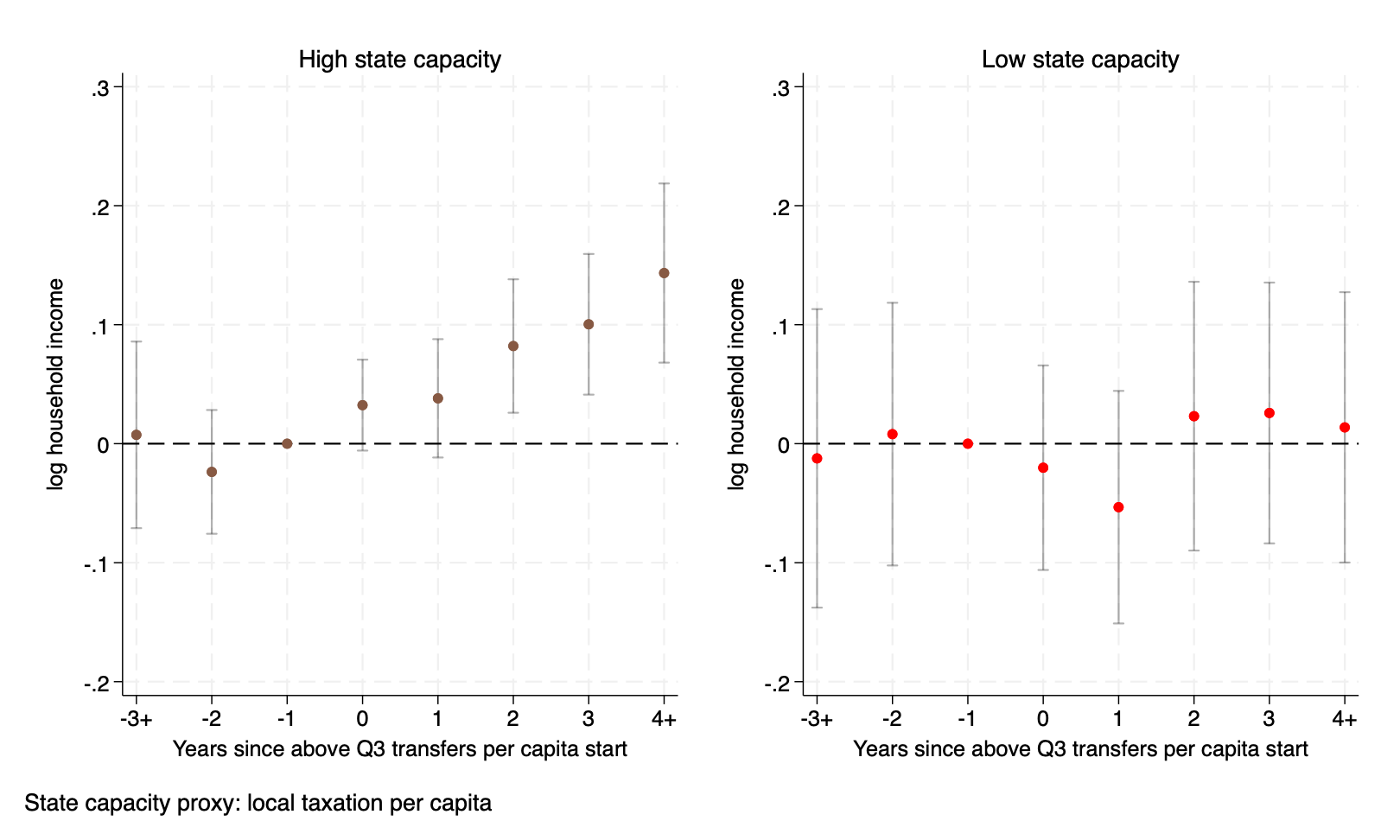}
    \caption{Staggered Event Study for Windfalls: Heterogeneous Effects by State Capacity}
    \label{fig:staggered_hte}
\end{figure}

We  run equation (\ref{eq:staggered}) but splitting the sample into high and low state capacity districts. Table \ref{tab:stagcanon}'s Column (1) shows the results with the high state capacity sample for five distinct time periods before and after event onset. Column (2) shows the same results for the low state capacity group.\footnote{See Figure \ref{fig:staggered_hte} for the event study coefficients for all periods. Panel \ref{fig:staggered_hte}(a)  plots the estimated $\theta_k$ event study coefficients from equation (\ref{eq:staggered}) for the high state capacity subsample, where the dependent variable is log household income. Household income increases gradually after the -1 period. There is a gradual increasing effect that becomes significant to the 95\% confidence level in period +1. Note that 4+ years after the mining production starts, the effect size is  around .2 log points. This is consistent with the findings in Table \ref{tab:main_hte} Columns 1-2 and Figure \ref{fig:results_logy_HH} Panel (a). On the other hand, Panel \ref{fig:staggered_hte}(b) shows the event study coefficients for low state capacity districts. Noticeably, these are not statistically distinguishable from 0, suggesting that only the high state capacity districts are benefiting from the large windfall transfers. Consistent with our triple difference design, gains in household income are only present in the high state capacity sample. } 

Column (3) reveals that the difference in living standards between high and low state capacity  mining localities (as measured by tax per capita) are indeed statistically significant. The difference is significant 1 year after the event onset and remains significant in the medium-to-long run (4+ years after receiving large windfalls for the first time). These results are consistent with those of in Table \ref{tab:main_hte} Columns 1-2.\footnote{We can only compare this design to the actual windfalls treatment since we cannot use predicted output as reliable event onsets.} This is reassuring because it suggests that our results are robust to changing our empirical strategy to a staggered event study.

\begin{landscape}
 \input{tables/TableC1_EventStudy}
 \clearpage

\end{landscape}

\section{Local Governments}\label{app:localgovt}
Local governments have several responsibilities and powers that they can exercise to promote social and economic development. They can invest windfalls in social and economic development public work projects, which helps improve local infrastructure. Additionally, local governments have the authority to collect local taxes and set and collect local contribution rates and fees. This financial autonomy allows them to generate revenue for local projects and services. Furthermore, they play a vital role in overseeing dialogue with the local community and civil society, ensuring that various stakeholders have a voice in local governance.

However, there are several areas where local governments do not have authority. They cannot give permission for mines to open; this responsibility lies with the Central Government. Similarly, local governments do not have the power to collect mining windfalls or set windfall tax rates, as these are also managed by the Central Government. Additionally, they cannot create, modify, or eliminate taxes, nor can they set tax rates, which are functions reserved for the Central Government. Local governments are also not authorized to limit, regulate, or sanction mining operations. 

Setting labor protection rules is beyond their jurisdiction, as is setting rules for mining contracting, such as local worker quotas, human capital investments, and environmental responsibility. These latter responsibilities are shared among the Central Government, civil society, indigenous communities, and the mining firms themselves, based on their goodwill.

We conducted interviews with mining sector stakeholders for them to define the relationship between  local governments and the mine operating within a district's borders. The takeaway from this qualitative work was that the mining firm is mainly concerned with the Central Government since they obtain their ability to operate from them. The local government plays a very minor role in talking with the mine.

\end{document}

%% file: tables/Table1_Balance.tex
{\footnotesize
\begin{tabular}{l*{7}c}
\hline\hline\\
 & (1) & (2) & (3) & (4) & (5) & (6) & (7) \\
 & All & Treated & Control & Diff. (2-3) & High SC & Low SC & Diff. (5-6) \\
\hline \\
\textit{\underline{Panel A: Household-level variables}} \\\\
Monthly household income (USD per capita) &63.528&44.291&70.297&-26.007***&85.594&40.235&45.359***\\
&(92.318)&(79.599)&(95.184)&(0.834)&(105.693)&(67.938)&(0.721)\\
[.2em]
Female head of household&0.199&0.192&0.201&-0.009***&0.214&0.181&0.034***\\
&(0.399)&(0.394)&(0.401)&(0.004)&(0.410)&(0.385)&(0.003)\\
[.2em]
Age in years&47.813&48.219&47.672&0.548***&48.234&47.370&0.864***\\
&(15.334)&(15.756)&(15.182)&(0.136)&(15.113)&(15.559)&(0.120)\\
[.2em]
\% employed in agriculture&36.139&53.951&29.429&24.523***&13.039&58.021&-44.982***\\
&(48.041)&(49.845)&(45.573)&(0.439)&(33.674)&(49.353)&(0.357)\\
[.2em]
\% employed in services&53.054&36.621&59.245&-22.624***&72.989&34.265&38.724***\\
&(49.907)&(48.178)&(49.139)&(0.458)&(44.403)&(47.460)&(0.388)\\
[.2em]
Access to sewage&0.633&0.469&0.691&-0.222***&0.822&0.427&0.396***\\
&(0.482)&(0.499)&(0.462)&(0.004)&(0.382)&(0.495)&(0.003)\\
[.2em]
Access to electricity&0.733&0.571&0.789&-0.219***&0.927&0.519&0.408***\\
&(0.443)&(0.495)&(0.408)&(0.004)&(0.260)&(0.500)&(0.003)\\\\ \hline\\
Observations & 66,433 & 21,810 & 44,623 & 66,433 & 28,182 & 37,302 & 66,433 \\\\\hline\\

\\
\textit{\underline{Panel B: District-level variables}}\\\\
Mining Windfalls (USD per capita) &1.221&3.823&0.318&3.506***&0.842&1.631&-0.789***\\
&(5.183)&(9.521)&(1.247)&(0.094)&(4.043)&(6.180)&(0.088)\\
[.2em]
Population&122,082&34,944&162,126&-127,200***&202,060&43,284&158,775***\\
&(160,571)&(47,213)&(183,828)&(3,025)&(186,038)&(66,040)&(1,926)\\
[.2em]
Municipality Taxes (USD per capita) &15.307&4.542&18.839&-14.298***&28.953&1.300&27.654***\\
&(45.520)&(18.704)&(50.889)&(0.864)&(60.927)&(1.579)&(0.596)\\
[.2em]
Stock of roads per capita ($m^2$)&0.304&0.433&0.259&0.174***&0.407&0.203&0.204***\\
&(1.421)&(1.789)&(1.265)&(0.038)&(1.062)&(1.724)&(0.034)\\
[.2em]
Unsatisfied basic needs (1993)&98.641&122.979&90.193&32.786***&57.043&140.001&-82.957***\\
&(62.009)&(58.856)&(60.833)&(1.139)&(28.353)&(58.649)&(0.770)\\\\\hline\\
Observations & 21,984 & 6,976 & 7,680 & 21,984 & 3,036 & 18,180 & 21,984 \\\\

\hline\hline
\end{tabular}
}

%% file: tables/Table2_FirstStageOutput.tex
{\begin{table}[h!]\centering \footnotesize
\def\sym#1{\ifmmode^{#1}\else\(^{#1}\)\fi}
\caption{Relationship between district's metal density and mining output}
\label{tab:firststage}
\begin{tabular}{l*{2}{c}}
\hline\hline\\
                    &\multicolumn{1}{c}{(1)}         &\multicolumn{1}{c}{(2)}         \\
                    &Mining output (logs, USD)         &Mining output (logs, USD)         \\\\
\hline \\
Copper density &       0.372\sym{***}&       0.383\sym{***}\\
(90th percentile, logs, ppm)                    &   (0.00415)         &   (0.00463)         \\
[1em]
Gold density (logs, ppm)&      -0.108\sym{***}&      -0.221\sym{***}\\
(90th percentile, logs, ppm)                    &   (0.00547)         &   (0.00667)         \\\\
\hline\\
Adjusted R$^2$      &       0.042         &        0.16         \\\\
Observations        &      205,042         &      205,042         \\\\
Year FEs            &         Yes         &         Yes         \\\\
Region FEs          &          No         &          No         \\\\
F-Statistic         &     4,081.10         &     3,657.42         \\\\
\hline\hline\\
\multicolumn{3}{l}{\footnotesize Standard errors in parentheses}\\
\multicolumn{3}{l}{\footnotesize \sym{*} \(p<0.10\), \sym{**} \(p<0.05\), \sym{***} \(p<0.01\)}\\
\end{tabular}
\end{table}
}

%% file: tables/Table3_FirstStageWindfalls.tex
\begin{table}[h!]\centering
{\footnotesize
\def\sym#1{\ifmmode^{#1}\else\(^{#1}\)\fi}
\caption{First stage of  reduced form geology IV}
\label{tab:firstIV}
\begin{tabular}{l*{4}{c}}
\hline\hline \\
                    &\multicolumn{1}{c}{(1)}         &\multicolumn{1}{c}{(2)}         &\multicolumn{1}{c}{(3)}         &\multicolumn{1}{c}{(4)}         \\
                    &Actual windfalls         &Actual windfalls         &High windfalls         &High windfalls         \\
                    &(logs, p/c)         &(logs, p/c)        &dummy         &dummy         \\\\
\hline \\
Predicted windfall (logs, pc)&       2.231\sym{***}&       1.492\sym{***}&                     &                     \\
                    &     (0.168)         &     (0.107)         &                     &                     \\
[1em]
High pred. windfall dummy&                     &                     &       0.371\sym{***}&       0.265\sym{***}\\
                    &                     &                     &    (0.0348)         &    (0.0298)         \\\\
\hline\\
Adjusted R$^2$      &        0.38         &        0.82         &        0.14         &        0.72         \\\\
Observations        &       23,616         &       23,616         &       36,640         &       36,640         \\\\
\# Districts        &        1,390         &        1,390         &        1,832         &        1,832         \\\\
UBN x Year Controls &          Yes           &            Yes         &          Yes           &        Yes             \\\\
Region FEs     &          No           &      Yes               &             No        &      Yes               \\\\
F-Statistic         &      176.91         &      196.11         &      113.75         &       79.03         \\\\
\hline\hline\\
\multicolumn{5}{l}{\footnotesize Standard errors in parentheses}\\
\multicolumn{5}{l}{\footnotesize \sym{*} \(p<0.10\), \sym{**} \(p<0.05\), \sym{***} \(p<0.01\)}\\
\end{tabular}
}
\end{table}

%% file: tables/Table4_HTE_HHinc.tex
\begin{table}
\caption{Triple Differences Result. Dep. Var.: Household Income (logs, per capita)}
\label{tab:main_hte}
\begin{tabular}{l*{5}{c}} \\ 
\hline\hline &     &       &    &     &        \\
Treatment:
& \multicolumn{2}{c}{Real} & & \multicolumn{2}{c}{Predicted} \\
\\
                  \cline { 2 - 3}\cline{5-6} \\
                                       &\multicolumn{1}{c}{(1)}         &\multicolumn{1}{c}{(2)}         & &\multicolumn{1}{c}{(3)}         &\multicolumn{1}{c}{(4)}         \\
                                       \\
\hline       \\
Windfalls $\times$ Post-Boom $\times$ SC      &       0.226\sym{***} &       0.239\sym{**} &    &   0.186\sym{**}  &       0.165\sym{*}         \\
                    &    (0.0843)         &    (0.0973)     &    &    (0.0898)         &    (0.0921)         \\
[1em]
Windfalls $\times$ Pre-boom $\times$ SC      &      0.0884         &       0.238\sym{**} &  &    0.120         &       0.117         \\
                    &    (0.0955)         &     (0.110)      &   &     (0.0978)         &     (0.100)         \\
[1em]
Windfalls $\times$ Post-Boom        &     -0.0689         &      -0.206\sym{**} &    & -0.0928         &      -0.215\sym{***} \\
                    &    (0.0539)         &    (0.0828)      &   &    (0.0588)         &    (0.0754)         \\
                    \\
Windfalls $\times$ Pre-boom         &     -0.0280         &      -0.203\sym{**}     &    &     -0.0902         &     -0.141         \\
                    &    (0.0635)         &     (0.0982)      &   &    (0.0683)         &    (0.0860)         \\
[1em]                    
\hline \\
Adjusted R$^2$      &        0.41         &        0.41     &    &        0.41         &        0.41         \\ \\
Observations      &      348,647       &     348,647       &  &      348,647         &      348,647         \\ \\
\# Clusters        &       10,444         &       10,444         &  &      10,444         &       10,444         \\ \\
Fixed Effects &         Time         &         Region-Time         & &         Time         &         Region-Time         \\ \\

\hline\hline 
\end{tabular}
\vspace{-.3cm}
\begin{changemargin}{0.25cm}{0.25cm} 
\begin{spacing}{0.125}
{\footnotesize \setstretch{.125}  Standard errors clustered at the  level of the primary sampling unit (enumeration areas known as \textit{conglomerados}). Columns (1) and (3) show our results from estimating equation (\label{eq:1S}). Columns (2) and (4) show a modified specification including region-time fixed effects (relying on both political and geographic regions). The 3rd row shows the coefficient $\beta^W$ (Large Windfalls $\times$ Post-Boom Period) which captures the effect of the mining boom’s large windfalls on low state capacity districts. The 1st row shows our coefficient of interest, $\beta^{WI}$, the triple interaction coefficient (Large Windfalls $\times$ Post-Boom Period $\times$ High State Capacity), which measures the windfalls' additional effects on high state capacity districts. The 2nd and 4th rows test for pre-trends in the pre-boom period.}  
\end{spacing}
\end{changemargin}
\end{table}

%% file: tables/Table5_Mecha1.tex
\begin{table}[p]\centering
\def\sym#1{\ifmmode^{#1}\else\(^{#1}\)\fi}
 \caption{Mechanisms: Windfall transfers and public good provision}
\label{tab:mecha1}
{\footnotesize
\begin{tabular}{l*{12}{c}}
\hline\hline \\ 
Dep. Var.: & \multicolumn{2}{c}{Windfalls (PEN, p/c)} & & \multicolumn{2}{c}{ Roads Stock (logs, p/c)}  & & \multicolumn{2}{c}{HH Has Sewage} & & \multicolumn{2}{c}{HH Has Electricity} \\ 
                  \cline { 2 - 3}\cline{5-6} \cline{8-9} \cline{11-12} \\
                    &\multicolumn{1}{c}{(1)}         &\multicolumn{1}{c}{(2)}      &   &\multicolumn{1}{c}{(3)}         &\multicolumn{1}{c}{(4)}       &  &\multicolumn{1}{c}{(5)}         &\multicolumn{1}{c}{(6)}        & &\multicolumn{1}{c}{(7)}         &\multicolumn{1}{c}{(8)}         \\ \\
\hline \\
Windfalls  $\times$ Post-Boom $\times$ SC          &      -61.44         &       1.694         &   &    0.173\sym{*}  &       0.285\sym{***}&  &     0.0907\sym{***}&      0.0759\sym{*}  &   &   0.0249         &      0.0118         \\
                                            &     (39.85)         &     (53.13)         & &    (0.0948)         &     (0.102)         &   & (0.0352)         &    (0.0432)         & &   (0.0291)         &    (0.0351)         \\
[1em]
Windfalls  $\times$ Pre-boom $\times$ SC      &       8.130         &      -3.091        & &      -0.114         &    -0.00445       &  &       0.173\sym{***}&       0.0773         &  &     0.141\sym{***}&      0.0426         \\
                                            &     (6.119)         &     (6.039)         & &    (0.0777)         &    (0.0549)         & &    (0.0414)         &    (0.0517)         & &   (0.0339)         &    (0.0424)         \\
[1em]
Windfalls  $\times$ Post-Boom                  &       260.3\sym{***}&       186.1\sym{***}&  &    0.0310         &     -0.00723         &  &  0.00595         &      0.0417\sym{*}  & &   0.00536         &     -0.0280         \\
                                         &     (30.91)         &     (29.89)         &  &   (0.0359)         &    (0.0374)         & &   (0.0221)         &    (0.0230)         & &   (0.0232)         &    (0.0237)         \\
[1em]
Windfalls  $\times$ Pre-boom             &      -22.42\sym{***}&      -6.442\sym{**}  &  &   0.00625         &      0.0196         &  &   -0.0816\sym{***}&     -0.0135         & &    -0.0970\sym{***}&     -0.0538\sym{*}  \\
                                        &     (3.397)         &     (3.209)         & &   (0.0248)         &    (0.0284)         & &   (0.0276)         &    (0.0282)         &  &  (0.0278)         &    (0.0284)         \\\\            
\hline \\
Adjusted R$^2$       &        0.44         &        0.43         &    &    0.59         &        0.59         &   &     0.37         &        0.37         &  &      0.49         &        0.49         \\\\
Observations        &       39,026         &       39,026         &   &    30,158         &       30,158         &    &  382,931         &      382,931         &   &   382,931         &      382,931         \\\\
\# Districts        &        1,774         &        1,774         &  &      1,774         &        1,774         &  &      1,774         &        1,774         &   &     1,774         &        1,774         \\\\
  Windfalls dummy ($>$p75)  &Actual         &      Predicted        &  &Actual         &      Predicted       & &Actual         &       Predicted     &   &Actual         &       Predicted     \\ \\
\hline\hline \\
\end{tabular}
}
\vspace{-.5cm}
\begin{changemargin}{0.05cm}{-0.4cm} 
\begin{spacing}{0.125}
{\footnotesize \setstretch{.125}  Standard errors clustered at the district level (level of treatment assignment).  Standard errors in parentheses. Odd numbered columns show results for large windfalls defined as actual windfalls above the (population-weighted) 75th percentile during the mining boom. Even numbered columns  define ``large windfalls'' as predicted windfalls (estimated from predicting output using district geology) above the (population-weighted) 75th percentile.}  
\end{spacing}
\end{changemargin}
\end{table}

%% file: tables/Table6_Mecha2.tex
\begin{table}[p]\centering
\def\sym#1{\ifmmode^{#1}\else\(^{#1}\)\fi}
 \caption{Mechanisms: Price convergence and structural transformation}
\label{tab:mecha2}
{\footnotesize
\begin{tabular}{l*{9}{c}}
\hline\hline \\ 
Dep. Var.: & \multicolumn{2}{c}{Rel. price deviation } & & \multicolumn{2}{c}{ \% in Agriculture}  & & \multicolumn{2}{c}{\% in Services}    \\ 
                  \cline { 2 - 3}\cline{5-6} \cline{8-9}  \\
                    &\multicolumn{1}{c}{(1)}         &\multicolumn{1}{c}{(2)}      &   &\multicolumn{1}{c}{(3)}         &\multicolumn{1}{c}{(4)}       &  &\multicolumn{1}{c}{(5)}         &\multicolumn{1}{c}{(6)}           \\ \\
\hline \\
Windfalls  $\times$ Post-Boom $\times$ SC       &     -0.0110         &   -0.000104         &  &    -2.072         &      -7.137\sym{**} &   &    3.032         &       10.53\sym{***}          \\
                    &    (0.0132)         &    (0.0155)    &     &     (2.995)         &     (3.585)    &     &     (3.142)         &     (3.399)               \\
[1em]
Windfalls  $\times$ Pre-boom $\times$ SC       &     0.0278         &     0.00564         & &     -7.553\sym{**}  &       0.393         & &      5.419         &       1.515             \\
                    &    (0.0220)         &    (0.0140)       &  &     (3.498)         &     (4.386)     &    &     (3.792)         &     (4.107)               \\
[1em]
Windfalls  $\times$ Post-Boom             &      0.0144         &      0.0139     &    &      -1.269         &       2.778      &   &       1.009         &      -2.674           \\
                    &   (0.00956)         &   (0.00927)   &      &     (1.945)         &     (1.867)       &  &     (1.839)         &     (1.718)             \\
[1em]
Windfalls  $\times$ Pre-boom          &    -0.00606         &    -0.00535       &  &       2.650         &      -0.0634      &   &      -2.622         &       0.612           \\
                    &   (0.00553)         &   (0.00721)    &     &     (2.348)         &     (2.311)      &   &     (2.250)         &     (2.152)             \\\\
\hline \\
Adjusted R$^2$       &        0.57         &        0.57    &     &        0.51         &        0.51         &     &   0.33         &        0.33             \\\\
Observations        &       25,275         &       25,275    &     &      333,095         &      333,095         &  &    333,095         &      333,095                 \\\\
\# Clusters       &        1,694         &        1,694     &    &        10,452         &        10,452         &  &      10,452         &        10,452               \\\\
  Windfalls $>$p75  &Actual         &      Predicted        &  &Actual         &      Predicted       & &Actual         &       Predicted          \\ \\
\hline\hline 
\end{tabular}
}
\vspace{-.3cm}
\begin{changemargin}{2.4cm}{2.63cm} 
\begin{spacing}{0.125}
{\footnotesize \setstretch{.125}  Standard errors clustered at the district level (level of treatment assignment) for specifications that use district-panel data and the level of the primary sampling unit (enumeration areas known as \textit{conglomerados}) for repeated cross-section survey data.  Standard errors in parentheses. Odd numbered columns show results for large windfalls defined as actual windfalls above the (population-weighted) 75th percentile during the mining boom. Even numbered columns define large windfall treatment as predicted windfalls (estimated from predicting output using district geology) above the (population-weighted) 75th percentile.}  
\end{spacing}
\end{changemargin}
\end{table}

%% file: tables/TableA1_DescriptiveStatistics_ALL.tex
\begin{table}[p]
    \centering
    \caption{Summary Statistics tables showing mean, median, minimum, maximum values and standard 
    deviation for variables}
    \label{tab:ds}
    \begin{tabular}{lcccccc}
        \toprule
        & & Mean & Median & Min & Max & SD \\
        \midrule
        \textit{\underline{Panel A: Household-level data}} \\
        \\
        \textit{(N = 302,526 households; 1,560 districts)} \\
        Monthly household income (USD per capita) & & 107.94 & 64.71 & 0.02 & 8,417.35 & 135.63 \\
        Head is female (dummy) & & 0.17 & 0.00 & 0.00 & 1.00 & 0.37 \\
        Head's age & & 46.90 & 46.00 & 14.00 & 98.00 & 13.58 \\
        Head's education & & 8.56 & 9.00 & 0.00 & 18.00 & 4.40 \\
        Located in urban area (dummy) & & 0.66 & 1.00 & 0.00 & 1.00 & 0.47 \\
        Access to sewage (dummy) & & 0.63 & 1.00 & 0.00 & 1.00 & 0.48 \\
        Access to electricity (dummy) & & 0.81 & 1.00 & 0.00 & 1.00 & 0.39 \\
        Access to drinking water (dummy) & & 0.69 & 1.00 & 0.00 & 1.00 & 0.46 \\
        Head works in agriculture (dummy) & & 0.39 & 0.00 & 0.00 & 1.00 & 0.49 \\\\\\
        \textit{\underline{Panel B: District-level data}} \\
        \\
        \textit{(N = 1,832 districts)} \\
        Roads built (total m2 per capita, 1994-2016) & & 4.94 & 1.15 & 0.00 & 599.66 & 20.66 \\
        Municipal taxes per capita (USD, 2002-2003) & & 5.30 & 0.23 & 0.00 & 1057.96 & 37.04 \\
        Mining output per capita  (yearly USD, 2004-2011) & & 1,546.66 & 411.97 & 0.00 & 109,941.96 & 5,344.34 \\
        Windfalls per capita (yearly USD, 2004-2011) & & 16,233.36 & 0.00 & 0.00 & 10,976,326 & 270,731.93 \\
        Poverty Index (1993 Unmet Basic Needs; 0-5 scale) & & 1.48 & 1.51 & 0.05 & 3.48 & 0.54 \\
        Conflicts per 10K people (2004-2017) & & 0.43  & 0 & 0 &  102.73 & 3.21\\
        \bottomrule
    \end{tabular}
\end{table}

%% file: figures/FigureA7_hypothesized_mech.tex
%TIKZ SETUP
    \tikzstyle{cloud}     = [rectangle, draw,fill=black!1, node distance=3cm, text width=3cm, rounded corners, minimum height=0.5cm]
    \tikzstyle{arrow}             = [draw, -latex']
    \tikzstyle{segment_dot} = [draw, densely dotted]
    \tikzstyle{segment}        = [draw]
    \tikzstyle{box_light}        = [rectangle, draw, fill=black!1, text width=2.2cm, rounded corners, minimum height=1.3cm]
    \tikzstyle{box_light_big}        = [rectangle, draw, fill=black!1, text width=3cm, rounded corners, minimum height=0.5cm]
   \tikzstyle{box_light_bigger}        = [rectangle, draw, fill=black!1, text width=3.5cm, rounded corners, minimum height=1cm]

\begin{tikzpicture}
 \node[box_light, ] (q_0) {{\begin{center}
 \vspace{-5mm}                {\footnotesize Commodity Boom}
 \end{center}}
            };
  \node[box_light,]          (q_1) [above right=of q_0] {\begin{center}\vspace{-7mm} 
      {\footnotesize High State Capacity}
  \end{center}};
  \node[box_light,]          (q_2) [below right=of q_0]  {\begin{center}\vspace{-7mm} 
      {\footnotesize Low State Capacity}
  \end{center}};
    \node[box_light_bigger,]          (q_11) [ right=of q_1]  {\begin{center}\vspace{-5mm} 
          {\footnotesize Windfalls Invested Well}
  \end{center}};
    \node[box_light_bigger,]          (q_21) [ right=of q_2]  {\begin{center}\vspace{-11mm} 
          \item {\footnotesize Windfalls Wasted}
  \end{center}};
   \node[box_light_bigger, ] (q_31)  [ right=of q_11] {{\begin{center}
 \vspace{-5mm}                {\footnotesize Infrastructure, \\  Market Access, \\ Structural Transformation, Private Investment, \textcolor{blue}{Development}}
 \end{center}}
            };
   \node[box_light_bigger, ] (q_32)  [ right=of q_21] {{\begin{center}
 \vspace{-5mm}                {\footnotesize No Development, \\ Political Grievances, \\ Unrest \& Conflict, \\ No Private Investment, \\ Possible \textcolor{red}{negative effects} in the long-run}
 \end{center}}
            };
  \path[-] (q_0) edge              node [above left]  {} (q_1)
                  edge              node [below left]  {} (q_2)
            ;
\path[->] (q_1) edge              node [ left]  {} (q_11)
            ;
\path[->] (q_2) edge              node [ left]  {} (q_21)
            ;
\path[->] (q_11) edge              node [ left]  {} (q_31)
            ;
\path[->] (q_21) edge              node [ left]  {} (q_32)
            ;
\end{tikzpicture}

%% file: tables/TableA2_ptpop_HHinc.tex
\begin{tabular}{l*{5}{c}} \\ 
\hline\hline &     &       &    &     &        \\
Treatment:
& \multicolumn{2}{c}{Actual Windfalls $>$ p75} & & \multicolumn{2}{c}{ Predicted Windfalls $>$p75} \\
\\
                  \cline { 2 - 3}\cline{5-6} \\
                                       &\multicolumn{1}{c}{(1)}         &\multicolumn{1}{c}{(2)}         & &\multicolumn{1}{c}{(3)}         &\multicolumn{1}{c}{(4)}         \\
                                       \\
\hline       \\
Windfalls $\times$ Post-Boom $\times$ SC       &       0.108\sym{*}  &       0.114\sym{*}  &    &   0.142\sym{**} &       0.110\sym{*}  \\
                    &    (0.0601)         &    (0.0614)        & &    (0.0625)         &    (0.0597)         \\
[1em]
Windfalls $\times$ Pre-boom $\times$ SC       &       0.118         &      0.0932      &   &       0.208\sym{**} &       0.249\sym{**} \\
                    &    (0.0955)         &    (0.0966)      &   &    (0.0980)         &     (0.102)         \\
[1em]
Windfalls $\times$ Post-Boom        &     -0.0663         &      -0.125\sym{**} &   &  -0.0746         &      -0.161\sym{***}\\
                    &    (0.0418)         &    (0.0612)       &  &    (0.0520)         &    (0.0561)         \\
[1em] 
Windfalls $\times$ Pre-boom        &      -0.114         &      -0.128       &  &      -0.114         &      -0.160\sym{*}  \\
                    &    (0.0719)         &     (0.119)      &   &    (0.0799)         &    (0.0886)         \\
[1em]                    
\hline \\
Adjusted R$^2$      &        0.41         &        0.41      &   &        0.41         &        0.41         \\\\
Observations        &      338798         &      338798     &    &      338798         &      338798         \\\\
\# Districts        &        1452         &        1452      &   &        1452         &        1452         \\\\
Fixed Effects                 &        Time         & Region-Time         & &       Time         & Region-Time         \\ \\

\hline\hline 
\end{tabular}

%% file: tables/TableA3_Lights.tex
\begin{tabular}{l*{5}{c}} \\ 
\hline\hline &     &       &    &     &        \\
Treatment:
& \multicolumn{2}{c}{Actual Windfalls $>$ p75} & & \multicolumn{2}{c}{ Predicted Windfalls $>$p75} \\
\\
                  \cline { 2 - 3}\cline{5-6} \\
                                       &\multicolumn{1}{c}{(1)}         &\multicolumn{1}{c}{(2)}         & &\multicolumn{1}{c}{(3)}         &\multicolumn{1}{c}{(4)}         \\
                                       \\
\hline       \\
Windfalls $\times$ Post-Boom $\times$ SC        &       0.196\sym{**} &      0.0635     &    &       0.148\sym{**} &     0.00945         \\
                    &    (0.0837)         &    (0.0809)        & &    (0.0748)         &    (0.0631)         \\
[1em]
Windfalls $\times$ Pre-boom $\times$ SC      &     -0.0730\sym{**} &     -0.0500\sym{*}  & &     -0.0387         &    -0.00719         \\
                    &    (0.0292)         &    (0.0290)    &     &    (0.0261)         &    (0.0247)         \\
[1em]
Windfalls $\times$ Post-Boom       &     -0.0577\sym{*}  &     -0.0332     &    &      -0.164\sym{***}&      -0.186\sym{***}\\
                    &    (0.0330)         &    (0.0554)     &    &    (0.0346)         &    (0.0446)         \\
[1em] 
Windfalls $\times$ Pre-boom       &      0.0253\sym{*}  &    0.000219      &   &     0.00683         &     -0.0221\sym{*}  \\
                    &    (0.0142)         &    (0.0227)    &     &    (0.0118)         &    (0.0129)         \\
[1em]                    
\hline \\\\
Adjusted R$^2$       &        0.98         &        0.98         &   &     0.98         &        0.98         \\\\
Observations        &       36,855         &       36,855         &   &    36,855         &       36,855         \\\\
\# Districts        &        1,755         &        1,755         &   &     1,755         &        1,755         \\\\
Fixed Effects                 &        Time         & Region-Time         & &       Time         & Region-Time         \\ \\

\hline\hline 
\end{tabular}

%% file: tables/TableA4_AltcutoffWindfall50.tex
\begin{tabular}{l*{5}{c}} \\ 
\hline\hline &     &       &    &     &        \\
Treatment:
& \multicolumn{2}{c}{Actual Windfalls $>$ p50} & & \multicolumn{2}{c}{ Predicted Windfalls $>$p50} \\
\\
                  \cline { 2 - 3}\cline{5-6} \\
                                       &\multicolumn{1}{c}{(1)}         &\multicolumn{1}{c}{(2)}         & &\multicolumn{1}{c}{(3)}         &\multicolumn{1}{c}{(4)}         \\
                                       \\
\hline       \\
Windfalls $\times$ Post-Boom $\times$ SC        &       0.114\sym{**} &       0.131\sym{**} &   &   0.0804         &       0.105\sym{*}  \\
                    &    (0.0578)         &    (0.0623)        & &    (0.0562)         &    (0.0557)         \\
[1em]
Windfalls $\times$ Pre-boom $\times$ SC       &     -0.0817         &      0.0494         &    &  0.0397         &      0.0958         \\
                    &    (0.0950)         &    (0.0895)       &  &    (0.0871)         &    (0.0791)         \\
[1em]
Windfalls $\times$ Post-Boom       &      0.0124         &    -0.00922       &  &      0.0146         &     -0.0284         \\
                    &    (0.0494)         &    (0.0775)       &  &    (0.0440)         &    (0.0455)         \\
[1em] 
Windfalls $\times$ Pre-boom       &      0.0758         &      -0.137       &  &     -0.0255         &     -0.0716         \\
                    &    (0.0815)         &    (0.0995)      &   &    (0.0699)         &    (0.0660)         \\
[1em]                    
\hline \\
Adjusted R$^2$      &        0.41         &        0.41      &   &        0.41         &        0.41         \\\\
Observations        &      348,180         &      348,180       &  &      348,180         &      348,180         \\\\
\# Districts        &        1,493         &        1,493       &  &        1,493         &        1,493         \\\\
Fixed Effects                 &        Time         & Region-Time         & &       Time         & Region-Time         \\ \\

\hline\hline 
\end{tabular}

%% file: tables/TableA5_AltcutoffWindfall90.tex
\begin{tabular}{l*{5}{c}} \\ 
\hline\hline &     &       &    &     &        \\
Treatment:
& \multicolumn{2}{c}{Actual Windfalls $>$ p90} & & \multicolumn{2}{c}{ Predicted Windfalls $>$p90} \\
\\
                  \cline { 2 - 3}\cline{5-6} \\
                                       &\multicolumn{1}{c}{(1)}         &\multicolumn{1}{c}{(2)}         & &\multicolumn{1}{c}{(3)}         &\multicolumn{1}{c}{(4)}         \\
                                       \\
\hline       \\
Windfalls $\times$ Post-Boom $\times$ SC         &     0.00877         &       0.126      &   &     -0.0494         &      0.0838         \\
                    &    (0.0858)         &    (0.0972)       &  &     (0.161)         &     (0.141)         \\
[1em]
Windfalls $\times$ Pre-boom $\times$ SC       &      0.0209         &       0.118      &   &       0.131         &       0.229         \\
                    &     (0.157)         &     (0.149)         &  &   (0.229)         &     (0.224)         \\
[1em]
Windfalls $\times$ Post-Boom         &      0.0817         &     -0.0908      &   &      0.0276         &      -0.161\sym{***}\\
                    &    (0.0544)         &    (0.0631)     &    &    (0.0539)         &    (0.0570)         \\
[1em] 
Windfalls $\times$ Pre-boom        &      0.0324         &      0.0547      &   &      -0.121         &      -0.149         \\
                    &     (0.113)         &     (0.131)     &    &     (0.102)         &     (0.111)         \\
[1em]                    
\hline \\
Adjusted R$^2$      &        0.41         &        0.41         &  &      0.41         &        0.41         \\\\
Observations        &      348,180         &      348,180         & &     348,180         &      348,180         \\\\
\# Districts        &        1,493         &        1,493         &  &      1,493         &        1,493         \\\\
Fixed Effects                 &        Time         & Region-Time         & &       Time         & Region-Time         \\ \\

\hline\hline 
\end{tabular}

%% file: tables/TableB1_SCValidation.tex
\begin{table}[htbp]\centering
\def\sym#1{\ifmmode^{#1}\else\(^{#1}\)\fi}
\caption{State Capacity Validation. Dep. Var.: \% Windfalls that are Spent}
\label{tab:SC_validation} 
\begin{tabular}{l*{4}{c}}
\hline\hline \\
                    &\multicolumn{1}{c}{(1)}         &\multicolumn{1}{c}{(2)}         &\multicolumn{1}{c}{(3)}         &\multicolumn{1}{c}{(4)}         \\  \\
\hline
\\
State Capacity      &     7.0906$^{***}$ &	4.3049$^{***}$ &	4.3227$^{***}$	&1.9391$^{**}$    \\
              &     (1.1164) &	(1.3214)	&(0.8805)&	(0.9513)      \\
[1em]
Baseline Poverty   &  --  &	-0.0409$^{***}$& -- &	-0.0503$^{***}$\\
$\times$ Year	 & -- &	(0.0106)	& -- &	(0.0085) \\
[1em]
\hline \\
Observations	&777	&777&	777&	777 \\ \\
Subregion FEs & No & No & Yes         &         Yes         \\ \\
R$^2$ &	0.0495	&0.0674	&0.5401&	0.5603  \\\\
\hline\hline 
\end{tabular}
\vspace{-.2cm}
\begin{changemargin}{2cm}{2.1cm} 
\begin{spacing}{0.125}
{\footnotesize \setstretch{.125} Standard errors in parentheses. Sample includes districts with windfall transfers above the (population-weighted) 75-th percentile during the mining boom. }  
\end{spacing}
\end{changemargin}
\end{table}

%% file: tables/TableC1_EventStudy.tex
   
\begin{table}[p]
\footnotesize
    \centering
      \caption{Staggered Event Study with Windfalls}
\begin{tabular}{lccc} \hline
& (1) & (2) & (3) \\
 & HTE - High SC & HTE - Low SC & Diff. 2-3  \\ \hline
 &  &  &   \\
Pre  (4   years before onset or more)  & -0.0188 & 0.0781 & -0.0707 \\
 & (0.0371) & (0.0649) & (0.0714) \\
Pre  (2-3 years before onset) & -0.00108 & -0.0570** & -.05808\\
  & (0.0327) & (0.0289) & (0.0442)  \\
Post (0-1 years after onset)  & 0.0516*** & -0.0400 & 0.0938**  \\
  & (0.0195) & (0.0338) & (0.0390)  \\
Post (2-3 years after onset) & 0.0858*** & -0.0272 & 0.113** \\
  & (0.0225) & (0.0431) & (0.0473)  \\
Post (4   years after onset or more)  & 0.131*** & -0.000923 & 0.130**  \\
  & (0.0304) & (0.0467) & (0.0536)\\
 \\
Observations  & 138,153 & 41,370 & 179,523  \\
R-squared  & 0.284 & 0.359 & 0.323  \\\hline\hline
\end{tabular} \label{tab:stagcanon}
\vspace{-.2cm}
\begin{changemargin}{4.6cm}{4.7cm} 
\begin{spacing}{0.125}
{\footnotesize \setstretch{.125} $\sym{*}$ $p<0.10$, $\sym{**}$ $p<0.05$, $\sym{***}$ $p<0.01$. The dependent variable is log household income. Standard errors are clustered at the district level. Data based on the ENAHO household survey. Table is presenting event study estimates in which each district's ``event'' is the year in which any mine starts receiving windfalls above the 75th percentile (referred to as Q3 in Figure \ref{fig:staggered_hte}).}  
\end{spacing}
\end{changemargin}
 
 \end{table}    